# Nano-Clustering Mediates Phase Transitions in a Diastereomerically-Stabilized Ferroelectric Nematic System

Hiroya Nishikawa[1]*, Koki Sano[1], Saburo Kurihara[2], Go Watanabe[2], Atsuko Nihonyanagi[1], Barun Dhara[1] and Fumito Araoka[1]*

[1] RIKEN Center for Emergent Matter Science, 2-1 Hirosawa, Wako, Saitama 351-0198, Japan

[2] Department of Physics, School of Science, Kitasato University, 1-15-1 Kitasato, Sagamihara, Kanagawa 252-0373, Japan

**Correspondence and requests for materials** should be addressed to

H.N. (hiroya.nishikawa@riken.jp); F.A. (fumito.araoka@riken.jp).

Present address (K.S.): Department of Chemistry and Materials, Faculty of Textile Science and Technology, Shinshu University, 3-15-1 Tokida, Ueda, Nagano 386-8567, Japan

**Abstract**

During the last half-decade, a new class of ferroic-fluid, ferroelectric nematic liquid crystals ($N_F$LCs), creates a noise owing to its exceptional properties such as a colossal polarization, high electro-optic activity plus high fluidity. Regardless of recent huge efforts on design and development of new $N_F$LC molecules based on molecular parameters, the control of $N_F$ phase transitions and the stabilization of $N_F$ phase are still challenging. Here we discuss the impact of mixing of DIO diastereomer [$^{trans}$DIO (**1**) and $^{cis}$DIO (**2**)] to the $N_F$ phase transition, in terms of the smectic cybotactic cluster formation examined by X-ray diffraction. Interestingly, the result suggests that smooth exchange of $N_F$LC **1** by non-LC **2** both with similar dipole and molecular backbone plays a role in the alteration of the $N_F$ phase transition.

**Introduction**

Since a discovery of a true 3D-fluid ferroelectric nematic ($N_F$) phase showing the field-induced polarization reversal via a domain wall motion in liquid crystals (LCs), great interest has been dedicated to both fundamental science and applications[1–35]. In the $N_F$ phase, the inversion symmetry of the nematic (N) director is broken, forming the macroscopic domains with a uniform polarization (***P***) along the director, **n** (Fig. 1a). The striking features of the $N_F$LCs are large spontaneous polarization, gigantic dielectric permittivity[3,8,13,14,18,21,24,28], high nonlinear-optical activity[3,9,24,32], electro-optic response with very low or zero threshold voltage[8,12,13,15,17,20,22,29] as well as high fluidity[3,8,27] that may engender a paradigm shift in materials science and revolutionize soft matter technologies. According to the simple model by Born, the electric-dipolar interaction (is proportional to the square of dipole moment, $\mu$), which should be strong enough to withstand thermal fluctuation ($k_BT$), is the most important to emerge the $N_F$ phase.[35,36] Indeed, such a unique $N_F$ phase was ascertained in the specific molecules as a 1,3-dioxane-tethered fluorinated molecule (DIO)[3], a pear-shaped molecule (RM734)[1,2], a fluorinated bearing a terminal cyano group molecule (UUQU-4-N)[14] and other generics[13,21,24], whose $\mu$ are very large (> 9 Debye). However, although the present systemized molecular parameters including $\mu$ (another are the oblique dipole angle and the geometrical aspect ratio of molecular shape etc.) are confirmed for the $N_F$ emergence, there is no consideration of the effects on the phase sequence passing the $N_F$ phase. Archetypal phase behavior of the $N_F$ molecules is as follows: (1) DIO experiences three mesophases: the N–M–$N_F$ phase transition (M is antiferroelectric N, $SmZ_A$ or $N_s$ with periodic density modulation in the direction perpendicular to **n**; the nomenclature is under debate[21,25,34]); (2) RM734 exhibited the N–$N_F$ phase transition which corresponds with ferroelectric–ferroelastic phase transition as seen in inorganic solids[9]; (3) the latest key is UUQU-4-N showing the direct $N_F$ phase transition from isotropic liquid (IL), i.e., IL–$N_F$ phase transition, which has been observed in some molecules[21,24]. Additionally, the $N_F$ phase is thermodynamically unstable below room temperature, in most cases irrespective of a single molecule or LC blends. With consideration of results of single crystal X-ray diffraction[4,21,24], the adjacent two $N_F$ molecules with synparallel arrangement may flip each other, destructing the $N_F$ state and undergoing crystallization below room temperature. Under such a circumstance, despite huge efforts on design and development of new $N_F$LC molecules based on molecular parameters, the control of phase sequences involving the $N_F$ emergence and the successful way for stabilizing $N_F$ are still ambiguous. To note, there has just emerged a newfound ferroelectric state, a uniaxial ferroelectric smectic A ($SmA_F$) phase[37,38], which appears just below the $N_F$ phase, offering an important clue. Since the $SmA_F$ looks correlating highly with the $N_F$ phase, the smectic cluster floated in the $N_F$ media may exert influence on the characteristic of the $N_F$ phase.

In this work, we report a new concept based on the dipole equivalent for controlling the $N_F$ phase transition and their possible impact to the cluster formation. Besides, very recently, almost the same diastereomeric system was coincidently introduced by another group, and yet the mechanism behind this well-controlled $N_F$ phase is unknown[39]. However, in the present study, we focus on the nanoscopic structure in the present diastereomeric binary system and discuss the alteration of the $N_F$ phase transition in terms of cybotactic cluster formation. The result suggests that the exchange of the molecules with almost equivalent dipole moment and analogous structure takes place, without reducing the macroscopic polarization. For this strategy, although two components including a $N_F$ molecule are treated, the counter molecule need not be of a liquid crystalline. Note that this point is obviously unique compared to the conventional approach[22,40,41]. We herein adopt DIO diastereomer, $^{trans}$DIO (**1**) and $^{cis}$DIO (**2**), which are a ferroelectric nematogen and a non-liquid crystalline, respectively (Fig. 1b), to investigate the effect of diastereomeric-controlling and structure of $N_F$ phase in view of X-ray diffraction (XRD) analysis and computer simulation.

**Results and Discussion**

**Control of the variant $N_F$ phase transition in $^{trans/cis}$C3DIO system.** A stereoisomer with an equivalent topological structure with yet different geometric configuration is a.k.a. diastereomer[42]. C3DIO has also a diastereomer because 1,3-dioxane moiety with alkyl chain in C3DIO can have two geometric configurations. Such a C3DIO diastereomer, $^{trans}$C3DIO (**1**) and $^{cis}$C3DIO (**2**), exhibit different physical properties (Fig. 1b and Supplementary Figs. 1–6 for the characterization data). **1** is typical ferroelectric nematogen and exhibits the ferroelectric nematic ($N_F$) phase on cooling[3] whereas **2** is not a LC material despite similar large dipole moment to **1** ($\mu_1$ = 9.36 D; $\mu_2$ = 9.04 D. See also Supplementary Fig. 7) but interestingly a polar crystal instead of the apolar crystal in the case of **1** (the single crystal X-ray crystallographic structures of **1** and **2** are Supplementary Fig. 8). If the diastereomeric ratio (*dr*) of **2** in **1/2** mixture is increased (i.e. exchanging **1** by **2**), intuitively, the $N_F$ phase is expected to be destabilized because the molecule (**2**) may collapse the strong dipole-dipole interaction between **1**–**1** molecule. Fig. 1c and Supplementary Fig. 9 shows the polarizing optical microscope images via the variant phase sequence in the diastereomeric mixture (**1/2**). For *dr* (**1/2** = 100/0), a uniform texture (in green-highlighted area) was changed to the inhomogeneous one (in blue-highlighted area), and finally to the stripe texture due to the polar defect, e.g. 2π twist-wall[12] (in yellow-highlighted area) during cooling. These three phases are characterized as the N, M and $N_F$ phases in the order of high to low of temperature. With increasing *dr*, the intermediate M phase was vanished at *dr* = 70/30. Thus, the mixture with *dr* (70/30) experienced the N–$N_F$ phase transition as in the common generic $N_F$ molecules including RM734[1,3]. Further increment of *dr* (> 60/40) eliminated the N and M phases, instead, the direct phase transition between an isotropic liquid and $N_F$ phase was occurred. For instance, in case of *dr* = 60/40, small droplets of the $N_F$ phase emerged in the isotropic liquid and conglomerated each other, resulting in the complete formation of $N_F$ phase. This direct IL–$N_F$ transition type is a scarce case which has been observed in UUQU-4-N etc.[14,21,24] Thus, against expectation, the doping of **2** probably contributes to the control of the phase transition related to the stability of the $N_F$ phase. To explore the stabilization of the $N_F$ phase by the diastereomeric control of DIO, the phase diagram of **1**/**2** system was constructed (Fig. 1d). At *dr* = 90/10, the temperature range of the M phase was reduced, and instead the $N_F$ phase regime was expanded. Furthermore, the $N_F$ phase regime expanded to a lower temperature range across room temperature. This trend was observed increasing doping level of **2** up to *dr* = 70/30, yielding the maximum temperature range of the $N_F$ phase. Interestingly, at *dr* = 70/30, the $N_F$ phase still stand at 0 °C and the wide regime was 80 °C, which is approximately four times larger than that of pure DIO (**1**). More interestingly, we found that the mixture with *dr* = 70/30 exhibited the enantiotropic $N_F$ phase, which was thermodynamically stable, unlike the monotropic $N_F$ phase in **1** (Fig. 1e). Similarly, at the regime with high doping level of **2** (*dr* ≥ 40/60), the $N_F$ phase persisted wide temperature range (~ 50–20 K) across room temperature. The dielectric permittivity of the $N_F$ and N phases in **1**/**2** system were comparable to those of pure DIO; $N_F$ and N phases showed dielectric permittivity of the order of $10^4$ and 10 at a frequency of 1 kHz, respectively (Fig. 1f, details discussed later). The complete DSC curves are shown in Supplementary Fig. 9 and the corresponding enthalpies were summarized in Supplementary Table 1.

**Structure analysis of the diastereomeric-controlled $N_F$ phase.** As mentioned above, the diastereomeric combination probably alters the nature of the $N_F$ phase transition. Here, we discuss how the doping level of **2** in **1** has an effect on the structure of $N_F$ phase mainly with the aid of X-ray diffraction analysis. Fig. 2a displays a 2D wide-angle XRD pattern of the $N_F$ phase ($T-T_c = -10$ °C) in **1** under the magnetic field ($B \sim 0.5$ T) (The complete data are shown in Supplementary Fig. 10). The unique diffraction patterns consist of: (i) a pair of skewed peaks (corresponds with the molecular length, ~ 2.2 nm) at a small-angle region on the equatorial direction (parallel to **n**); (ii) a series of weak overtone ones spanning the small- to wide-angle region on the equatorial direction; (iii) the halo peaks due to the intermolecular stacking on the meridional direction (normal to **n**). To analyzed the XRD profile in detail, the horizontal and vertical scan were carried out within the angle $\varphi_1 = 80°$, $\varphi_2 = 60°$, generating the 1D XRD pattern. For of the fitted halo/skewed peaks (Supplementary Figs. 11 and 12), the relative intensity ($I_{rel}$) and the full width at half maximum (FWHM) in $q$-space (i.e., $q_{FWHM}$) as a function of $dr$ are shown in Fig. 2b and 2c. For the halo peak, the all $q_{FWHM}$ was comparable while the intensity was slightly decreased with increasing doping level of **2**. On the other hand, the $q_{FWHM}$ of the skewed peaks slightly decreased as $dr$ increased. Notably, the intensity showed a significant dependence on $dr$. It suggests that the doping of **2** may promote the growth of cybotactic cluster in the $N_F$ phase. In addition, the peak separation on the equator direction was performed, providing the six peaks (Fig. 2d). The data of the primary peak (p1) are displayed in Fig. 2c. As shown in Fig. 2e and 2f, for the $q_{FWHM}$ and intensity of the peaks (p2–p6), although both were on a downward trend at $dr = 60/40$, there was no remarkable difference on any $dr$. Let us consider the unusual anisotropic diffraction pattern in the $N_F$ phase of **1/2** system. For the diffraction on (i) (vide supra), the skewed peak is reflected on the normal cybotaxis, in which SmC-like stratification within clusters of mesogens, floating in the N phase[43,44]. The diffraction on (ii) is maybe due to the anomalously large correlation length in cybotactic clusters, which has been observed in the N phase in a rigid lath molecule[45]. Notably, Mandle et al. confirmed that a series of weak overtone observed in the $N_F$ phase of RM734 was a consequence of polar nematic order with the aid of MD calculation[18]. By considering these important tips, in **1**/**2** system, the SmC-type cybotactic cluster may coexist in the $N_F$ phase (Fig. 3a). In that case, the anomalous changes in intensity of diffractions may indicate growth of the size of clusters with polar order, depending on the doping level of **2**.

To gain more information on the cluster in the $N_F$ phase in detail, we analyze 2D small-angle XRD pattern in the $N_F$ phase. Fig. 3a shows the schematic illustration connecting the structural parameters of a SmC-type cybotactic domain. The magnetic field aligns the clusters, in which the director n is parallel to **B**. In the magnetic field, the normal to the smectic layer, *k*, is randomly distributed on **B** at the tilt angle *β*. Consequently, the XRD profile resulting from such distribution of microscopic SmC-type clusters is observed (Fig. 3b and Supplementary Figs. 13–20). According to reference.40,41, we estimated the average size of the cybotactic clusters in the $N_F$ phase and its temperature and *dr* dependence. The average size of the clusters can be estimated from the longitudinal (∥ **B**) and transversal (⊥ **B**) intensity profiles of the skewed (four-spot) pattern (Fig. 3b). Fig. 3c shows an example of diffraction intensity as a function of $\Delta q_{\parallel,\perp} = \pm|q-q_{\parallel,\perp}|$ of the **1**/**2** mixture with *dr* (70/30). The full width at half maximum of $\Delta q_{\parallel,\perp}$ correlates with the anisotropic short-range positional order of the cluster, which can be characterized by the correlation length $\xi_{\parallel,\perp}$, i.e., $\xi_{\parallel,\perp} = \alpha/q_{\text{FWHM }\parallel,\perp}$, where $q_{\text{FWHM}}$ is full width at half maximum at the H-/V-scan profiles and $\alpha = 2$ (for Lorentzian fitting case) (Fig. 3b and 3c). Besides, the longitudinal and transversal dimension of the cluster, $L_\parallel$ and $L_\perp$ are given as $L_\parallel = 3\xi_\parallel$ and $L_\perp = 3\xi_\perp$. Fig. 3d shows the temperature evolution of the small-angle XRD pattern with various *dr*. The four-spot pattern emerged below the critical point ($T_c$) for **1**/**2** mixtures with various *dr* (100/0–50/50), indicating the presence of the SmC-type cybotactic cluster over the entire range of the $N_F$ phase. With increasing doping level of **2**, the contrast of the four-spot tends to be strong and its intensity drastically increased. From H-/V-scan profiles **1**/**2** mixtures with various $dr$ (100/0–50/50), we estimated the corresponding $L_\parallel$ and $L_\perp$ as a function of temperature which are shown in Supplementary Fig. 21. For the **1/2** mixtures with high $dr$, $L_\parallel$ and $L_\perp$ monotonically increase with similar slope toward low temperature whereas the value of $L_\parallel$ elevated significantly compared to $L_\perp$ in a series of **1/2** with low $dr$ (> 65/35). The fact indicates that the growth anisotropy of the cluster ($L_\parallel/L_\perp$) should be different with respect to $dr$ of **1/2** mixtures. Fig. 3e clearly evidences the striking difference in the grow process of the cluster. In case A, up to $T-T_c \sim -30$ °C, the cluster almost isotropically grew and then tended to grow transversally beyond the temperature. On the contrary, in case B, a tendency of longitudinal growth of the cluster was observed over the entire range of the $N_F$ phase. It is noted that the case B occurred in the **1/2** mixture with $dr$ (> 65/35), which experienced the direct IL–$N_F$ phase transition. A similar tendency was observed in UUQU-4-N exhibiting the direct IL–$N_F$ phase transition, in which SmA-like cybotactic cluster coexists in the $N_F$ phase (Supplementary Fig. 22). Thus, the cybotactic cluster in a series of molecules showing the direct IL–$N_F$ phase transition may grow inherently, irrespective cybotactic cluster types via the case B. Next, we investigated the internal molecular environment in the cluster. The left panel of Fig. 4a shows a schematic illustration of the SmC-like stratification within the cluster in the $N_F$ phase with the $dr$ variation at a fixed temperature ($T-T_c = -10$ °C). At $dr$ = 100/0, the constituent molecules (**1**) arranged in a smectic layer with tilt angle $\beta$ = 24.9° are occupied in the cluster. With increasing doping of **2**, substitution of **1** leads to generation of the scrambled dipoles of **1** and **2** with changing in the cluster anisotropy (see Fig. 4b, discuss later). At $dr$ = 50/50, the majority of cluster is colonized by the equimolar molecules. Although $\beta$ increased nonlinearly by cooling, reflecting on the SmC feature at each $dr$, at a fixed temperature, the value of $\beta$ was reduced slightly up to 19.5° with $dr$ increases (Supplementary Fig. 23). With considering the volume of the cybotactic cluster (i.e. $L_\parallel \times L_\perp \times L_\perp$), we estimated the average molecular numbers in the cluster of a series of **1/2** mixture. The average molecular numbers in the longitudinal dimension ($N_\parallel$) or transversal one ($N_\perp$) are given as follows:

$N_\parallel = L_\parallel / l = 3\xi_\parallel / l$

$N_\perp = L_\perp / w = 3\xi_\perp / w$

, where $l$ and $w$ are the molecular length (ca. 2.23 nm) of DIO and typical intermolecular distance (ca. 0.47 nm), respectively. Thus, the average molecular numbers ($\bar{N}$) in the cluster are expressed as $N_\parallel \times N_\perp \times N_\perp$. For instance, in case of **1** at $T-T_c = -10$ °C, $\bar{N}$ were calculated to be 1036, i.e., one thousand molecules are occupied in the cluster. The estimated $\bar{N}$ for all **1/2** series as a function of temperature are summarized in Fig. 4c. At low $dr$ regime (< 70/30), the $\bar{N}$ increased monotonically and reached to be approximately two thousand. By contrast, the **1/2** series with high $dr$ regime ($\geq$ 70/30) showed tendency of the nonlinear increment of $\bar{N}$, reaching to ca. 2,000–4,000. The relationship at $T-T_c = -10$ °C is highlighted in Fig. 4d. The value of $\bar{N}$ increased at the threshold of $dr$ (85/15), being 1.5–2 times at maximum $dr$. Interestingly, the $dr$ dependence of enthalpy related with phase transition via the $N_F$ phase showed quite similar relationship between $\bar{N}$ and $dr$ (Fig. 4e). It suggests that the anomalous latent heat have a strong association with the formation of the cybotactic cluster and its size. Besides, we found out that at threshold of $dr$ (70/30), two regimes (IL–N–M–$N_F$ or IL–(N)–$N_F$) were classified. The threshold of $dr$ was good agreement with $dr$ classifying cases, the trend of cluster growth as shown in Fig. 3e.

To assess the validation of the cluster model in the $N_F$ phase for **1/2** system, we performed molecular dynamics (MD) simulation using all-atom models using GROMACS 2020.5. Here, we simulated four systems, the compound **1** arranged all in parallel (**System 1**) and half in anti-parallel (**System 2**), and 50:50 compositions of **1** and **2** all in parallel (**System 3**) and half in anti-parallel (**System 4**). The initial structures were built by six layers, each of which contains 100 molecules, and hence the total number of the molecules in the simulation box was 600 (See Supplementary Note 1 for the detail). Fig. 4b compares the time-averaged total energy for these four systems and the corresponding MD snapshots in the $N_F$ regime ($T-$ $T_c$ = −15 °C). The calculated energies ($G$) of **Systems 1** and **2** were ~ 6087 kcal mol$^{-1}$ and ~ 7285 kcal mol$^{-1}$, respectively. Thus, the parallel (polar) orientation may predominately occur in the **1**-only system. This result corresponds with previous reports.[8,18] On sharp contrast, the time-averaged total energies of **Systems 3** and **4** were extremely lower, that is, $G_{(System\ 3)}$ = −6879 kcal mol$^{-1}$ and $G_{(System\ 4)}$ = −6306 kcal mol$^{-1}$, so that it is natural to consider that the doping of **2** stabilizes the system. Interestingly, by closely looking at the MD simulation snapshots, we see pairing of **1** and **2**, meaning plausible dimerization (Supplementary Fig. 33), which is confirmed also by the differences of the peaks on the two-dimensional radial distribution function profiles of **1** only, **2** only, and both **1** and **2** in the one layer for **System 3** (Supplementary Fig. 37). In addition, the possibility of dimerization is discussed by the density-functional theory (DFT) calculation[46] (Supplementary Note 1 and Supplementary Fig. 30). A rough estimate of the energetic difference of the pair interaction in **System 3** and **4** gives 573/300 ~ 2 kcal mol$^{-1}$, which is comparable with the thermal agitation energy above the room temperature (e.g. $n_A k_B T$ ~ 2.4 kcal mol$^{-1}$ at 300 K). Since the total energy in MD tends to be reduced due to the size limitation of the simulation box, it looks reasonable that the polar orientation of **1**–**2** dimers in a cluster was energetically preferable than the apolar **1**–**2** configuration. It is noteworthy that the tilt angles ($β$) of **System 1** and **System 3** were well accorded with them obtained by X-ray analysis (Fig. 4b and Supplementary Fig. 23). Hence, the results of MD simulation strongly support our model of the polar configuration.

**Polarization behavior of the diastereomeric-controlled N$_F$ phase.** As early prediction, the replacement of N$_F$ molecule (**1**) with non-LC molecule (**2**) probably causes destruction of the strong dipole-dipole interaction between **1**–**1** molecule so that the N$_F$ phase should be destabilized. Contrary to expectation, by diastereomeric control using **2**, we succeeded in realizing the N$_F$ state operating over an extended temperature range from 80 °C to 0 °C and control the various phase transition sequence, as mentioned above. If the diastereomeric-controlled $N_F$ phase is truly stabilized state, the corresponding polarization properties also should be equal to or greater than original ones. Therefore, we investigated the polarization behavior of the $N_F$ phase **1/2** system by dielectric relaxation, *P-E* hysteresis as well as SHG studies. The comparable temperature was fixed to be −10 °C of $T-T_c$. Fig. 5a shows the dielectric permittivity as a function of frequency of **1/2** mixture with various *dr*. All mixtures exhibited similar order of the dielectric permittivity (i.e. $10^4$) and the corresponding relaxation peak moved toward low frequency range (complete data are shown in Supplementary Figs. 24 and 25). It is noted that with increasing doping level of **2**, although the relaxation frequency ($f_r$) decreased, the order of the dielectric strength was maintained (Fig. 5b). The *dr* dependence of frequency is discussed in Supplementary Note 2. For *P-E* hysteresis variation, a typical parallelogram-like P-*E* loops, which often appears in ferroelectrics, were obtained in all mixtures (Fig. 5c). The corresponding coercive electric field was increased toward high *dr*. Notably, polarization density ($P \sim 4$ μC cm$^{-2}$) of the **1/2** mixture was increased and the maximum *P* was marked at *dr* = 70/30 (Fig. 5d). Complete data on *P-E* hysteresis are summarized in Supplementary Fig. 26. The SHG is quickly recognized as a powerful tool to confirm the macroscopic polar order and its symmetry of the individual nanostructures. The SHG as a function of temperature under the electric field (0.7 V μm$^{-1}$) for all mixtures are shown in Fig. 5e (The optical setup is shown in Supplementary Fig. 27). For all mixtures, the high SHG activity was observed in the range of $N_F$ phase. For example, the SHG profile at *dr* = 70/30 was set as a master curve, the other SHG profiles were coincided well with the master curve. The *dr*-dependent SHG of all mixtures is summarized in Fig. 5f. The SHG intensities at *dr* ≥ 90/10 were larger than that at *dr* = 100/0. Hence, the combined results provide evidence that the diastereomeric-controlled $N_F$ phase is surly stabilized $N_F$ state without sacrificing polarization behavior. The counter molecule **2** against **1** plays an important role as a $N_F$ phase stabilizer.

Finally, we demonstrated that the potential of a stabilizer (**2**) for boosting temperature range of the $N_F$ phase even in another host $N_F$ molecule. Here, we selected transC4DIO (**3**) exhibiting a monotropic $N_F$ phase of which temperature range is very narrow (~ 7 K) (Fig. 5g, 5h and Supplementary Fig. 28a). The molecule **3** was blended with a stabilizer **2** to yield the **3/2** mixture with *dr* (=70/30). As a result, the temperature range of the $N_F$ phase was extremely extended from 7 K to 40 K across room temperature. Surprisingly, the $N_F$ phase was stably operated with temperature range of 26 K on heating, i.e. enantiotropic $N_F$ phase (Fig. 5g, 5h and Supplementary Fig. 28b). The characterization of $N_F$ phase in the **3/2** mixture with *dr* (=70/30) are summarized in Supplementary Fig. 29.

In conclusion, in this work, we demonstrated alteration of the $N_F$ phase transition and nano-cluster formation in the diastereomeric mixture of $^{trans}$DIO (**1**) and $^{cis}$DIO (**2**). In the **1/2** mixtures with various diastereomeric ratio, we successfully tuned three types of the phase sequences passing the $N_F$ phase: 1) IL–N–M–$N_F$, 2) IL–N–$N_F$ and 3) IL–$N_F$ phase transitions. The **1/2** mixture with *dr* = 70/30 exhibited the enantiotropic $N_F$ phase with wide temperature range (~50 K) across room temperature. The XRD analysis of **1/2** mixture clearly evidenced that the cybotactic cluster existed in the $N_F$ phase and the anisotropic growth of the cluster dominated the phase sequence types. The computer simulation clarified that in the nano-cluster, the substitution of **1** with **2** without energetic penalty in the polar configuration was allowed and **1**–**2** (parallel polar arrangement) was the energetically most stable configuration in another pair case. Furthermore, the combination of dielectric, *P-E* hysteresis as well as SHG studies confirmed that the diastereomeric-controlled $N_F$ phase had similar macroscopic polarization, indicating that the replacing **1** with **2** stabilized the $N_F$ phase without sacrificing its macroscopic polarization. This model was also adapted in the monotropic $N_F$ molecule, transC4DIO collaborating with **2**, giving rise to the enantiotropic nature and an extended temperature range of the $N_F$ phase (24 K) on heating. We believe that this new approach allows new scope for $N_F$ matter engineering and/or new theories to be explored and conceptualized.

**Method**

**Fabrication methods for liquid crystalline (LC) cells.** Sandwich-type electrical cell (5 and 13 μm-thickness): Pre-treated ITO-coated glass plates (EHC model D-type, electrode area: 5 × 10 mm$^2$) were silanized with a silane coupling reagent (octadecyltriethoxysilane, TCI) at 120 °C for 2 hours, and then were rinsed with EtOH and ultrapure water. The two resulting glass plates were fixed with drops of an UV-curable glue using polymeric beads (micropearl, SEKISUI) as a spacer. The cell gap was estimated by capacitance of the empty cell. For SHG measurement, we used ITO-coated glass plates (GEMOATEC, electrode area: 4 × 5 mm$^2$), which was pre-baked at 400 °C for 1 hour in the electronic furnace (SUPER100T, SHIROTA) prior to treatment.

**Preparation and measurement methods for LC cells.** Preparation of LC mixtures (**1/2** or **3/2**): the protocol is as follows: 1) a compound was added in a vial (6 mL) and precisely weighted using an electronic balance (MSE2.7S, sartorius); 2) a chloroform was added into this vial, preparing a mother liquor; 3) Two kind of mother liquors (chloroform solution of **1** (or **3**) and **2**) were mixed with appropriate concentration using an electronic micro-pipet (eVol, SGE Analytcical Science); 4) the solution mixture was vortexed and then evaporated at 40 °C in a jet oven overnight; 5) the resulting mixture was dried in vacuo at room temperature for 3 hours and then mechanically stirred at 120 °C by a magnetic stirrer followed by cooling to room temperature.

For experiments: The LC mixture was injected by capillary action into a LC cell (5 μm- and 13 μm-thickness).

**Polarized optical microscopy.** Polarized optical microscopy were performed on a polarizing microscope (Eclipse LV100 POL, Nikon) with controlling the temperature using a temperature controller and a hot stage (mK2000, INSTEC). Unless otherwise noted, the sample temperature was controlled using INSTEC model mK2000 temperature controller and a liquid nitrogen cooling system pump (LN2-P/LN2-D2, INSTEC).

**Differential scanning calorimetry.** Differential scanning calorimetry (DSC) was performed on a calorimeter (DSC30, Mettler-Toledo). Cooling/heating profiles were recorded and analyzed using the Mettler-Toledo STARe software system.

**Dielectric spectroscopy.** Dielectric relaxation spectroscopy was performed ranging between 1 Hz and 1 MHz using an impedance/gain-phase analyzer (SI 1260, Solartron Metrology) and a dielectric interface (SI 1296, Solartron Metrology). Prior to starting measurement of the LC sample, the capacitance of the empty cell was determined.

*P-E* **hysteresis measurements.** *P-E* hysteresis measurements were performed in the temperature range of the $N_F$ phase under a triangular-wave electric field (10 kV cm$^{-1}$, 200 Hz) using FCE system (TOYO Corporation), which equipped with an arbitrary waveform generator (2411B), a IV/QV amplifier (model 6252) and a simultaneous A/D USB device (DT9832).

**SHG measurement.** The SHG investigation was carried out using a Q-switched Nd: YAG laser (FQS-400-1-Y-1064, Elforlight) at $\lambda = 1064$ nm with a 5 ns pulse width (pulse energy: 400 µJ) and a 10 kHz repetition rate. The primary beam was incident on the LC cell following by the detection of the SHG signal. The electric field was applied normal to the LC cell. The optical setup is shown in Supplementary Fig. 27.

**Wide- and small-angle X-ray scattering (WAXS, SAXS) analysis.** Two-dimensional WAXS and SAXS measurements were carried out at BL38B1 in the SPring-8 synchrotron radiation facility (Hyogo, Japan). The samples held in a glass capillary (1.5 mm in diameter) were measured under a magnetic field at a constant temperature using a temperature controller and a hot stage (mk2000, INSTEC) with high temperature-resistance neodymium magnets (~ 0.5 T,

MISUMI). The scattering vector $q$ ($q = 4\pi\sin\theta\ \lambda^{-1}$; $2\theta$ and $\lambda$ = scattering angle and wavelength of an incident X-ray beam [1.0 Å (for WAXS) and 0.95 Å (for SAXS)], respectively) and position of an incident X-ray beam on the detector were calibrated using several orders of layer diffractions from silver behenate ($d$ = 58.380 Å). The sample-to-detector distances were 2.5 m (for WAXS) and 0.29 m (for SAXS), where acquired scattering 2D images were integrated along the Debye–Scherrer ring by using software (Igor Pro with Nika-plugin), affording the corresponding one-dimensional profiles.

**Data availability**

The authors declare that the data supporting the findings of this study are available within the paper and its supplementary information files. All other information is available from the corresponding authors upon reasonable request.

**Acknowledgements**

The authors are grateful to Dr. T. Hikima (RIKEN, Spring-8 Center) for supporting XRD measurement and Dr. H. Sato (RIKEN, CEMS) for allowing us to use a QTOF compact (BRUKER). We wish to thank Mr. M. Kuwayama (RIKEN, CEMS) for the QTOF-HRMS measurement, and Dr. D. Miyajima (RIKEN, CEMS), Dr. Y. Sasaki (Hokkaido University), Mr. Z. Li (RIKEN, CEMS), and Mr. A. Manabe (ex. Merck KGaA for fruitful discussions. We would like to thank Merck KGaA for sample contribution (UUQU-4-N). This work was supported by JSPS KAKENHI Grant Numbers JP19K15438 (H.N.), JP19H02537 (G.W.) and JP21H01801 (F.A.), Incentive Research Projects in RIKEN (No. 100689; H.N.) and JST CREST (Grant Number JPMJCR17N1; F.A.), and JST PRESTO (Grant No. JPMJPR20A6; K.S.). The small-/wide-angle X-ray scattering measurements were performed at BL38B1 in SPring-8 with the approval of the RIKEN SPring-8 Center (proposal 20210080). The computations were partially performed at the Research Center for Computational Science, Okazaki, Japan (Project: 21-IMS-C043, 22-IMS-C043).

**Author Contributions**

H.N. conceived the project and designed the experiments. F.A. co-designed the experiments. H.N. performed all the experiments. K.S., and F.A supported XRD measurements. S.K. and G.W. performed MD and DFT calculation. A.N. synthesized all compounds. B.D. measured and analyzed single crystal XRD. H.N., K.S., and F.A. analyzed data and discussed the results. H.N. and F.A wrote the manuscript and all authors approved the final manuscript.

**Competing Interests**

The authors declare no competing interests.

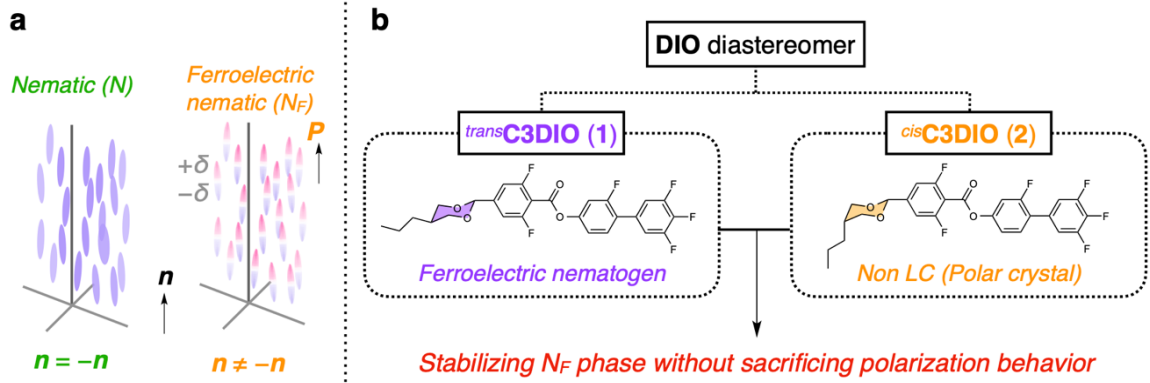

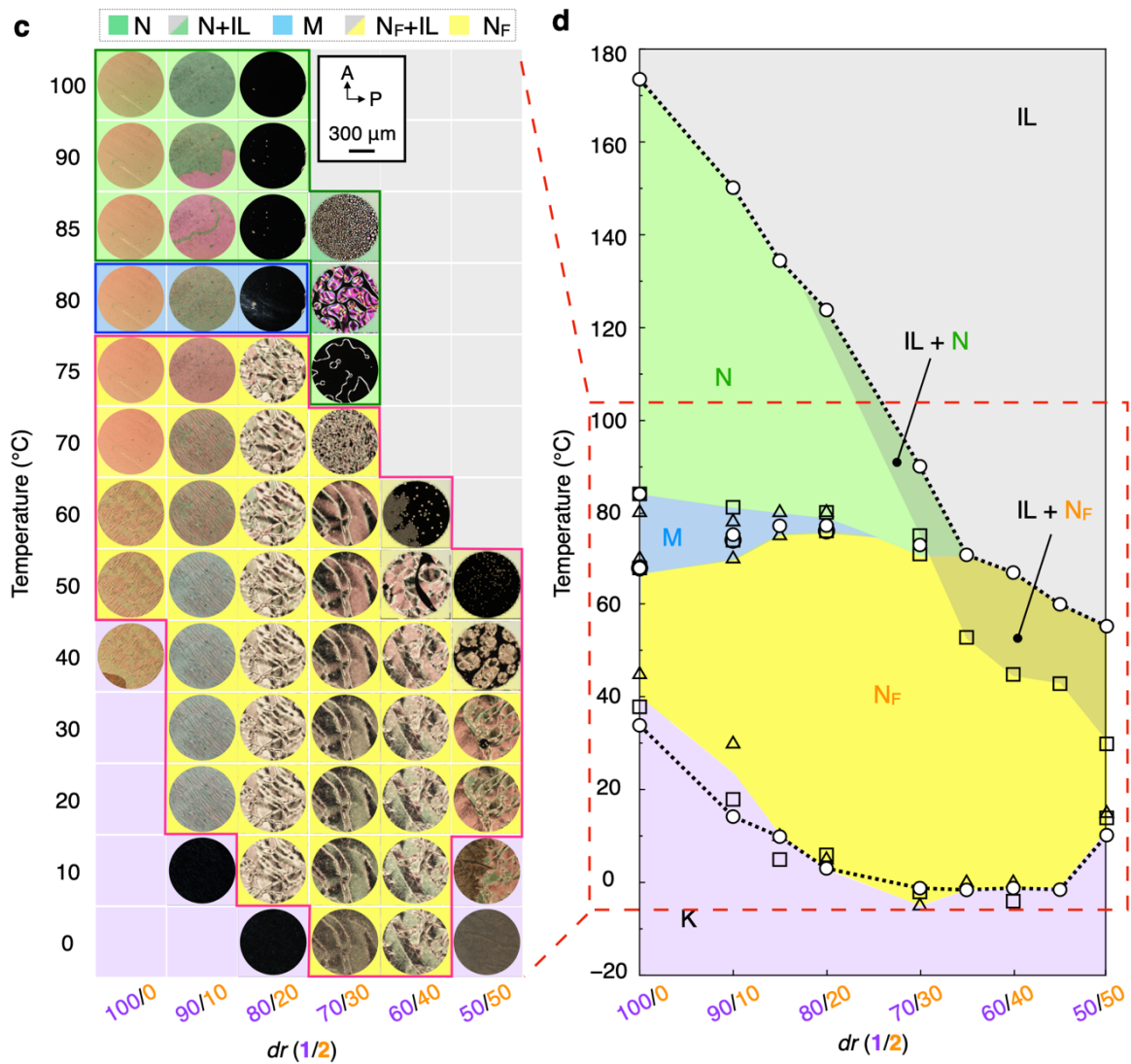

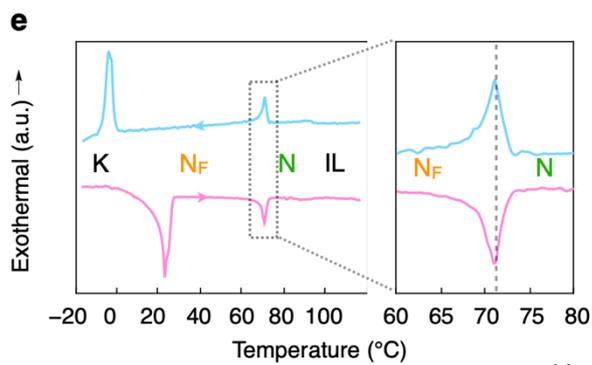

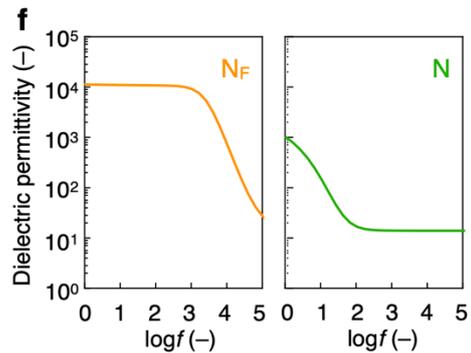

**Fig. 1 | Diastereomeric-controlled ferroelectric nematic system. a**, Schematic illustration of nematic and ferroelectric nematic phases. **b**, Chemical structure of DIO diastereomer: $^{trans}$C3DIO (**1**) and $^{cis}$C3DIO (**2**). **1** and **2** has dipole moment of 9.36 and 9.04 D, respectively. **c**, Evolution of polarized optical microscope images, which were taken under the crossed polarizers, of the mixture **1/2** with various *dr* (100/0–50/50). Thickness: 13 μm. **d**, A phase diagram of the mixture **1/2** as a function of *dr* and temperature. Symbols represent the phase transition temperature determined by DSC (circle), POM (square) and XRD (triangle) studies. Abbrev.: IL (isotropic liquid); N (nematic); M (mesophase); $N_F$ (ferroelectric nematic); K (crystal). The coexistence area is indicated by gray shadow. **e**, DSC curves of the mixture **1/2** with *dr* (70/30) on cooling (upper line) and heating (bottom line). The bottom panel denotes the enlarged temperature range between 60–80 °C, in which a good match of temperatures due to the $N_F$–N phase transition (ca. 72 °C) on cooling and heating because of a very weak 1st order or 2nd order phase transition. Temperature recorded on DSC was calibrated using a reference (8CB) exhibiting 2nd order phase transition (SmA–N). **f**, Dielectric permittivity as a function of frequency in $N_F$ phase (44 °C, left) and N phase (120 °C, right) for the mixture **1/2** with *dr* (90/10).

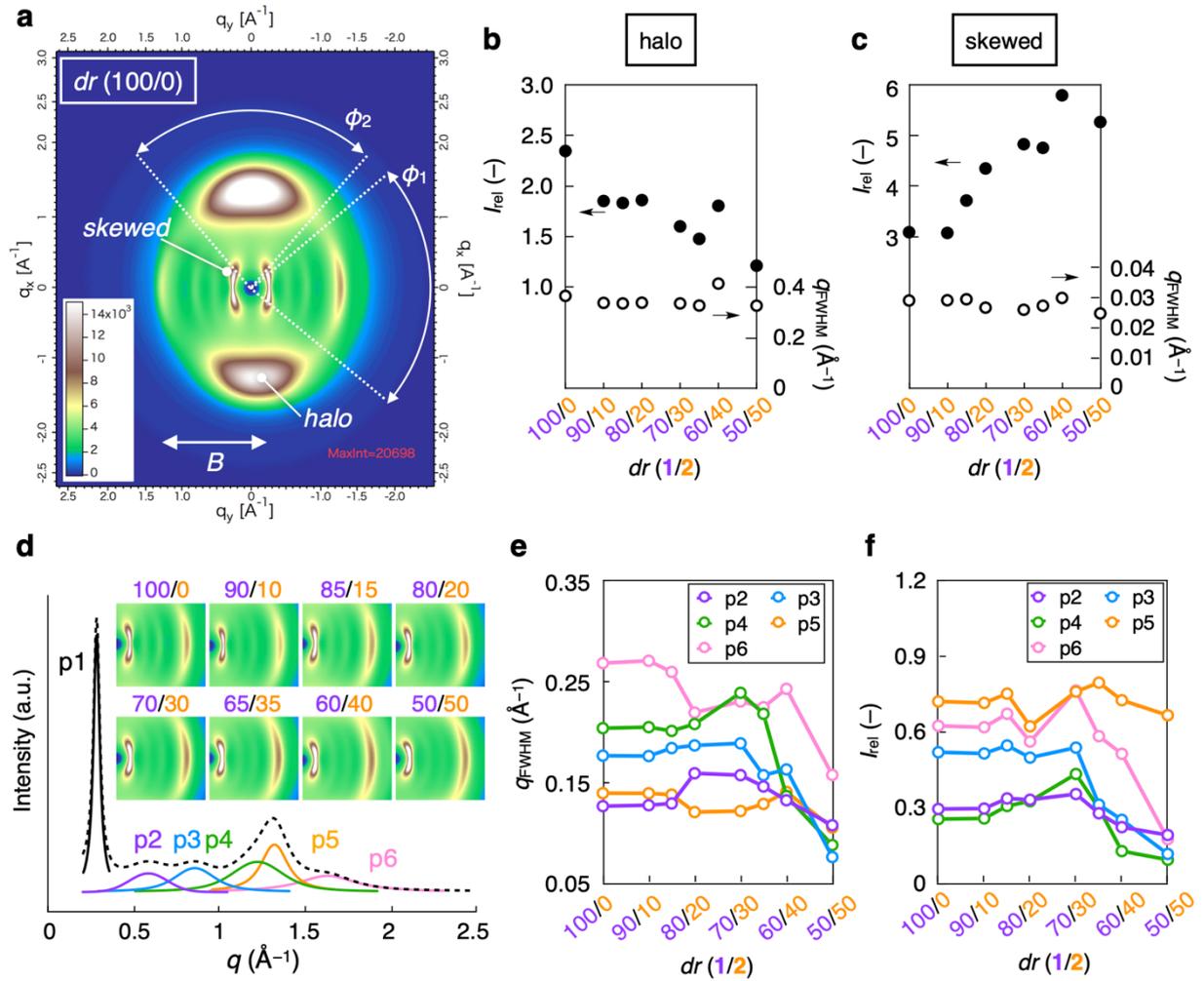

**Fig. 2 | Wide-angle XRD analysis of the 1/2 mixture with various *dr*. a**, Wide-angle 2D XRD profile of the **1/2** mixture (*dr* = 100/0). The azimuth angle $\varphi_1 = 80°$, $\varphi_2 = 60°$ represents the scanning range for generation of the 1D XRD profile. The value of $q_{\text{FWHM}}$ and relative intensity ($I_{\text{rel}}$) as a function of *dr* for halo (**b**) and skewed peaks (**c**). $I_{\text{rel}}$ was calculated by dividing $I$ by $I_{\text{IL}}$ for halo and skewed peaks. **d**, 1D XRD profile of the **1/2** mixture (*dr* = 70/30) with separated peaks (p1–p6). Insets denote the analyzed area for all entries (*dr* = 100/0–50/50). The value of $q_{\text{FWHM}}$ (**e**) and relative intensity ($I_{\text{rel}}$) (**f**) as a function of *dr* for multiple peaks (p2–p6). $I_{\text{rel}}$ was calculated by dividing $I_{\text{obs.}}$ by the corresponding $I_{\text{IL}}$.

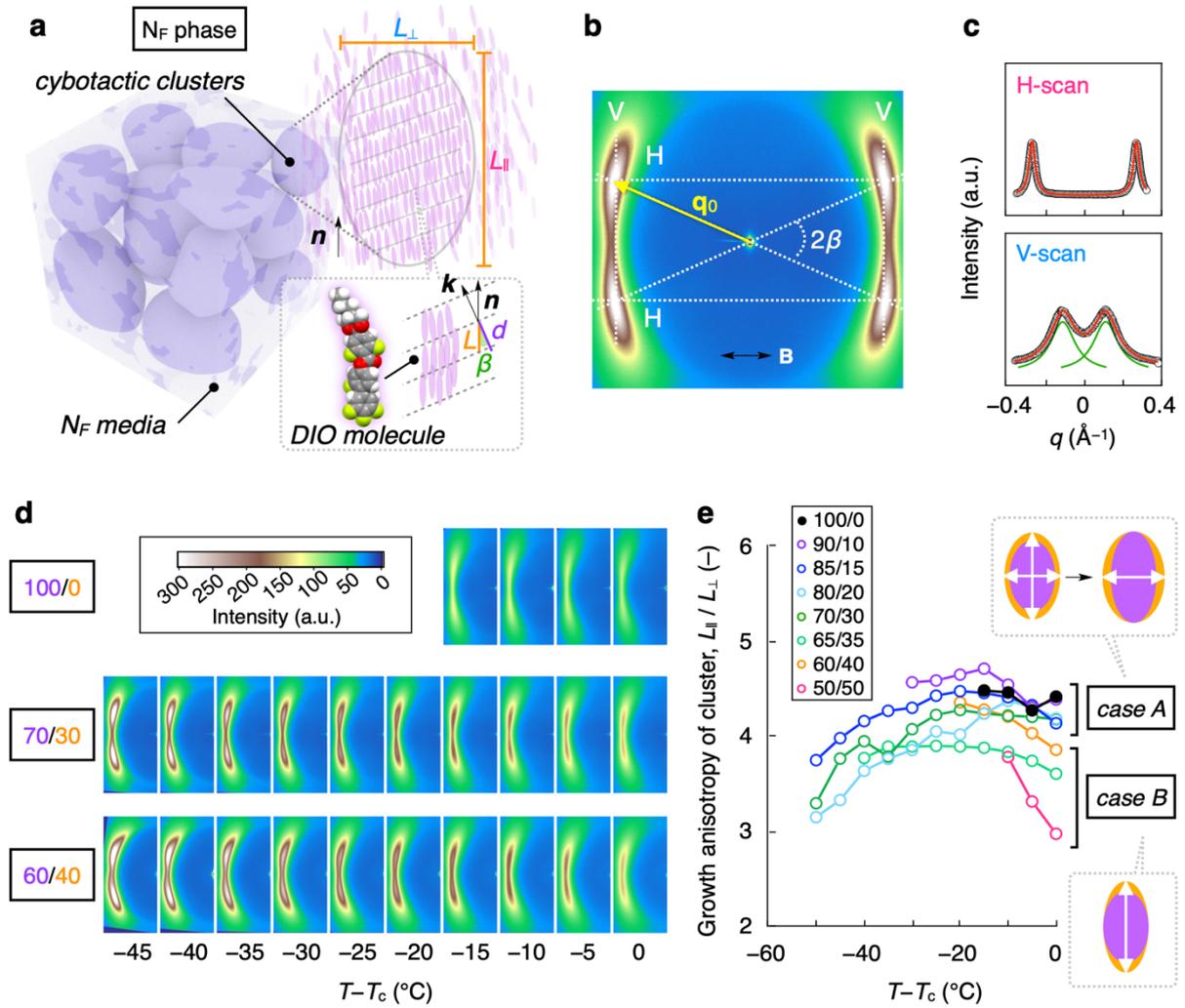

**Fig. 3 | Characterization of cybotactic cluster in the diastereomeric-controlled $N_F$ phase.**
**a**, Schematic illustration of the $N_F$ phase of **1** or **1/2** mixture. $L_\parallel$ and $L_\perp$ are longitudinal and transversal dimension of the cybotactic cluster, respectively. $L$, $d$, $\beta$ denote the molecular length, $d$-spacing between layers and tilt angle, respectively. **b**, Small-angle XRD profile. A white-colored arrow denotes the direction of the applied magnetic field. The symbols are defined in the text. **c**, The horizontal (H) and vertical (V) scans of the four-spot pattern along the H-/V-dash lines indicated in the panel (**b**). Intensity profiles $I(\Delta q_\parallel)$ (upper) and $I(\Delta q_\perp)$ (bottom) of the small-angle skewed spots measured via the maxima at $q_0$ long the longitudinal ($\parallel \mathbf{B}$) and transversal ($\perp \mathbf{B}$) direction, respectively. **d**, Evolution of the skewed peaks of the **1/2** mixture with $dr$ = 100/0, 70/30 and 60/40, as a function of the temperature, $T-T_c$. **e**, The relationship between growth anisotropy of cluster and $dr$, and its temperature-dependent. Inset cartoon depicts the two variation (case A and case B) of the growth of the cybotactic cluster in the $N_F$ phase.

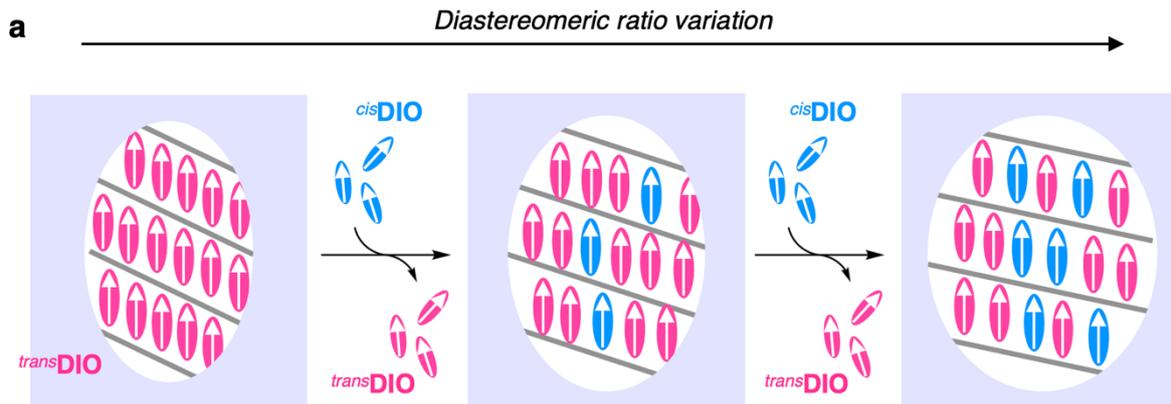

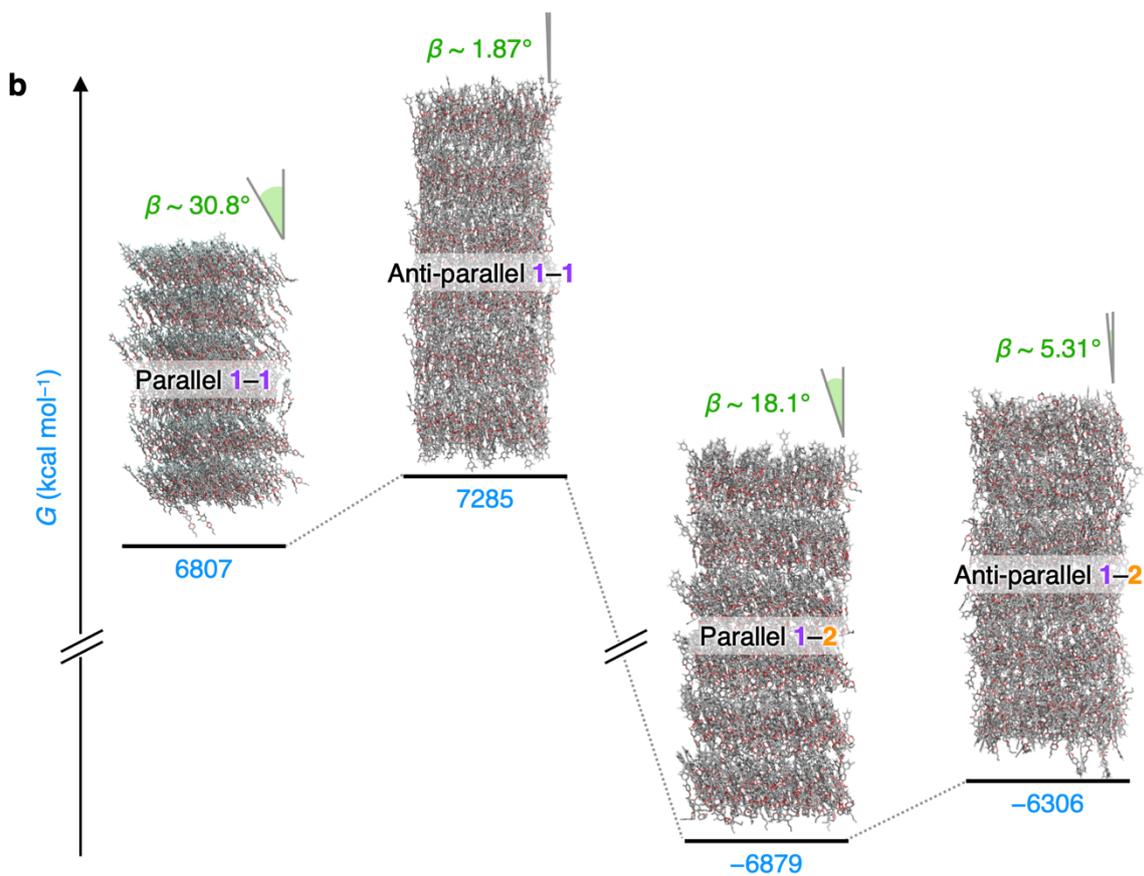

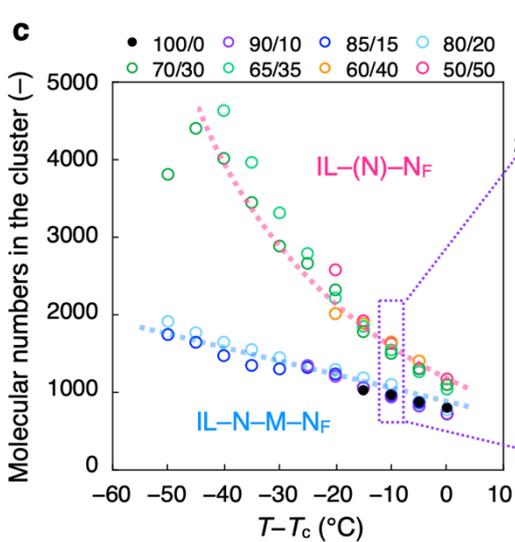 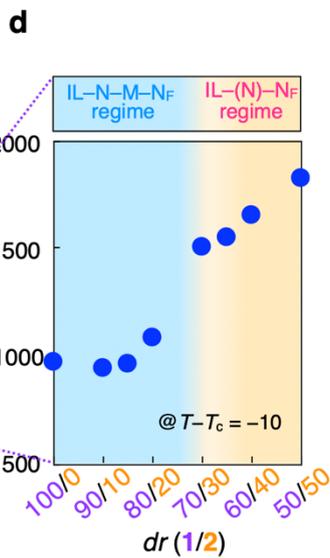 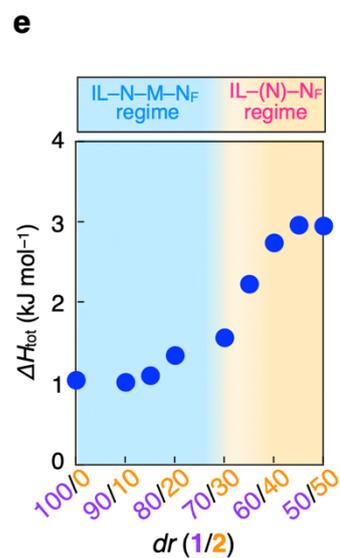

**Fig. 4 | Clustering effect on the evolution of phase sequence. a**, The schematic illustration of the growth process of the cybotactic cluster in the $N_F$ phase for **1/2** mixture. The exchange of cisDIO with transDIO promotes the anisotropic cluster growth. **b**, Energy diagram of the $N_F$ phase in **1/2** system generated by MD simulation. The snapshots (after 300 ns equilibration run) of four system consisting of parallel **1–1**, anti-parallel **1–1**, parallel **1–2** and anti-parallel **1–2** dimers obtained by all-atom MD simulation (for **1–1** dimer, 323 K; for **1–2** dimer, 288 K) are inserted. *β* denotes the average tilt angle of molecules in a smectic layer. The bottom values are the average total energy (simulation time: 200–300 ns). **c**, The average molecular numbers in the cluster ($\bar{N}$) as a function of $T-T_c$ of the **1/2** mixture with various *dr* (100/0–50/50). **d**, The corresponding $\bar{N}$ as a function of *dr* ($T-T_c = -10$ °C). **e**, The relationship between latent heat ($\Delta H$) and *dr* at $T-T_c = -10$ °C. Note: two distinct section of the phase transition types are indicated by color bands: IL–N–M–$N_F$ regime (blue), IL–(N)–$N_F$ regime (orange) in the panel (**d**) and (**e**).

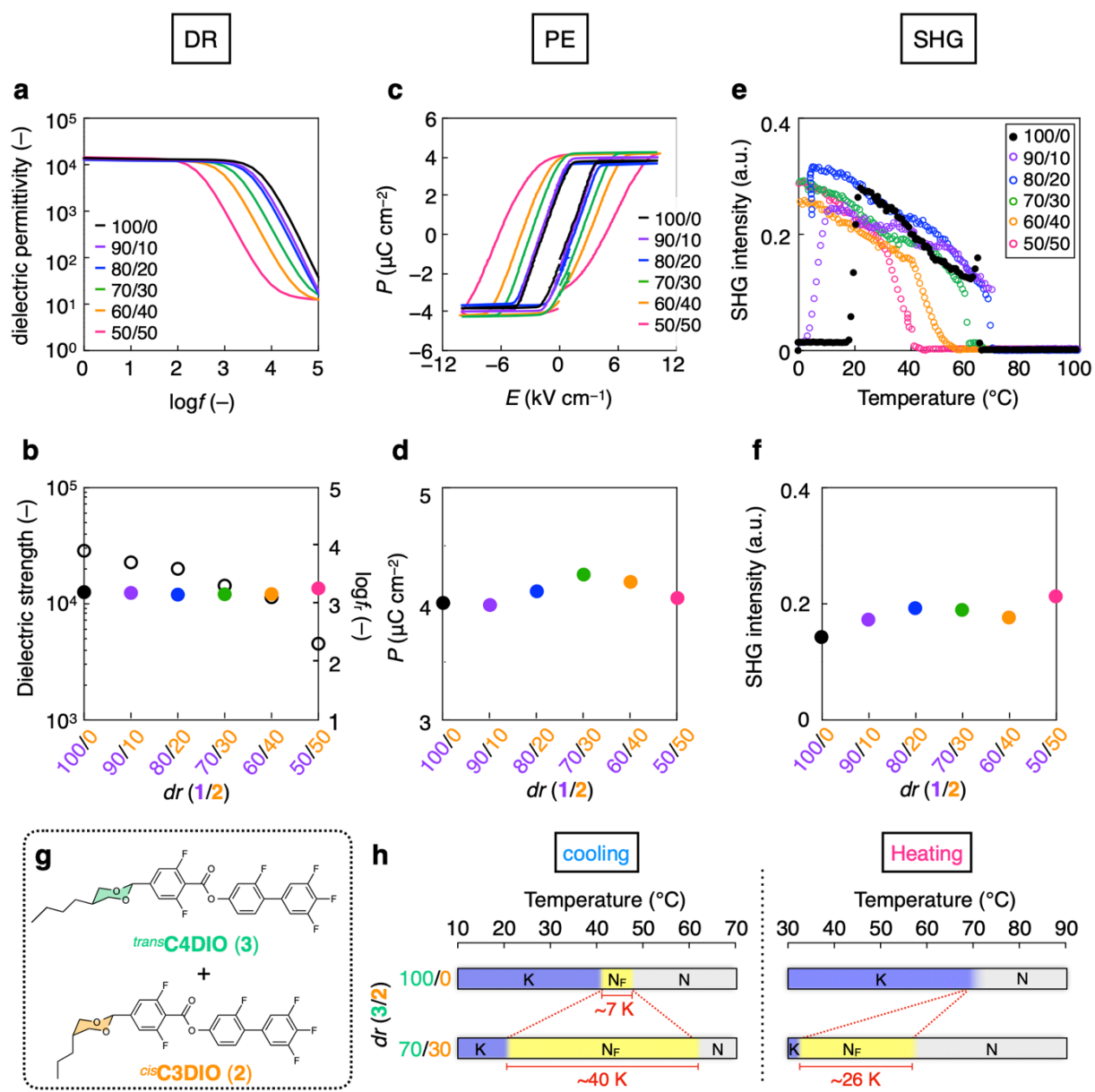

**Fig. 5 | Polarization behavior of the 1/2 mixture with various *dr*. a**, Dielectric spectra. **b**, Dielectric strength and the corresponding relaxation frequency (log$f_r$) vs. *dr* ($T-T_c = -10$ °C). **c**, *P-E* hysteresis loops ($f = 200$ Hz, $V_{max} = 10$ kV cm$^{-1}$). **d**, Polarization density vs. *dr* ($T-T_c = -10$ °C). **e**, Field-induced *p*-in/*p*-out SHG at a 45° incident angle as a function of temperature (0.7 V μm$^{-1}$). **f**, The SHG intensity *vs. dr* ($T-T_c = -10$ °C). Notes: LC-cell conditions: thickness of 13 μm; silanized coated ITO electrode (5 × 10 mm) for dielectric and *P-E* hysteresis studies); thickness of 5 μm; silanized coated baked-ITO electrode (4 × 5 mm) for SHG study. **g**, Chemical structures of $^{trans}$C4DIO (**3**) and $^{cis}$C3DIO (**2**). **h**, Temperature-dependent phase

behaviors of the **3**/**2** mixture with $dr$ = 100/70 (upper) and $dr$ = 70/30 (bottom) on cooling and heating.

# Supplementary Information

# Nano-Clustering Mediates Phase Transitions in a Diastereomerically-Stabilized Ferroelectric Nematic System

Hiroya Nishikawa*, Koki Sano, Saburo Kurihara, Go Watanabe, Atsuko Nihonyanagi, Barun Dhara and Fumito Araoka*

*To whom correspondence should be addressed.

E-mail: hiroya.nishikawa@riken.jp (H.N.) and fumito.araoka@riken.jp (F.A.)

## Table of contents

## Methods

### 1. General and materials

**General**: Analytical thin layer chromatography (TLC) was performed on silica gel layer glass plate Merck 60 F254 and visualized by UV irradiation (254 nm). Column chromatography was performed on a Biotage Isolera™ Prime flash system (Biotage) using Biotage SNAP Ultra (particle size 25 μm; HP-spherical silica) column cartridge. $^1$H and $^{13}$C nuclear magnetic resonance (NMR) spectra were recorded on Ascend 600 (600 MHz, BRUKER) operating at 600.00 MHz and 150.00 MHz for $^1$H and $^{13}$C NMR, respectively, using the TMS (trimethylsilane) as an internal standard for $^1$H NMR and the deuterated solvent for $^{13}$C NMR. The absolute values of the coupling constants are given in Hz, regardless of their signs. Signal multiplicities were abbreviated by s (singlet), d (doublet), t (triplet), q (quartet), quint (quintet), sext (sextet), dd (double–doublet), respectively. The quadrupole time-of-flight high-resolution mass spectrometer (QTOF-HRMS) was performed on COMPACT (BRUKER). The calibration was carried out using LC/MS tuning mix, for APCI/APPI (Agilent Technologies).

**Materials**: All reagents and solvents were purchased from Kanto Chemical Co., Inc., Tokyo Chemical Industry Co., Ltd., FUJIFILM Wako Pure Chemical Corporation, Sigma-Aldrich Co., LLC. and Combi-Blocks Inc. and used without further purification. A series of DIO molecule was synthesized in our laboratory.

## 2. Synthesis and Characterization of $^{trans}$CnDIO (n = 3,4) and $^{cis}$C3DIO.

A series of DIO (**1**,**2**) were synthesized as follows:

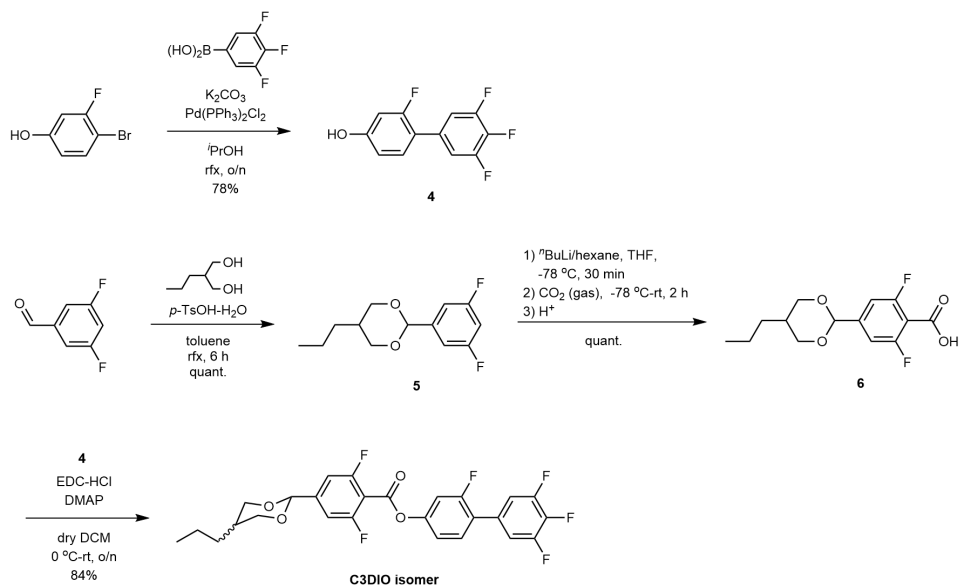

### 2.1. 2,6-Difluoro-4-(5-propyl-1,3-dioxan-2-yl)benzoic acid (4)

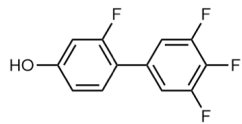

To a flask 4-bromo-3-fluorophenol (955 mg, 5.00 mmol), 3,4,5-trifluorophenylboronic acid (967 mg, 5.50 mmol), Pd(dppf)Cl$_2$·CH$_2$Cl$_2$ (204 mg, 5 mol%, 0.25 mmol), K$_2$CO$_3$ (2.07 g, 15.0 mmol), and DME/H$_2$O (15.0 mL/3.00 mL) were added. After stirring under reflux overnight, the reaction mixture was allowed to cool to room temperature, filtered through Celite, and concentrated. The residue was purified by column chromatography on silica gel with DCM as eluent and recrystallized from toluene to give **4** as a white solid. Yield: 93% (1.13 g).

$R_f$ = 0.33 (DCM)

### 2.2. 2-(3,5-Difluorophenyl)-5-propyl-1,3-dioxane (5)

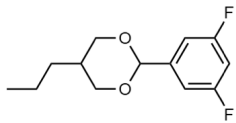

To a solution of 2-propylpropane-1,3-diol (7.28 mL, 60.0 mmol) and 3,5-difluorobenzaldehyde (5.48 mL, 50.0 mmol) in toluene (150 mL) was added *p*-TsOH·H$_2$O (861 mg, 5.00 mmol) under Ar atmosphere, and the resulting mixture was refluxed for 6 h. The H$_2$O generated during the condensation was azeotropically removed by using a Dean Stark apparatus. Then the reaction mixture was allowed to cool to room temperature and concentrated. The residue was purified by column chromatography on silica gel using EtOAc : *n*-hexane (gradient from 1% to 3% EtOAc) as eluent to give **5** as a colorless oil (*trans*/*cis* mixture = 3.9/1). Yield: quant. (11.8 g).

$R_f$ = 0.31 (*n*-hexane:EtOAc = 30:1)

**2.3. 2,6-Difluoro-4-(5-propyl-1,3-dioxan-2-yl)benzoic acid (6)**

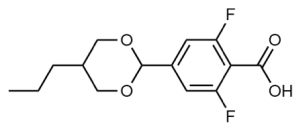

To a stirred solution of **5** (4.85 g, 20.0 mmol) in dry THF (70.0 mL) was added dropwise a solution of *n*-butyllithium (15.1 mL of 1.59 M in *n*-hexane, 24.0 mmol) at −78 °C over 10 min under Ar. After stirring for 30 min at −78 °C, the balloon with $CO_2$ gas was equipped with the flask. The reaction was stirred for 2h at the temperature under $CO_2$ atmosphere, quenched with 1N HCl *aq.*, extracted with EtOAc, and dried over $Na_2SO_4$. The solvent was evaporated, and dried in vacuo to give crude acid **6** as a white solid (*cis*/*trans* = 1/4). Yield: quant. (6.10 g). The product was used directly for the next step without further purification.

**2.4. 2,3',4',5'-Tetrafluoro-[1,1'-biphenyl]-4-yl 2,6-difluoro-4-(5-propyl-1,3-dioxan-2-yl)benzoate (C3DIO isomer)**

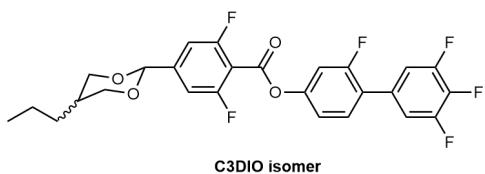

C3DIO isomer

To a solution of **4** (3.44 g, 12.0 mmol) and **6** (2.42 g, 10.0 mmol) in DCM (30.0 mL) were added EDC·HCl (2.30 g, 12.0 mmol) and DMAP (122 mg, 10 mol%, 1.00 mmol) at 0 °C under Ar atmosphere. After stirring overnight, the reaction mixture was quenched with $H_2O$, extracted with DCM, and dried over $Na_2SO_4$. The solvent was evaporated, and the residue was purified by column chromatography on silica gel using EtOAc : *n*-hexane (gradient from 3% to 5% EtOAc) as eluent to give **C3DIO isomer** (*trans*/*cis* = 5.6/1). Yield: 84% (4.28 g). The resulting mixture was completely separated by column chromatography again to afford $^{trans}$**C3DIO** (**1**) and $^{cis}$**C3DIO** (**2**). Both compounds were recrystallized slowly from DCM/*n*-hexane to afford a white needle crystal.

### *trans*C3DIO (1)

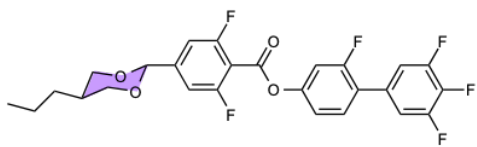

Yield: 71.1% (3.63 g), a white needle crystal. $R_f$ = 0.27 (*n*-hexane:EtOAc = 15:1)

$^1$H NMR (600 MHz, CDCl$_3$): δ = 7.428 (t, *J* = 8.7 Hz), 7.197–7.148 (m), 5.399 (s), 4.254 (dd, *J* = 11.7, 4.2 Hz), 3.539 (t, *J* = 11.4 Hz), 2.179-2.103 (m), 1.379–1.317 (m), 1.122-1.084 (m), 0.936(t, *J* = 7.2 Hz) ppm

$^{13}$C NMR (150 MHz, CDCl$_3$): δ = 161.06 (dd), 159.41 (s), 159.50 (d), 151.32 (ddd), 151.12 (d), 145.78 (t), 139.67 (dt), 131.04–130.90 (m), 130.79 (d), 124.42 (d), 118.27 (d), 113.35 (dt), 110.78 (t), 110.43 (dd), 109.57 (t), 98.93 (s), 72.72 (s), 34.03 (s), 30.36 (s), 19.66 (s), 14.31 (s) ppm.

QTOF-HRMS (*m/z*, [M+H]$^+$) Calcd for C$_{26}$H$_{21}$F$_6$O$_4$: 511.1344; found: 511.1332.

### *cis*C3DIO (2)

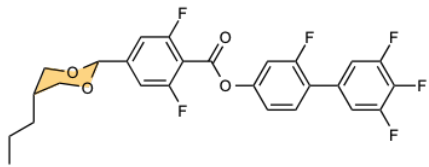

Yield: 12.7 % (0.65 g), a white needle crystal. $R_f$ = 0.20 (*n*-hexane:EtOAc = 15:1)

$^1$H NMR (600 MHz, CDCl$_3$): δ = 7.428 (t, *J* = 8.7 Hz), 7.194–7.149 (m), 5.499 (s), 4.114-4.063 (m), 1.735 (dd, *J* = 15.6, 7.8 Hz), 1.496–1.472 (m), 1.379–1.317 (m), 0.962 (t, *J* = 7.5 Hz) ppm.

$^{13}$C NMR (150 MHz, CDCl$_3$): δ = 161.09 (dd), 159.40 (s), 159.49 (d), 151.33 (ddd), 151.12 (d), 145.98 (t), 139.66 (dt), 131.05–130.90 (m), 130.78 (dd), 124.42 (d), 118.27 (dd), 113.35 (td), 110.76 (t), 110.47 (td), 109.57 (t), 99.20 (s), 70.80 (t), 34.09 (d), 31.68 (t), 20.66 (t), 14.20 (q) ppm.

QTOF-HRMS (*m/z*, [M+H]$^+$) Calcd for C$_{26}$H$_{21}$F$_6$O$_4$: 511.1344; found: 511.1350.

*trans*C4DIO (**3**) was synthesized as follows:

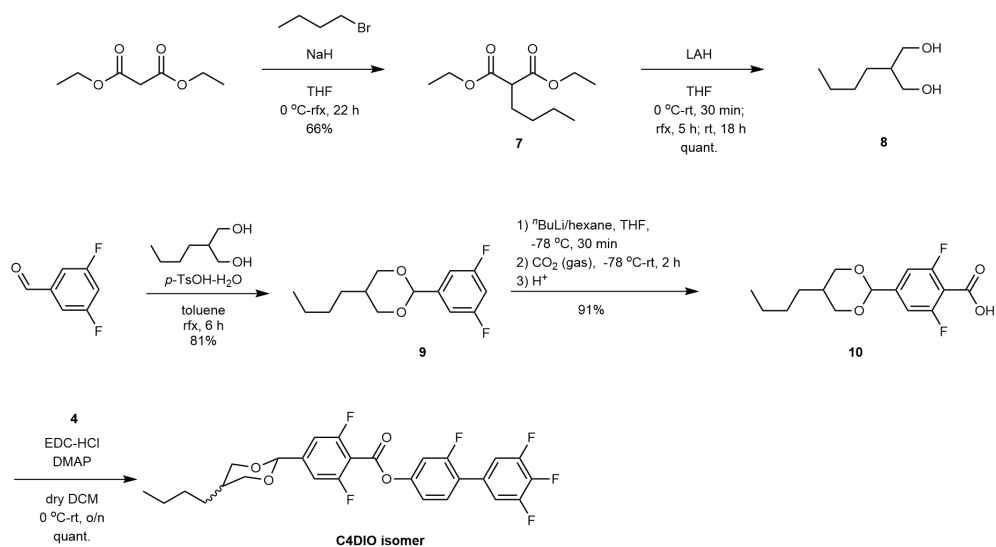

## 2.5. Diethyl 2-butylmalonate (7)

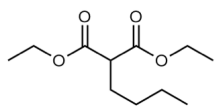

To a solution of NaH (4.00 g, 100 mmol) in THF (250 mL) was added diethylmalonate (21.2 mL, 140 mmol) at 0 °C under Ar atmosphere and stirred for 1h at the temperature. *n*-Butylbromide (18.0 mL, 168 mmol) was added dropwise, and the mixture was heated under reflux. After refluxing for 22 h, the reaction mixture was allowed to cool to room temperature and poured into ice-water, and extracted with petroleum ether. The organic extract was dried over $Na_2SO_4$, concentrated, and purified by column chromatography on silica gel using EtOAc : *n*-hexane (gradient from 3% to 10% EtOAc) as eluent to give **7** as a colorless oil. Yield: 66% (19.8 g).

$R_f$ = 0.38 (*n*-hexane:EtOAc = 10:1)

## 2.6. 2-Butylpropane-1,3-diol (8)

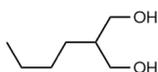

Lithium aluminum hydride (2.10 g, 55.5 mmol) was added portionwise to a solution of **7** (6.00 g, 27.7 mmol) in THF (150 mL) at 0 °C under Ar atmosphere. After stirring for 30 min at room temperature, 5 h under reflux, and 18 h at room temperature, the reaction mixture was cooled to 0 °C, and quenched with dropwise 2.10 mL water followed by 2.10 mL of 1N NaOH and, finally, 6.30 mL of water. After stirring for 45 min at 0 °C and 2 h at room temperature, the resulting mixture was filtered through Celite and concentrated. The residue was dried under reduced pressure to give **8** as a colorless oil. Yield: quant. (3.53 g).

**2.7. 5-Butyl-2-(3,5-difluorophenyl)-1,3-dioxane (9)**

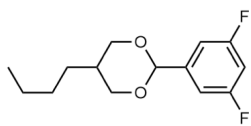

To a solution of **8** (10.96 g, 83.0 mmol) and 3,5-difluorobenzaldehyde (7.58 mL, 69.2 mmol) in toluene (200 mL) was added *p*-TsOH·H$_2$O (861 mg, 5.00 mmol) under Ar atmosphere, and the resulting mixture was refluxed for 6h. The H$_2$O generated during the condensation was azeotropically removed by using a Dean Stark. Then the reaction mixture was allowed to cool to room temperature, quenched with water, and extracted with EtOAc and DCM. The organic extracts were purified by column chromatography on silica gel using DCM : hexane (gradient from 5% to 15% DCM) as eluent to give **9** as a colorless oil (*trans*/*cis* = 3/1). Yield: 81% (17.2 g).

$R_f$ = 0.43 (*n*-hexane: DCM = 2:1)

**2.8. 4-(5-Butyl-1,3-dioxan-2-yl)-2,6-difluorobenzoic acid (10)**

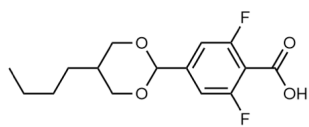

To a stirred solution of **9** (5.20 g, 20.4 mmol) in THF (80.0 mL) was added dropwise a solution of *n*-butyllithium (15.4 mL of 1.59 M in *n*-hexane, 24.5 mmol) at −78 °C over 15 min under Ar. After stirring for 30 min at −78 °C, the balloon with CO$_2$ gas was equipped with the flask. The reaction was stirred for 4 h at the temperature under carbon dioxide atmosphere, poured into H$_2$O with ice, extracted with petroleum ether, and dried over Na$_2$SO$_4$. The solvent was evaporated, and dried in vacuo to give crude acid **10** as a white solid (*trans*/*cis* = 2.5/1). Yield: 91% (5.58 g). The product was used directly for the next step without further purification.

**2.9. 2,3',4',5'-Tetrafluoro-[1,1'-biphenyl]-4-yl 4-(5-butyl-1,3-dioxan-2-yl)-2,6-difluorobenzoate**

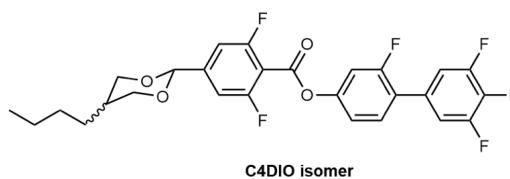

**C4DIO isomer**

To a solution of **10** (7.80 g, 26.0 mmol) and **4** (4.84 g, 20.0 mmol) in DCM (50.0 mL) were added EDC・HCl (4.98 g, 26.0 mmol) and DMAP (244 mg, 2.0 mmol) at 0 °C under Ar atmosphere. After stirring overnight, the reaction mixture was quenched with $H_2O$, extracted with DCM, and dried over $Na_2SO_4$. The solvent was evaporated, and the residue was purified by column chromatography on silica gel using EtOAc:*n*-hexane (gradient from 5% to 10% EtOAc) as eluent to give **C4DIO isomer** as a white solid (*trans*/*cis* mixture = 3/1). Yield: 96% (10.0 g). The resulting mixture was completely separated by column chromatography again to afford $^{trans}$**C4DIO** (**3**) and $^{cis}$**C4DIO**. Both compounds were recrystallized slowly from *n*-hexane/EtOAc to afford a white powder.

$R_f$ = 0.29 (*n*-hexane:EtOAc = 10:1)

**transC4DIO (3)**

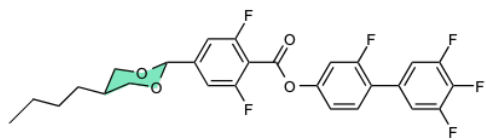

¹H NMR (600 MHz, CDCl₃): δ = 7.427 (t, *J* = 8.7 Hz), 7.196–7.148 (m), 5.397 (s), 4.255 (dd, *J* = 11.7, 4.8 Hz), 3.537 (t, *J* = 11.4 Hz), 2.144–2.076 (m), 1.362–1.267 (m), 1.118 (dd, *J* = 15.3, 7.2Hz) ppm.

¹³C NMR (150 MHz, CDCl₃): δ = 161.04 (dd), 159.40 (s), 159.49 (d), 151.32 (ddd), 151.12 (d), 145.78 (t), 139.66 (dt), 131.04–130.90 (m), 130.77 (d), 124.42 (d), 118.26 (d), 113.35 (dt), 110.76 (t), 110.43 (dd), 109.55 (t), 98.93 (s), 72.74 (s), 34.25 (s), 28.58 (s), 27.90 (s), 22.93 (s), 13.99 (s) ppm.

QTOF-HRMS (*m/z*, [M+H]⁺) Calcd for $C_{27}H_{23}F_6O_4$: 525.1501; found: 525.1503.

**Supplementary Notes (Supplementary Notes 1–2)**

**Supplementary Note 1 | Discussion of molecular packing using DFT calculation and MD simulation**

The intermolecular interaction energies of the dimers were estimated by density-functional theory (DFT) using Gaussian 09* program. [S1] The interaction energy for each dimer was calculated as a relative difference energy between the dimer and the total of the isolated monomers by correcting for the basis set superposition error (BSSE) according to the counterpoise correction method. The DFT calculation was performed with B3LYp/6-311G(d,p) with the GD3 dispersion correction. The intermolecular distance ($d_{i-j}$) of the dimer was adjusted to the corresponding halo peak in the WAXD data (for $dr$ = 100/0, $d_{1-1}$ = 0.450 nm; for $dr$ = 50/50, $d_{1-2}$ = 0.473 nm) for the $N_F$ phase. The molecular conformations are chosen as shown in supplementary Fig. 7 based on the crystallography data measured for the crystals of the isolated **1** (100/0) and **2** (0/100) samples.

Molecular dynamics (MD) simulation using all-atom models were performed by GROMACS 2020.5. The simulated systems are following four types: **Systems 1**, **2**, **3**, and **4** consisting of dimers of **1/1** in parallel, **1/1** in anti-parallel, **1/2** in parallel, and **1/2** in anti-parallel. The initial structure of the system was composed of six layers of which 100 molecules were contained in each layer and the total number of molecules were 600. The generalized Amber parameters [S2] were used for the force field parameters of the molecules to calculate the inter- and intramolecular interactions. The partial atomic charges of each molecule were calculated using the restrained electrostatic potential (RESP) methodology [S3], based on DFT calculations (B3LYP/6-31G(d,p)) with the GAUSSIAN 09 program. After the steepest energy minimization, pre-equilibration and equilibration runs were carried out at 333 K and 288 K for **Systems 1** and **2** and **Systems 3** and **4**, respectively. During the pre-equilibration run of 5 ns, the temperature of the system was kept constant using the velocity rescaling thermostat [S4] with the relaxation time $\tau_t$ of 0.2 ps and the pressure was maintained at 1 bar in all directions using the Berendsen semi-isotropic barostat [S5] with the relaxation time $\tau_p$ of 2.0 ps and the compressibility of 4.5 × $10^{-5}$ bar$^{-1}$. The equilibration run was performed for 300 ns at the constant temperature and pressure using the Nosé-Hoover thermostat [S6] with $\tau_t$ of 1.0 ps and Parrinello-Rahman semi-isotropic barostat [S7] with $\tau_p$ of 5.0 ps and the compressibility of 2.0 × $10^{-5}$ bar$^{-1}$, respectively. The time step was set to 2 fs and all bonds connected to hydrogen atoms were constrained with LINCS algorithm [S8]. The long-range Coulomb interactions were calculated with the smooth particle-mesh Ewald (PME) method [S9] with a 0.30 nm. The real space cut-off for both Coulomb and van der Waals interactions was 1.4 nm.

The time dependence of the size of the MD simulation box as shown in Supplementary Fig. 34 indicates that the system reached the equilibrium after about 200 ns. The time-averaged values of total energy, molecular tilt angle, and layer spacing of each system are listed in Supplementary Table 2. The two-dimensional radial distribution functions (RDFs) of the layer for **Systems 1** and **3** were analyzed as shown in Supplementary Figs. 35 and 36. Supplementary Fig. 37 shows that the two-dimensional RDFs of **1** only, **2** only, and both **1** and **2** in the bottom layer for **System 3**.

The average minimum distance between nearest neighboring molecules was calculated according to the following equation:

$$\langle d_{min} \rangle = \langle d_{xy} \rangle \cdot \cos\langle \theta \rangle$$

where $\langle d_{xy} \rangle$ represents the average distance parallel to the layer between the molecules and $\langle \theta \rangle$ denotes the average tilt angle of the molecular long axis from the layer normal. The values of $\langle d_{min} \rangle$ for **Systems 1** and **3** were 0.49 nm and 0.50 nm, respectively.

Moreover, the density profile along the layer normal for **Systems 1** and **3** were shown in Supplementary Figs. 38 and 39. It indicates that the periodic layer structures were confirmed in both systems.

**Supplementary Note 2 | Discussion of dielectric behaviour for 1/2 system**

Supplementary Fig. 30a shows the relaxation frequency ($f_r$) in the temperature range of $N_F$ phase for **1/2** mixture with various $dr$ (0/100–50/50). In their $N_F$ phase, the relaxation frequency followed a Vogel-Fulcher (VF) dependence, $f_r \equiv f_{FV} = f_0 \exp(-A(T-T_{FV}))$, where $f_0$ and $A$ are constant and $T_{FV}$ is VF temperature (is larger than the glass transition temperature, $T_g$), respectively. The VF model is a generalization of the Arrhenius one and has been extensively used in the description and characterization of cooperative molecular motion strongly depending on temperature. [S10] Interestingly, the relaxation profiles were superimposed, constructing a pseudo-master curve (Supplementary Fig. 40b). Thus, it suggests that the major relaxation mode in the $N_F$ phase for **1/2** system exhibits similar temperature dependence and approaches freezing of the cooperative molecular motion, irrespective of the doping level of **2**.

# Supplementary Figures (Supplementary Figs. 1–40)

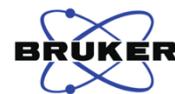

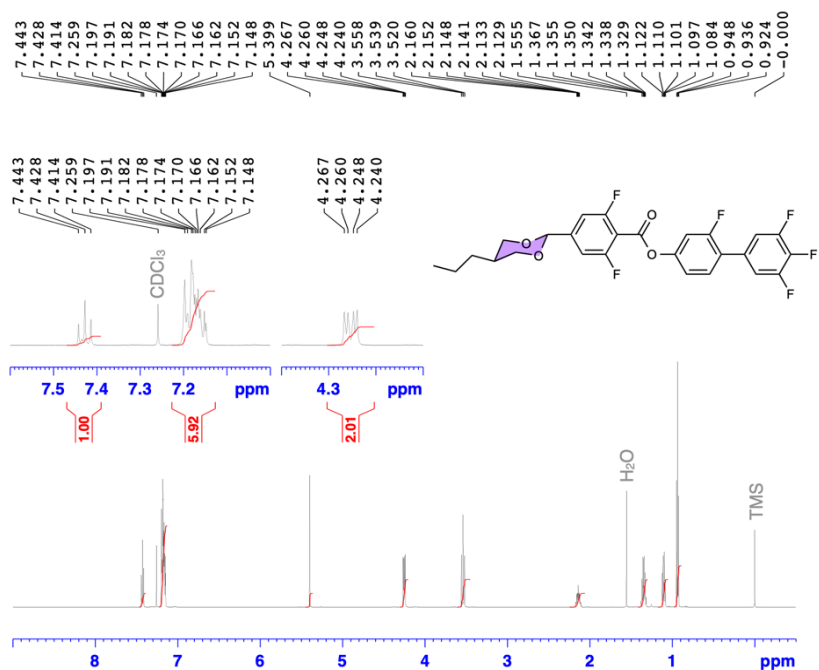

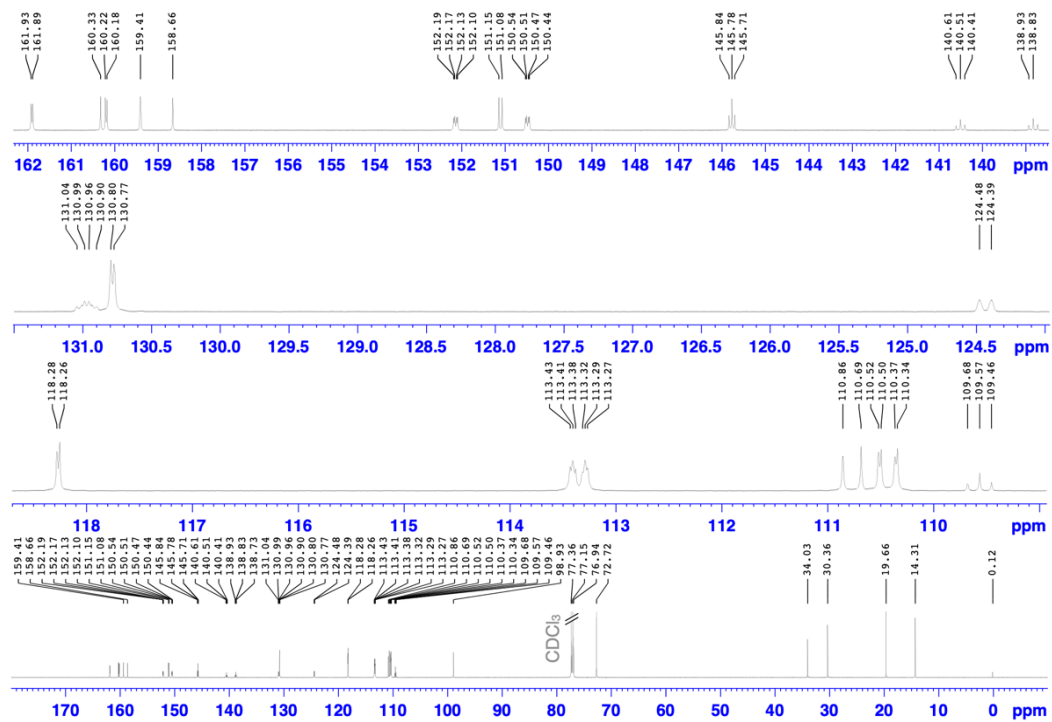

**Supplementary Fig. 1 | $^1$H NMR and $^{13}$C NMR spectra of compound 1.**

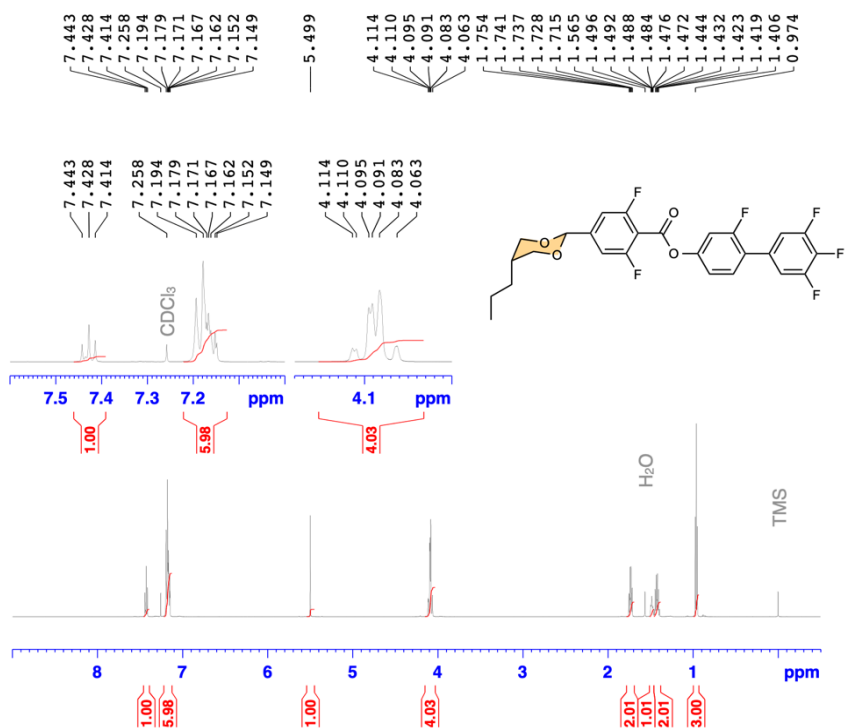
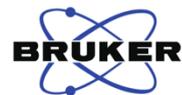
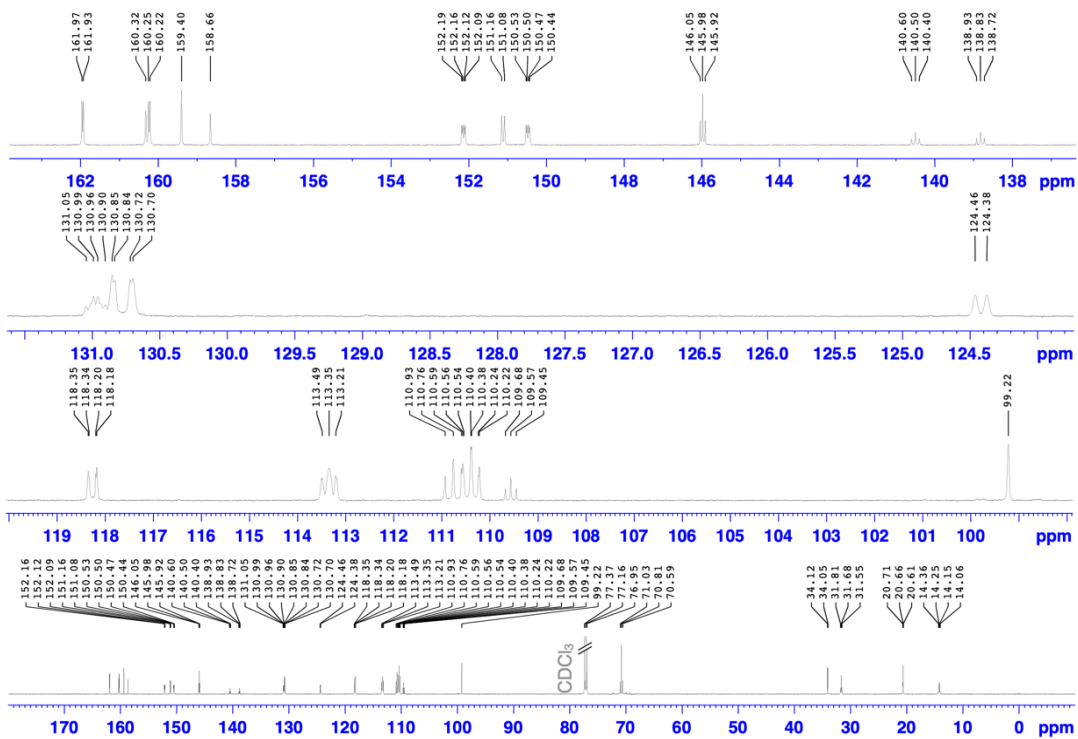

**Supplementary Fig. 2 | $^1$H NMR and $^{13}$C NMR spectra of compound 2.**

**Supplementary Fig. 3 | ¹H NMR and ¹³C NMR spectra of compound 3.**

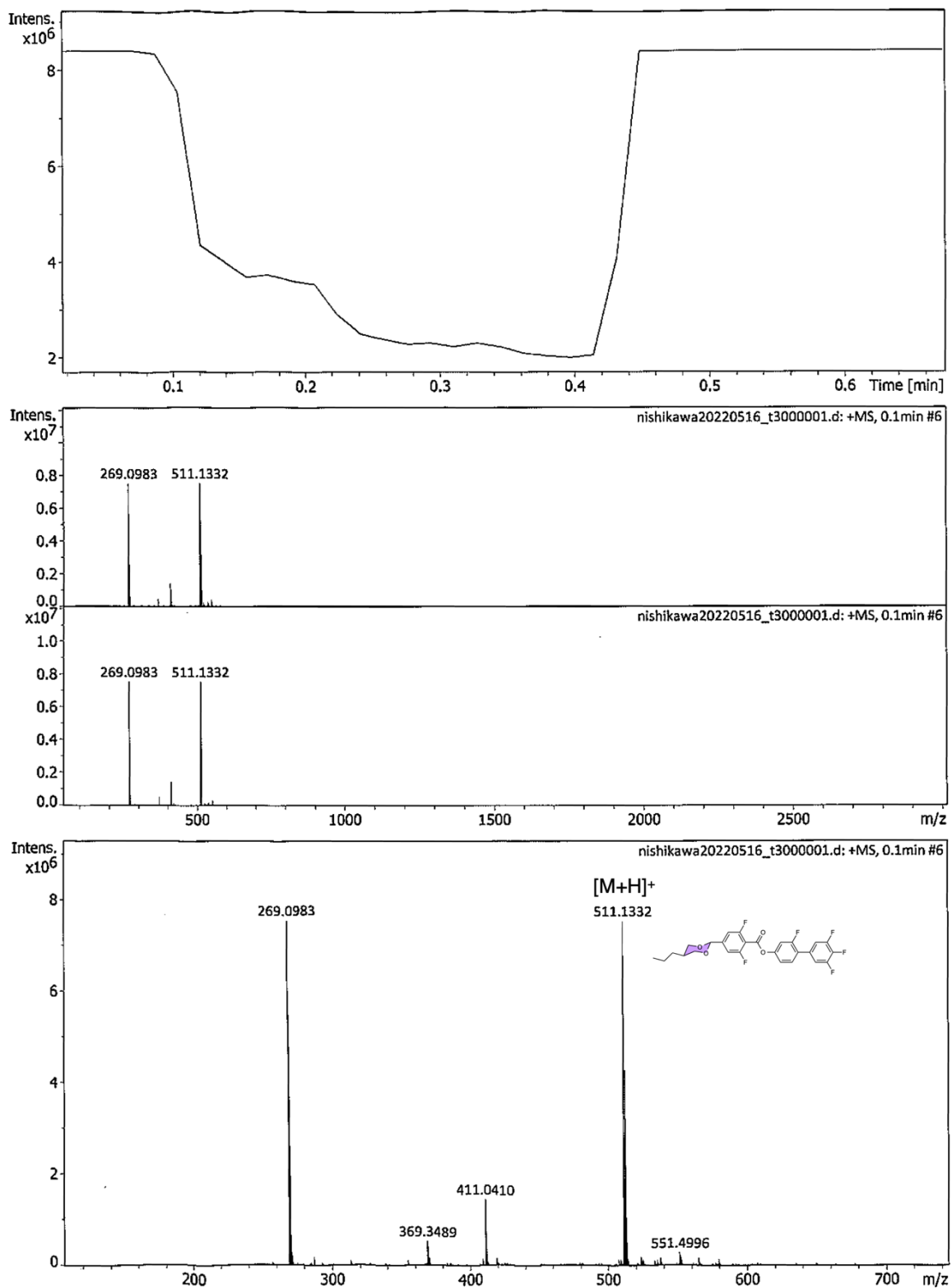

**Supplementary Fig. 4 | High-resolution mass spectra (HR-TOFMS) of compound 1.**

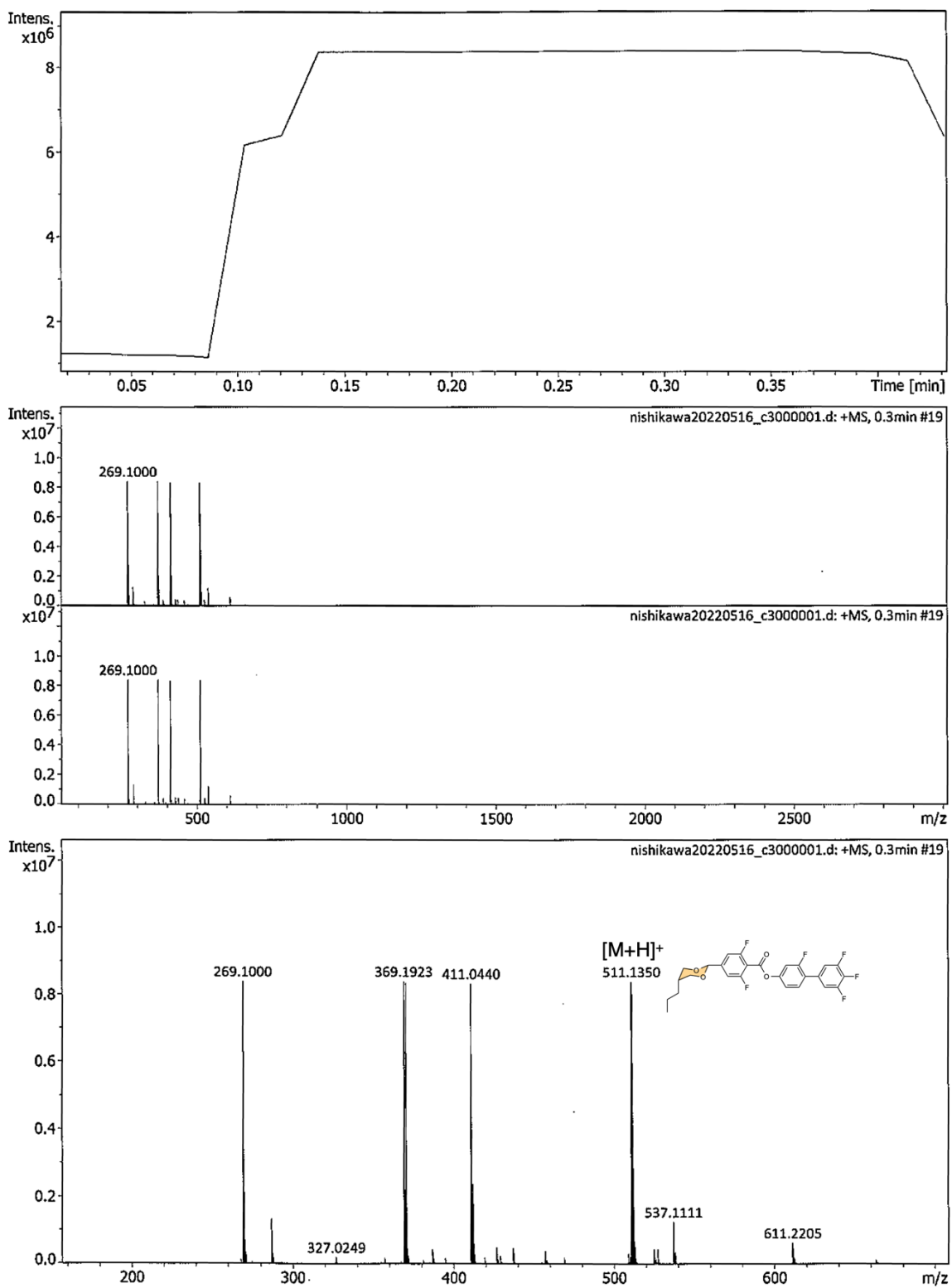

**Supplementary Fig. 5 | High-resolution mass spectra (HR-TOFMS) of compound 2.**

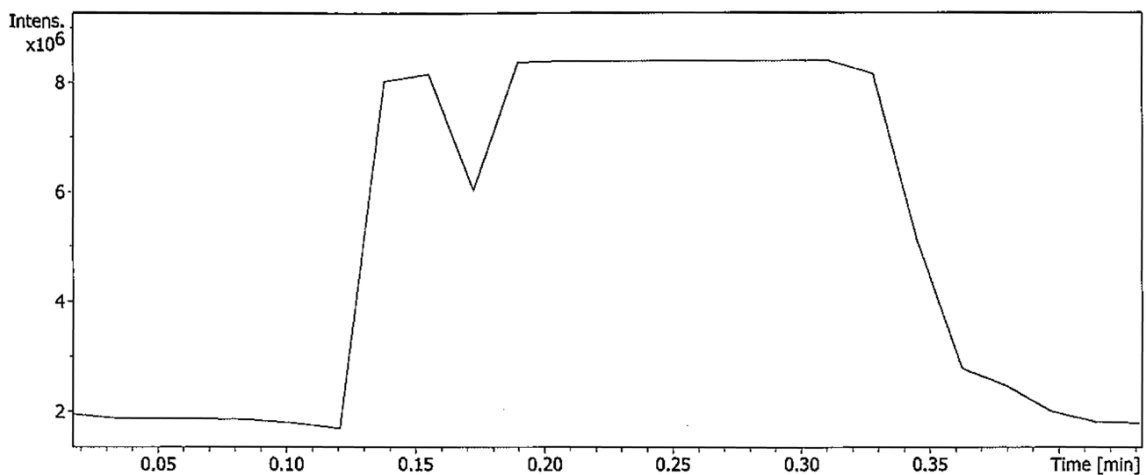
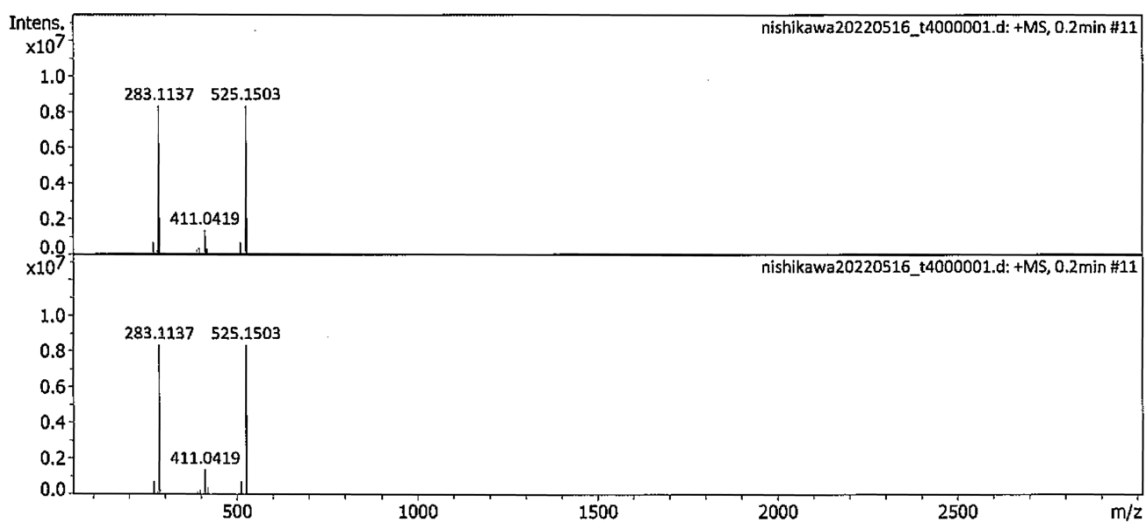
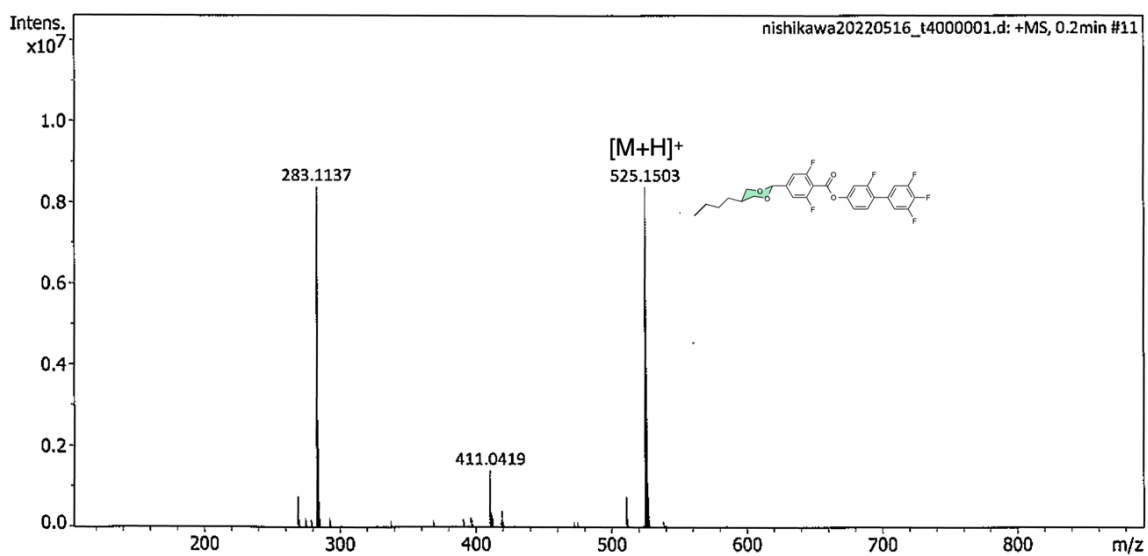

Bruker Compass DataAnalysis 5.2     printed: 5/16/2022 10:09:41 AM     by: demo     Page 1 of 1

**Supplementary Fig. 6 | High-resolution mass spectra (HR-TOFMS) of compound 3.**

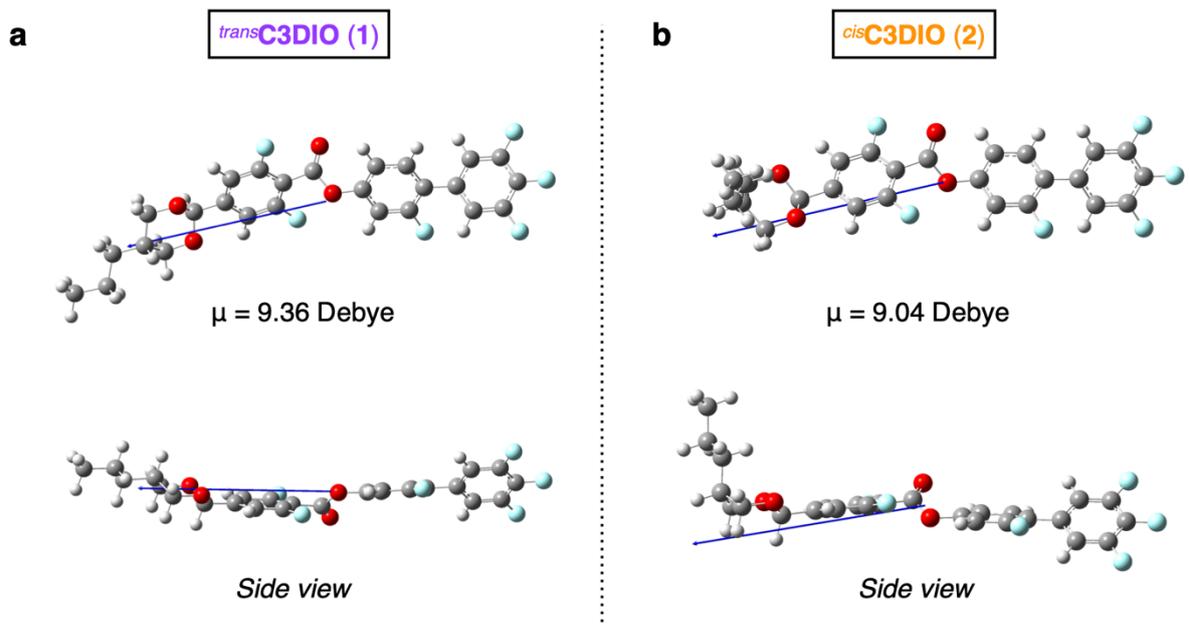

**Supplementary Fig. 7 | Optimized molecular structures of 1 and 2 based on the crystallography data in Supplementary Fig. 8, used for the DFT calculation.**

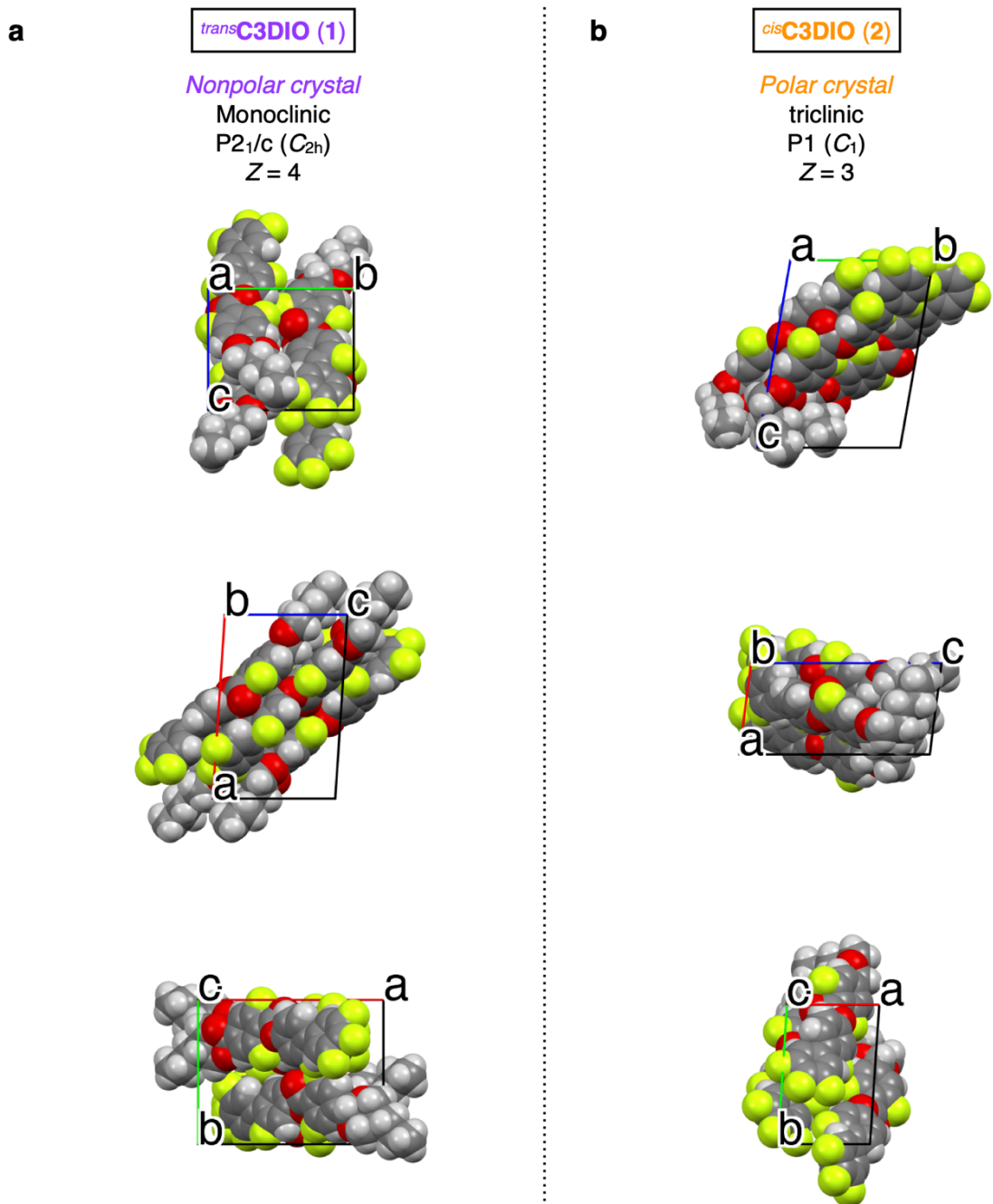

**Supplementary Fig. 8 | Single crystal X-ray crystallographic structures of 1 and 2.** $Z$ denotes the molecular numbers in a unit cell. For **1**, crystal system: monomeric, space group: P2$_1$/c, cell length: **a** = 16.3434(4) Å, **b** = 12.8126(3) Å, **c** = 10.7829(3) Å, cell angle: **α** = 90°, **β** = 93.736(2)°, **γ** = 90°, cell volume: **V** = 2253.16 Å$^3$, **Z** = 4, R-factor = 6.07 %; for **2**, crystal system: monomeric, space group: P1, cell length: **a** = 8.16600(10) Å, **b** = 12.5798(2) Å, **c** = 17.0008(2) Å, cell angle: **α** = 99.5410(10)°, **β** = 96.7250(10)°, **γ** = 92.5840(10)°, cell volume: **V** = 1706.71 Å$^3$, **Z** = 3, R-factor = 4.41 %.

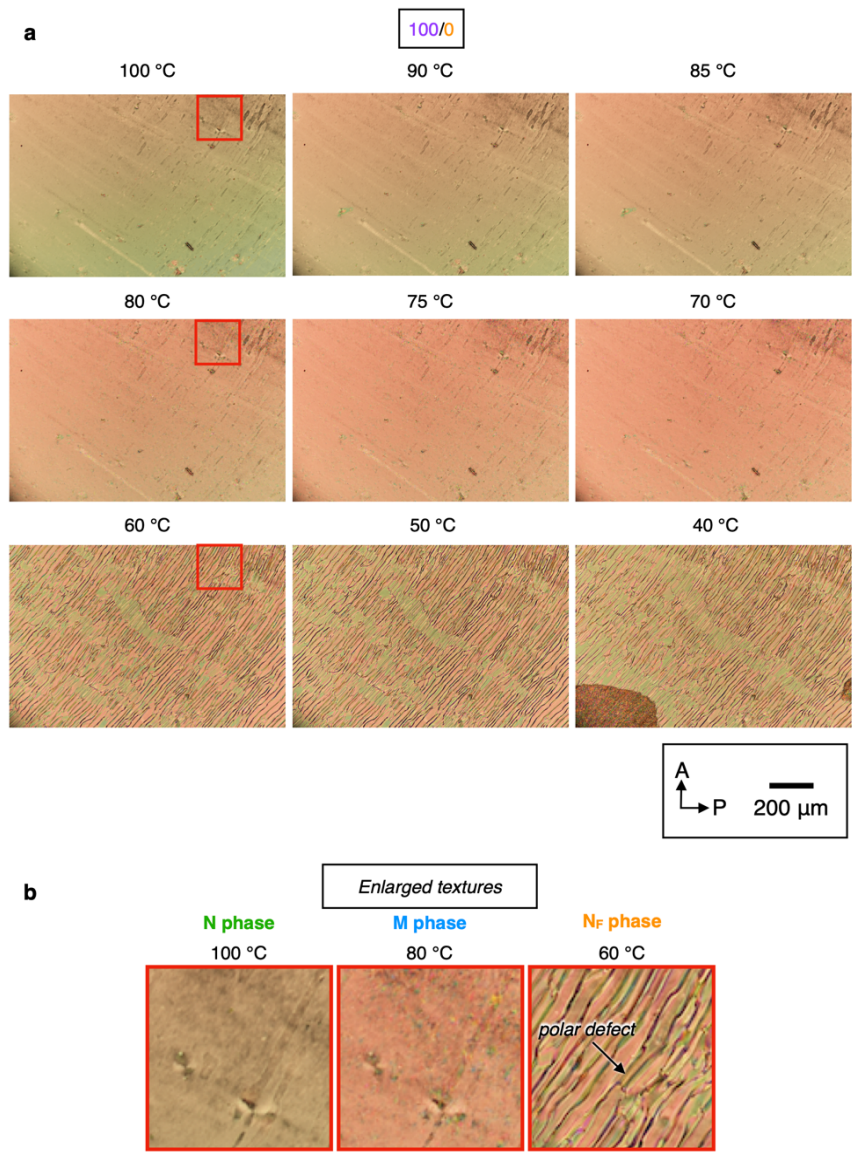

**Supplementary Fig. 9 | The original unedited version of POM images for the mixture 1/2 with *dr* = 100/0.** Evolution of POM images (a) and the corresponding enlarged textures (b) indicated in the red square in the panel (a).

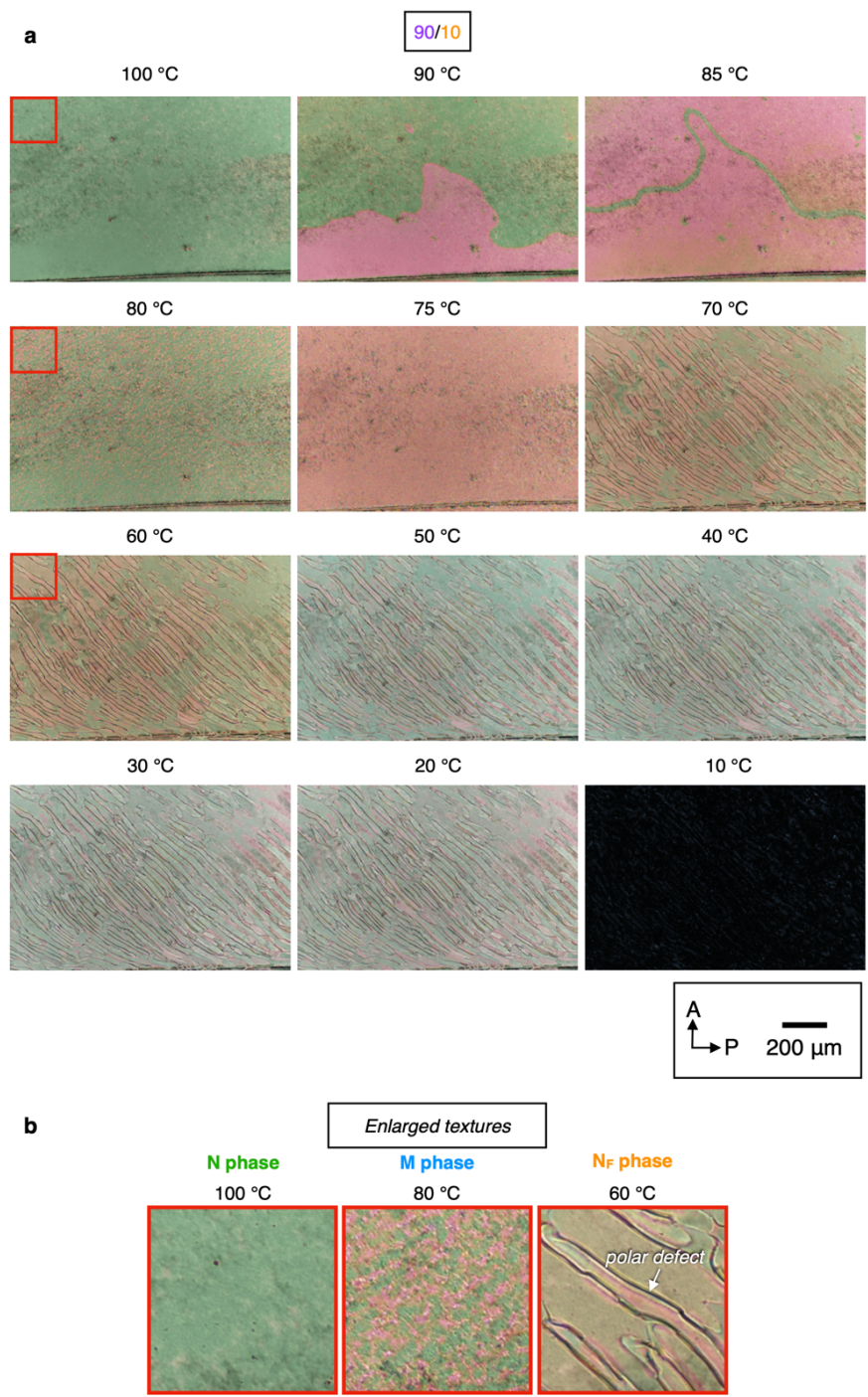

**Supplementary Fig. 10 | The original unedited version of POM images for the mixture 1/2 with *dr* = 90/10.** Evolution of POM images (a) and the corresponding enlarged textures (b) indicated in the red square in the panel (a).

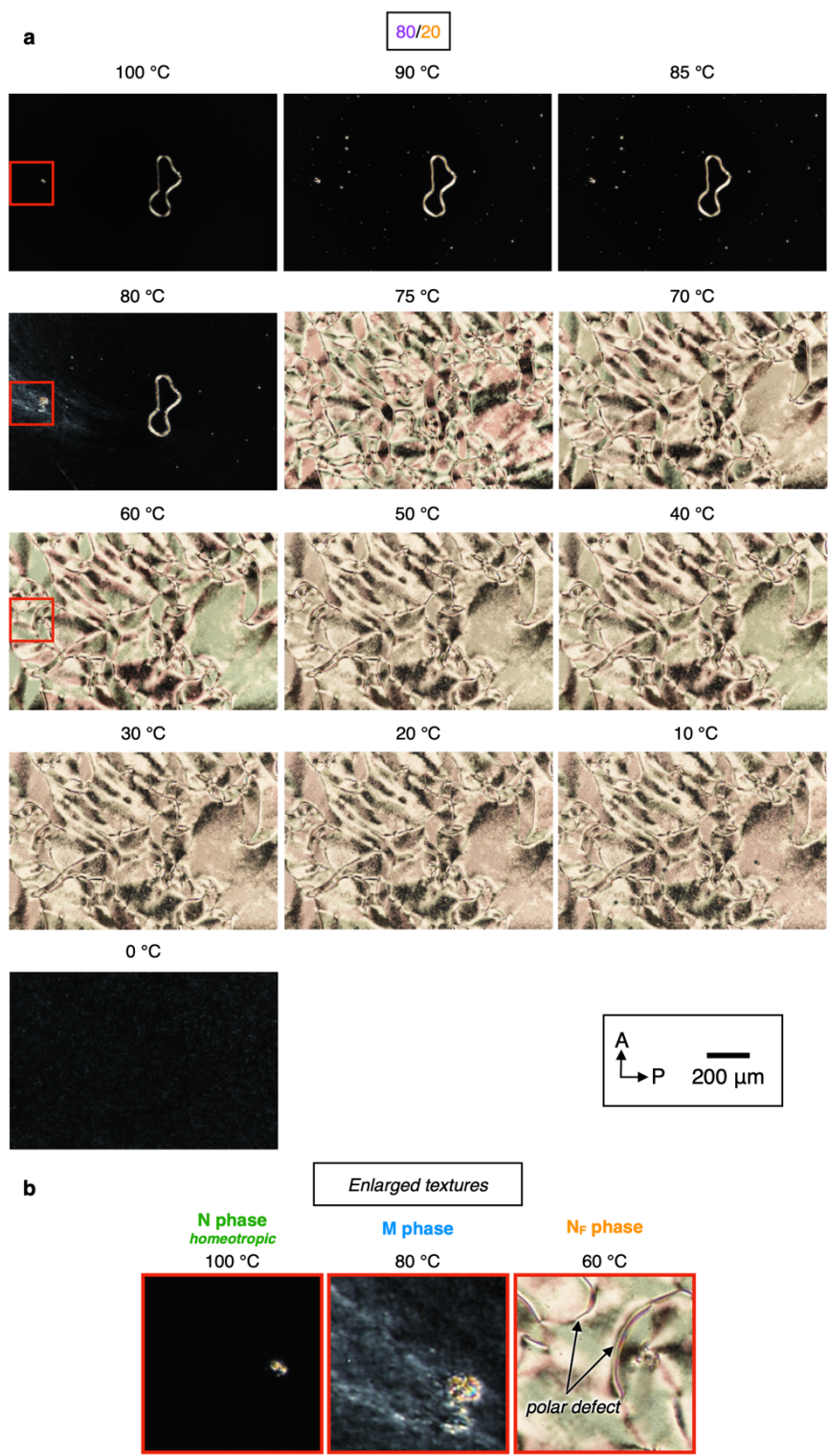

**Supplementary Fig. 11 | The original unedited version of POM images for the mixture 1/2 with *dr* = 80/20.** Evolution of POM images (a) and the corresponding enlarged textures (b) indicated in the red square in the panel (a).

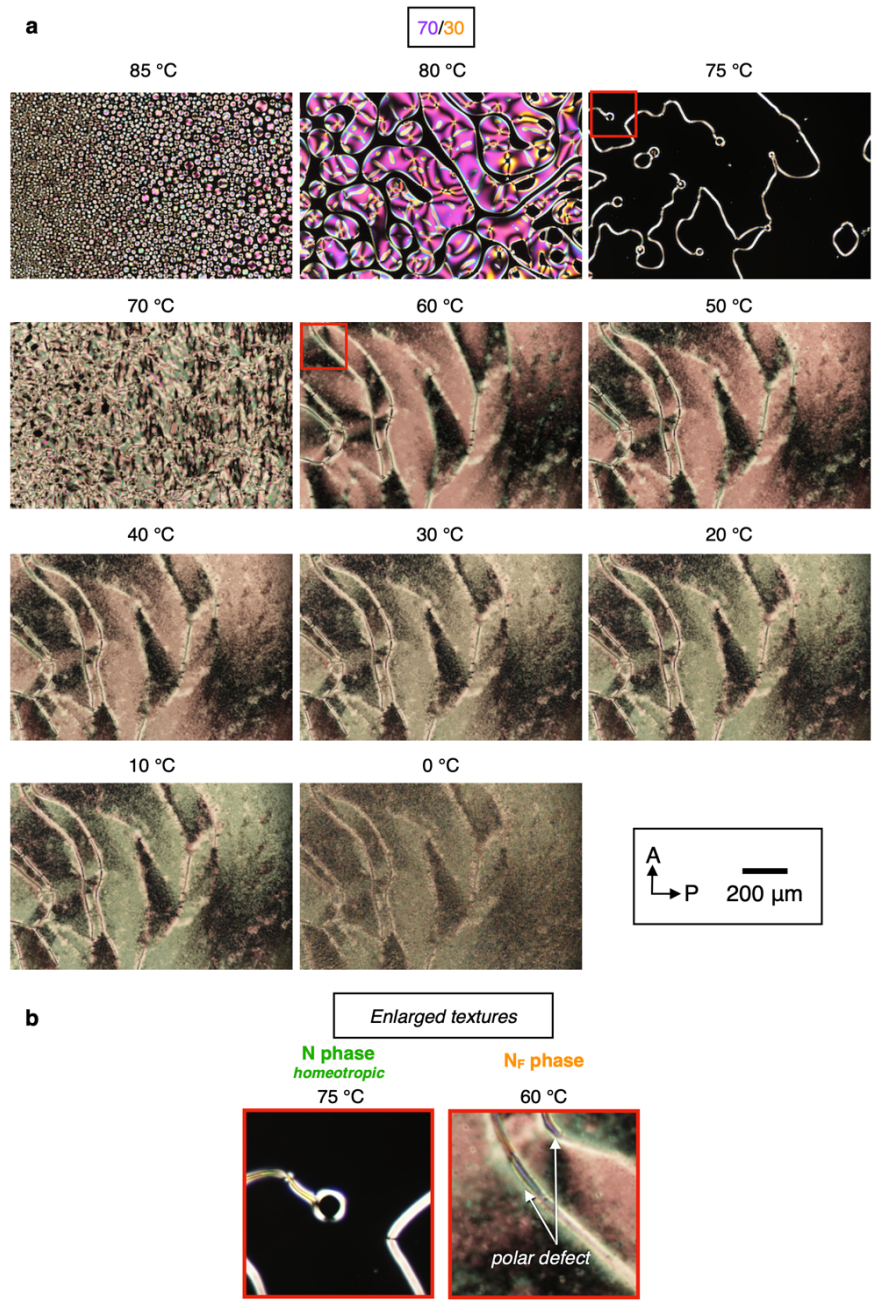

**Supplementary Fig. 12 | The original unedited version of POM images for the mixture 1/2 with *dr* = 70/30.** Evolution of POM images (a) and the corresponding enlarged textures (b) indicated in the red square in the panel (a).

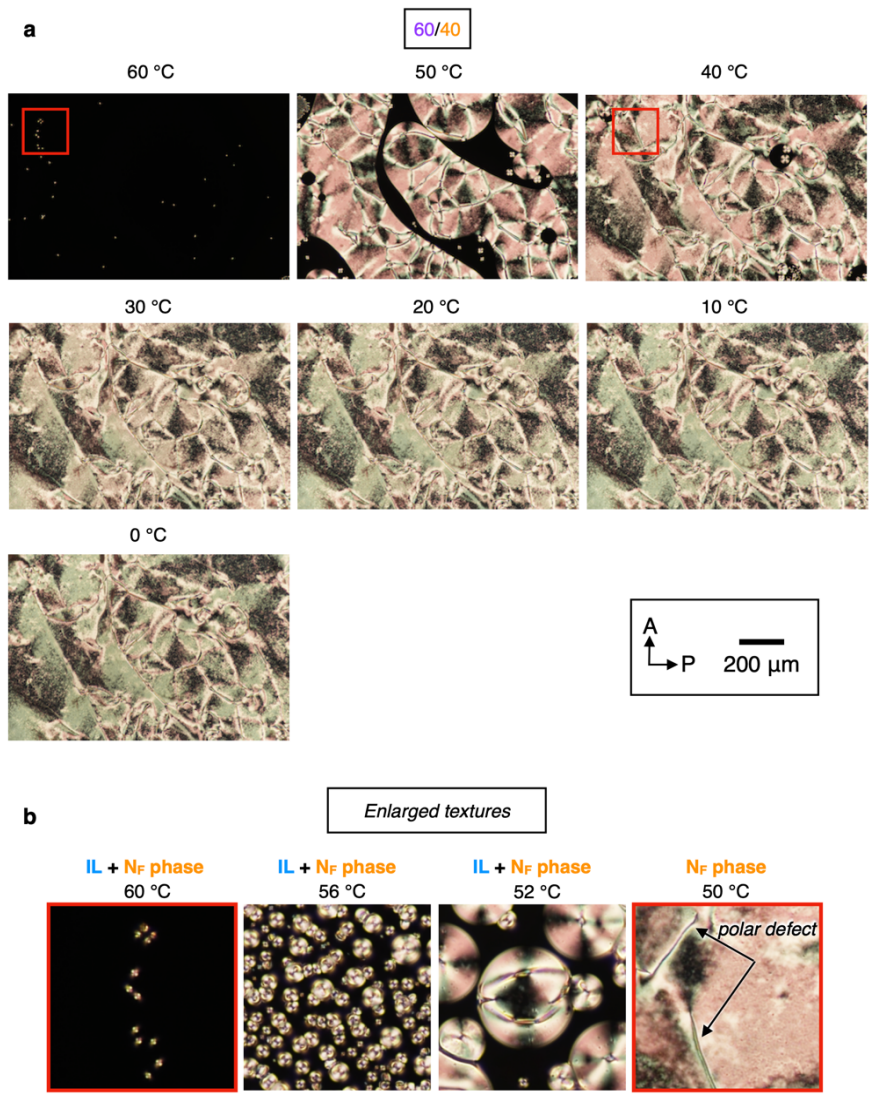

**Supplementary Fig. 13 | The original unedited version of POM images for the mixture 1/2 with *dr* = 100/0.** Evolution of POM images (a) and the corresponding enlarged textures (b) indicated in the red square in the panel (a). In the panel (b), the additional enlarged textures were inserted (*T* = 56 and 52 °C).

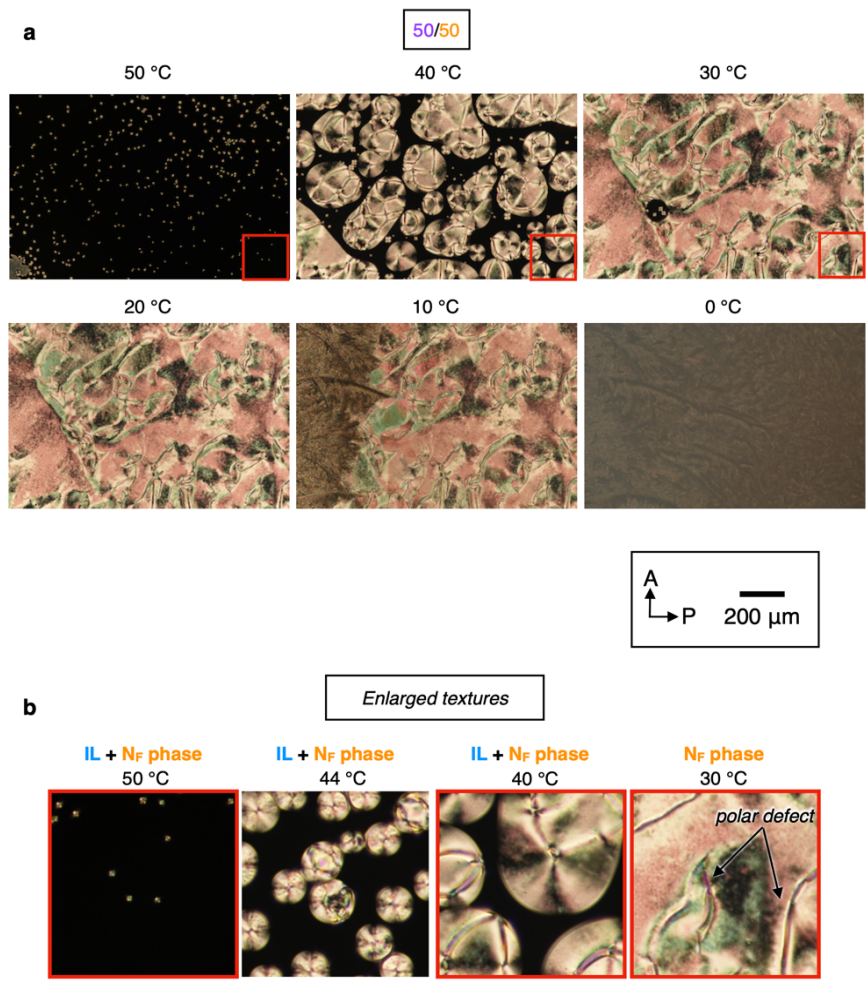

**Supplementary Fig. 14 | The original unedited version of POM images for the mixture 1/2 with *dr* = 100/0.** Evolution of POM images (a) and the corresponding enlarged textures (b) indicated in the red square in the panel (a). In the panel (b), the additional enlarged texture was inserted (*T* = 44 °C).

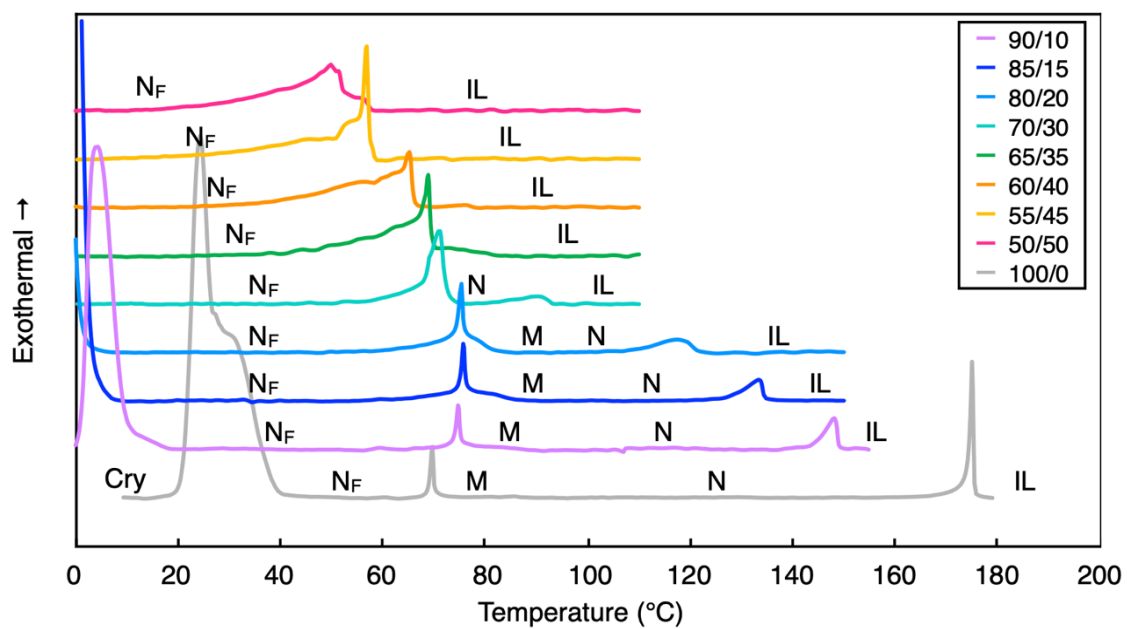

**Supplementary Fig. 9 | DSC curves of 1/2 mixture with various *dr*.** The baseline estimated by a polynomial fitting was subtracted from the original DSC curve.

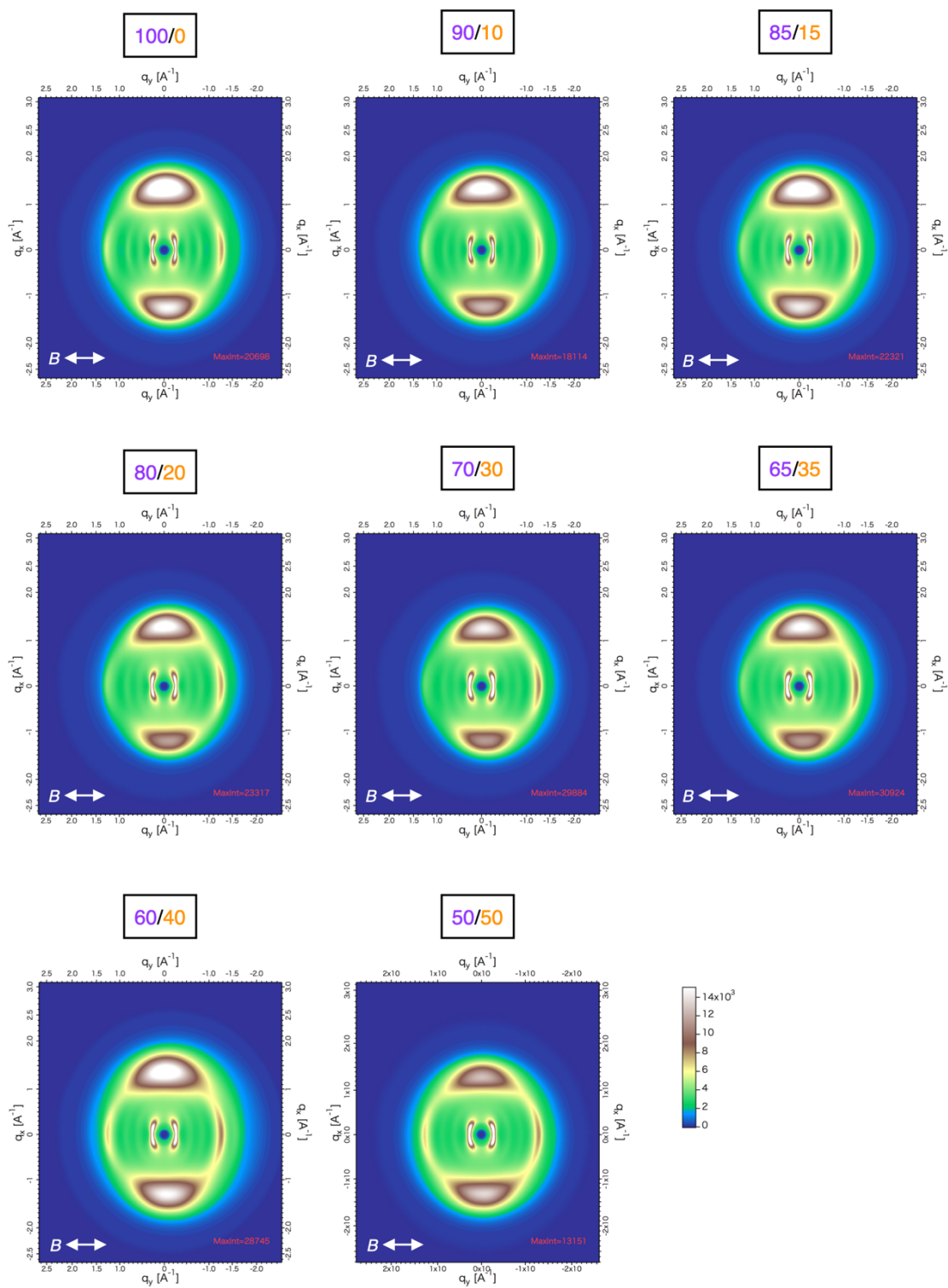

**Supplementary Fig. 10 | 2D wide-angle XRD profiles in the $N_F$ phase (10 °C below the transition point) of 1/2 mixture with various *dr*.** The arrow denotes the direction of the applied magnetic field.

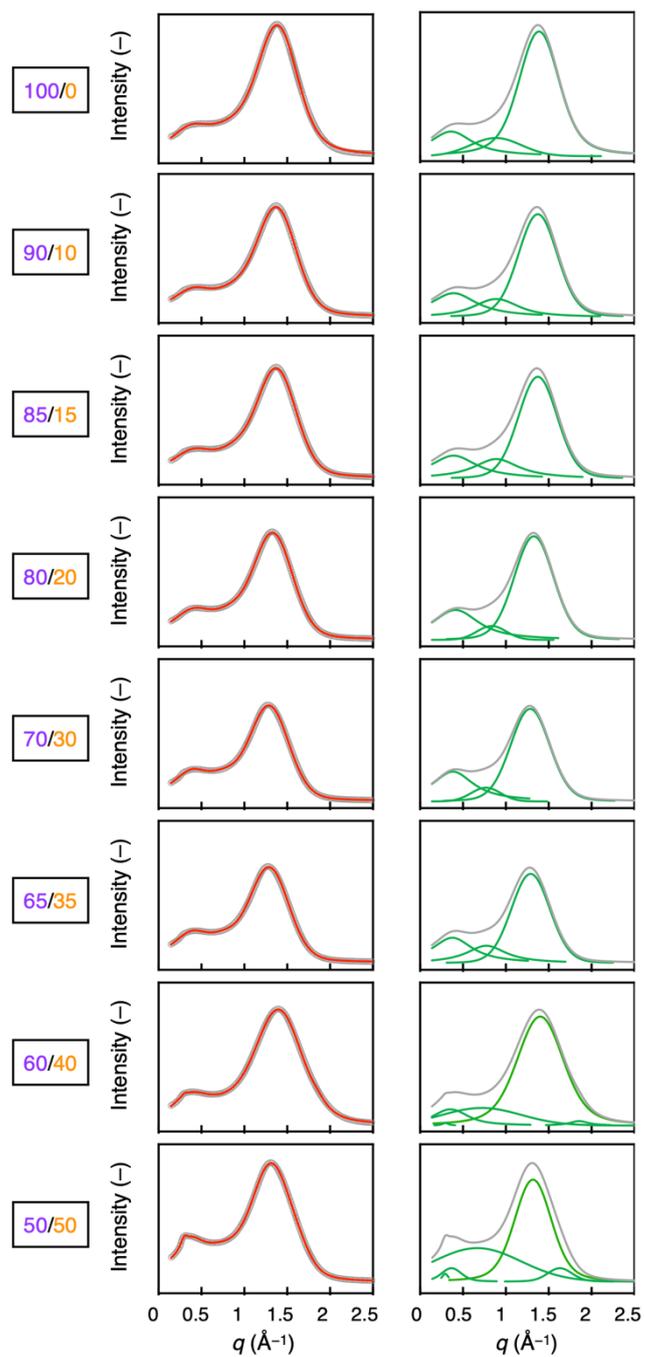

**Supplementary Fig. 11 | XRD analysis of 1/2 mixture with various *dr*.** The X-ray diffractogram on the meridional direction (normal to **n**) were fitted by Voigt function.

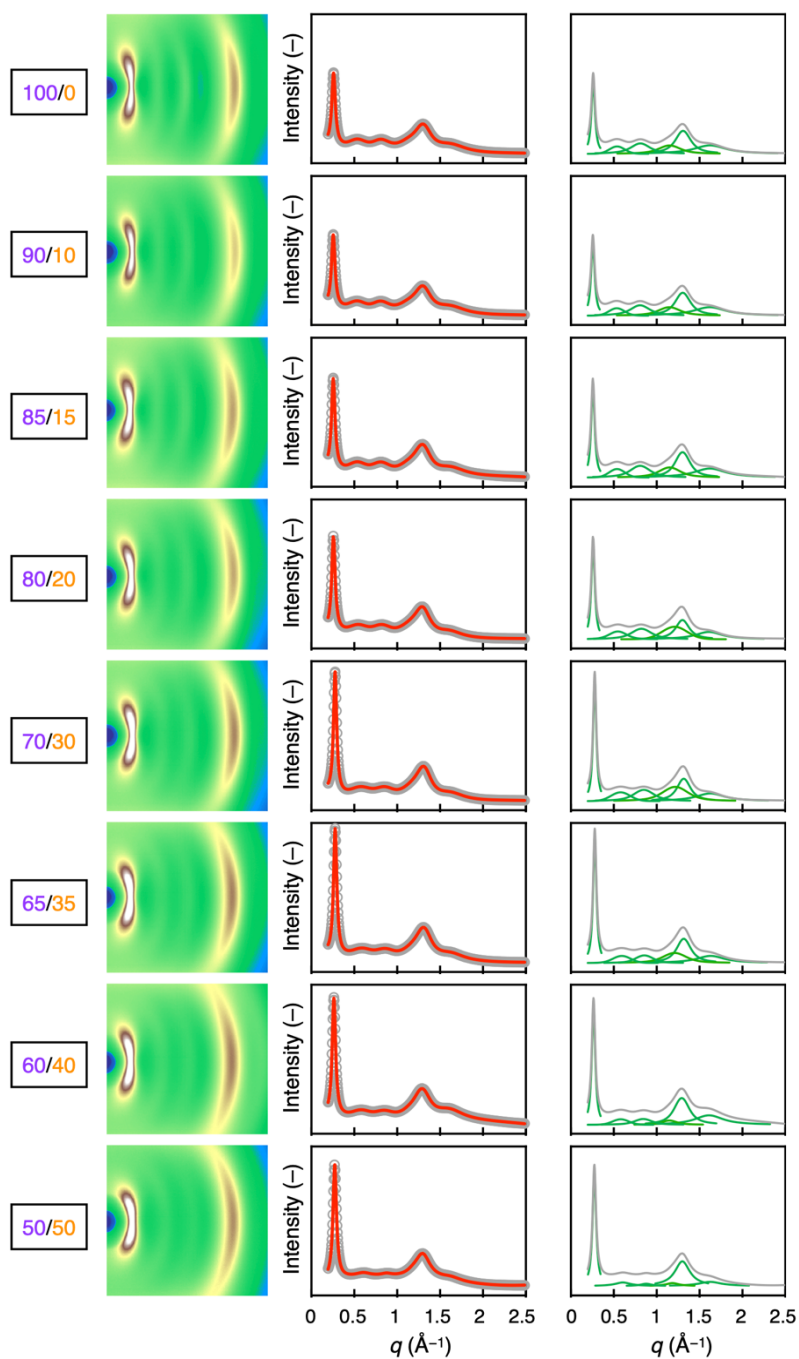

**Supplementary Fig. 12 | XRD analysis of 1/2 mixture with various *dr*.** The X-ray diffractogram on the equatorial direction (parallel to **n**) were fitted by Voigt function (except for *dr* = 50/50, which was applied by Lorentzian model).

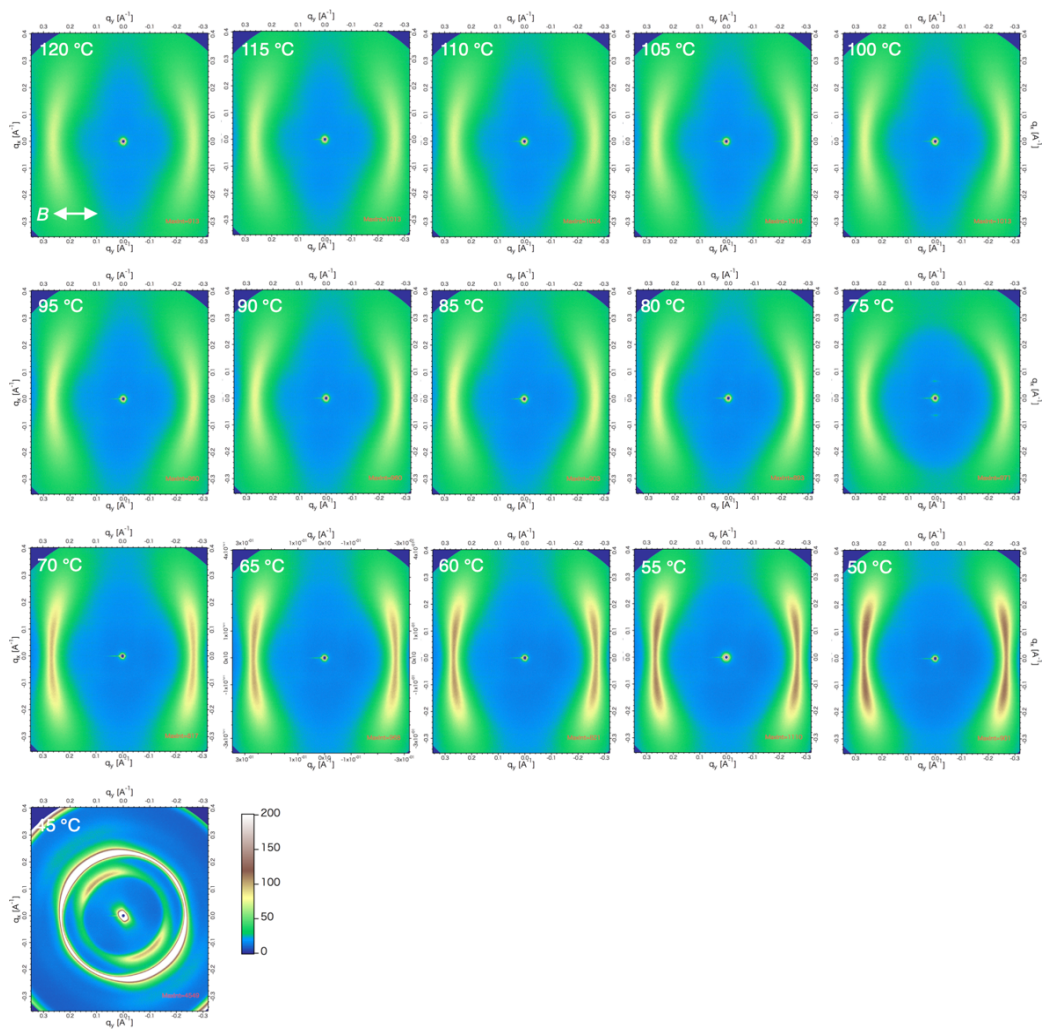

**Supplementary Fig. 13 | 2D XRD profiles of 1/2 mixture with *dr* = 100/0 at various temperatures.** The arrow denotes the direction of the applied magnetic field. Two small peaks on the meridional direction (normal to **n**) at 80–70 °C are due to the periodic modulation of the director orientation involving splay [S11].

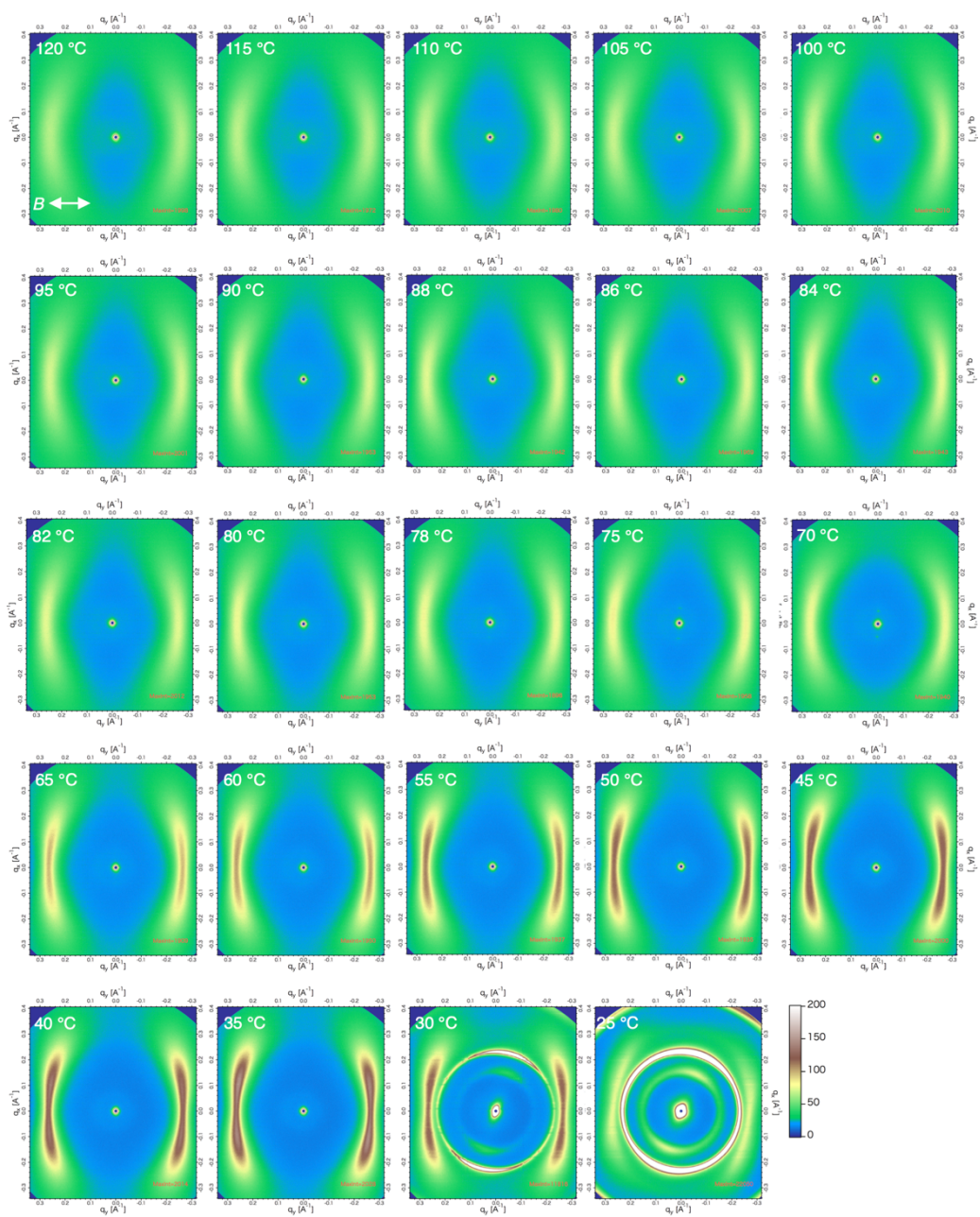

**Supplementary Fig. 14 | 2D XRD profiles of 1/2 mixture with *dr* = 90/10 at various temperatures.** The arrow denotes the direction of the applied magnetic field. Two small peaks on the meridional direction (normal to **n**) at 78–70 °C are due to the periodic modulation of the director orientation involving splay [S11].

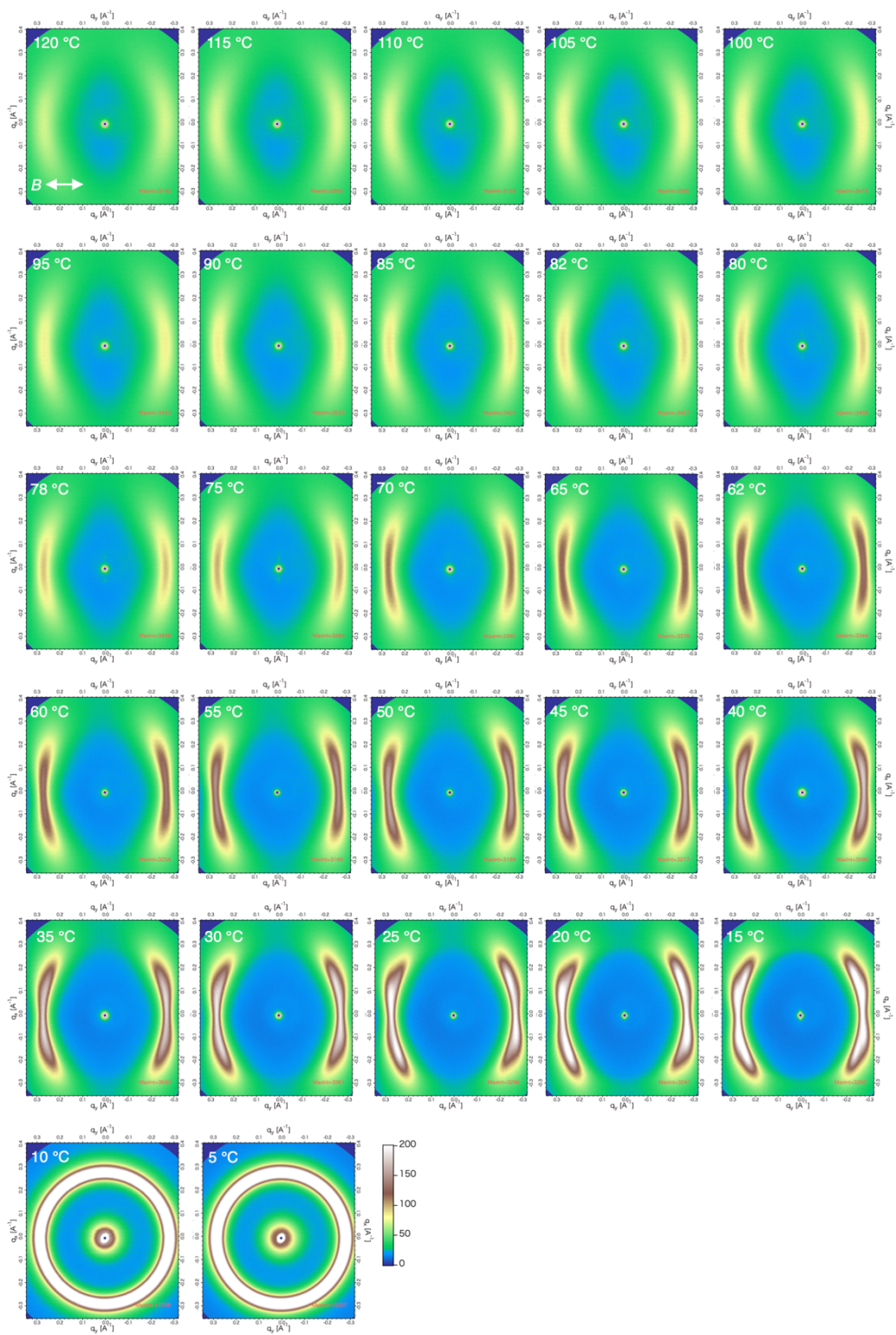

**Supplementary Fig. 15 | 2D XRD profiles of 1/2 mixture with *dr* = 85/15 at various temperatures.** The arrow denotes the direction of the applied magnetic field. Two peaks on the meridional direction (normal to **n**) at 80–75 °C are due to the periodic modulation of the director orientation involving splay [S11].

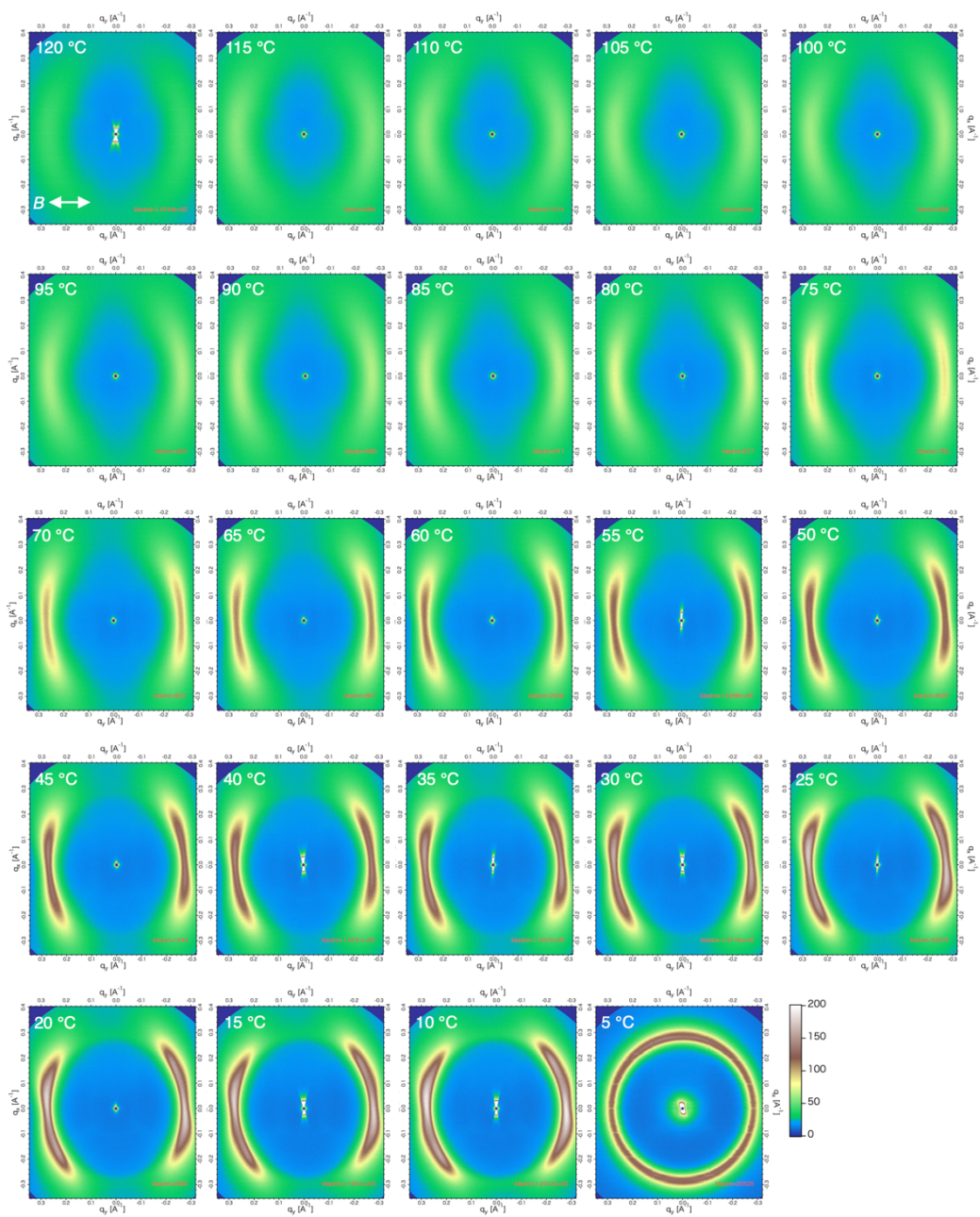

**Supplementary Fig. 16 | 2D XRD profiles of 1/2 mixture with *dr* = 80/20 at various temperatures.** The arrow denotes the direction of the applied magnetic field. Two small peaks on the meridional direction (normal to **n**) at 80 °C are due to the periodic modulation of the director orientation involving splay [S11].

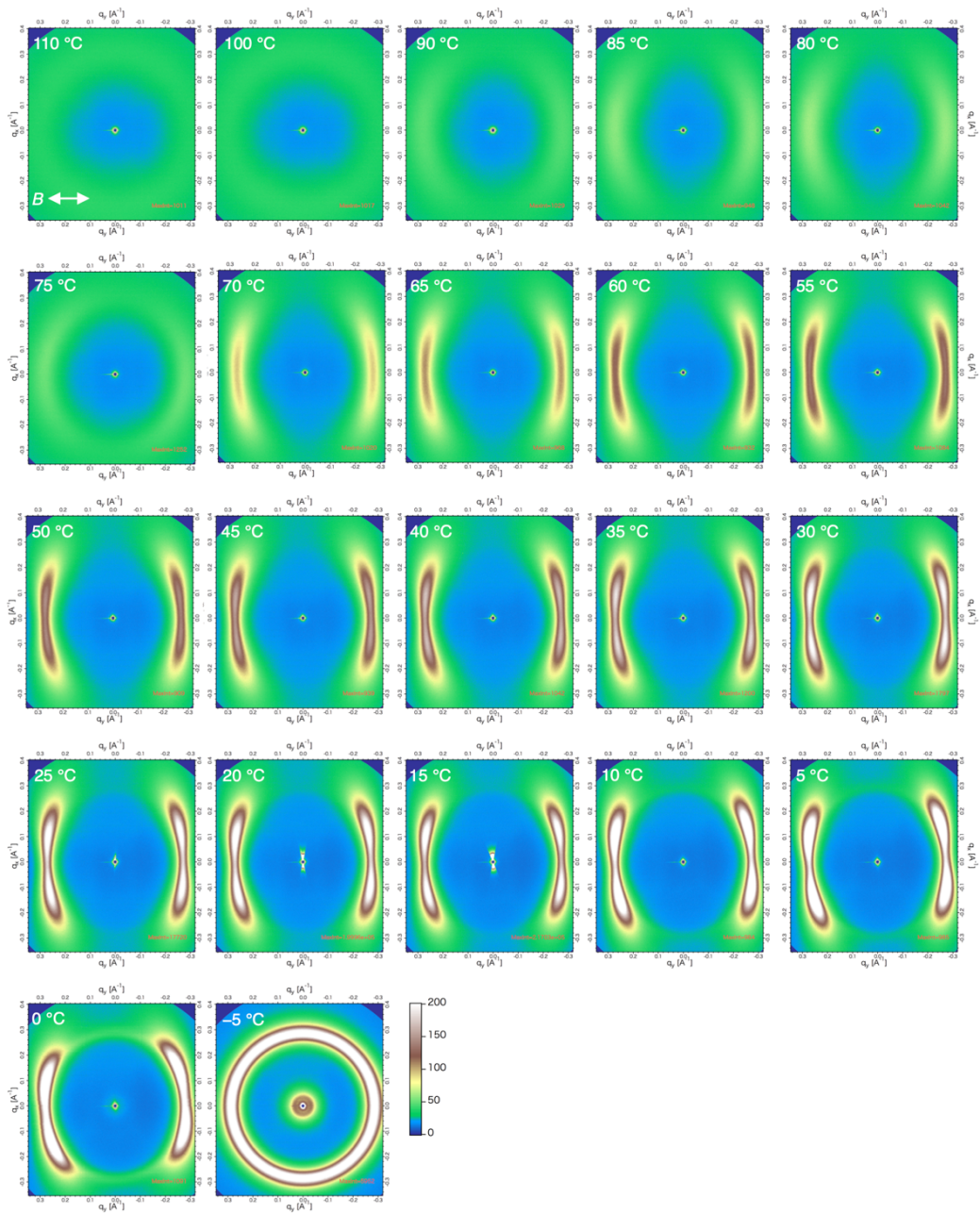

**Supplementary Fig. 17 | 2D XRD profiles of 1/2 mixture with *dr* = 70/30 at various temperatures.** The arrow denotes the direction of the applied magnetic field.

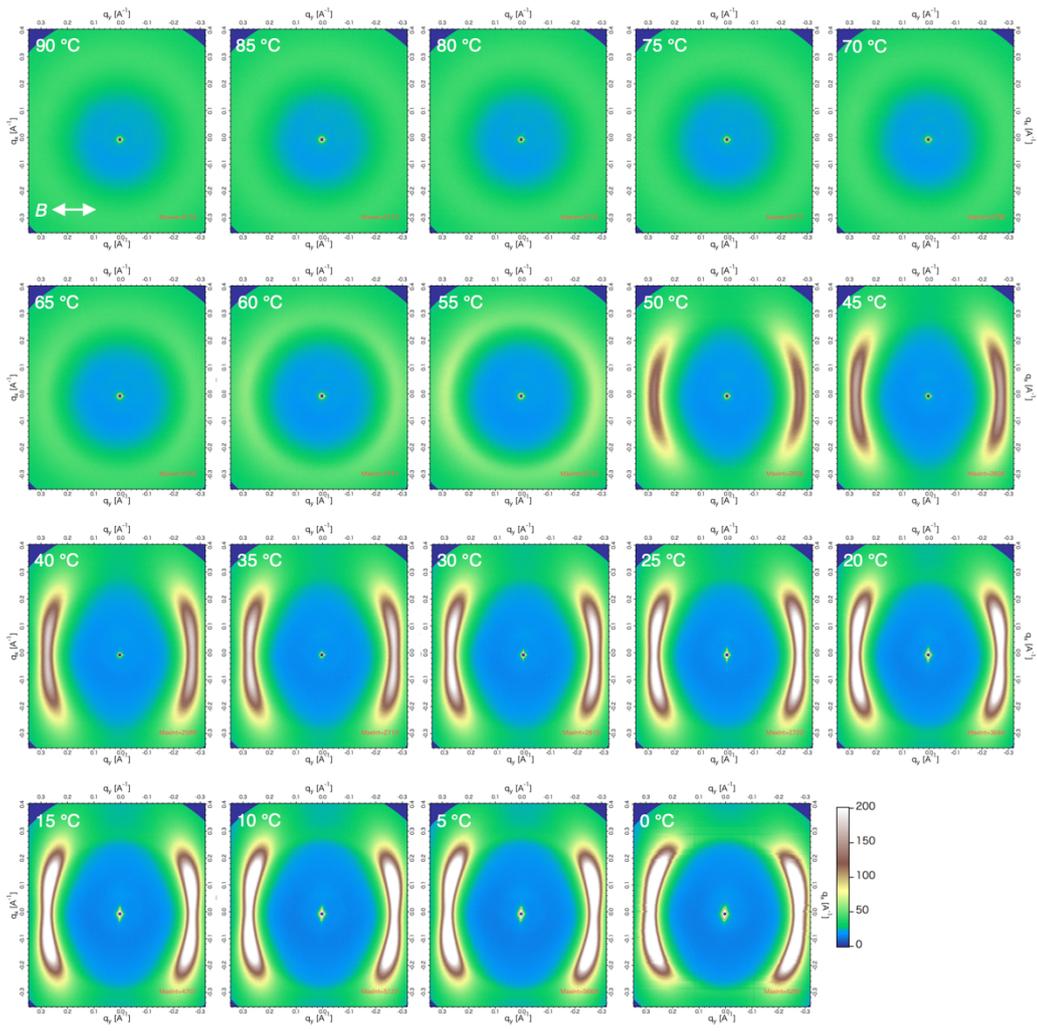

**Supplementary Fig. 18 | 2D XRD profiles of 1/2 mixture with *dr* = 65/35 at various temperatures.** The arrow denotes the direction of the applied magnetic field.

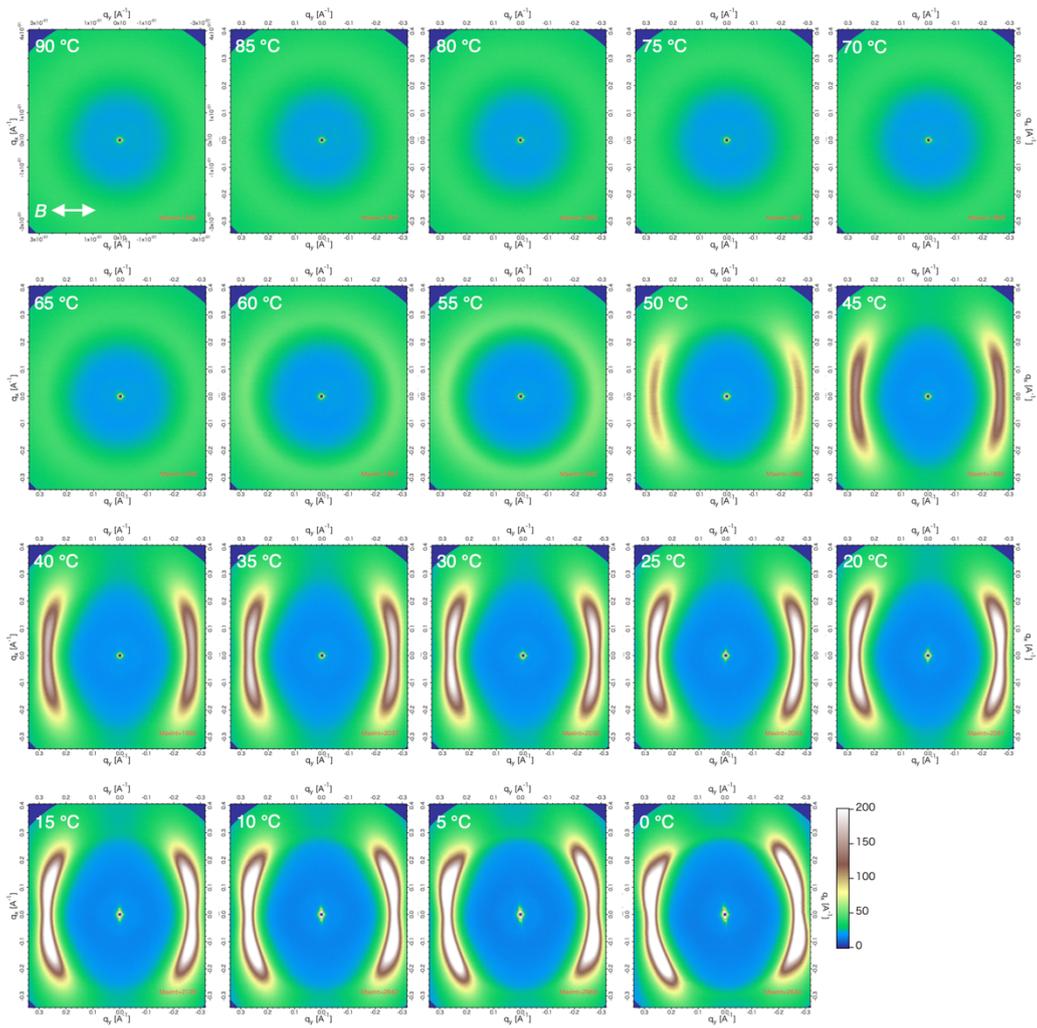

**Supplementary Fig. 19 | 2D XRD profiles of 1/2 mixture with *dr* = 60/40 at various temperatures.** The arrow denotes the direction of the applied magnetic field.

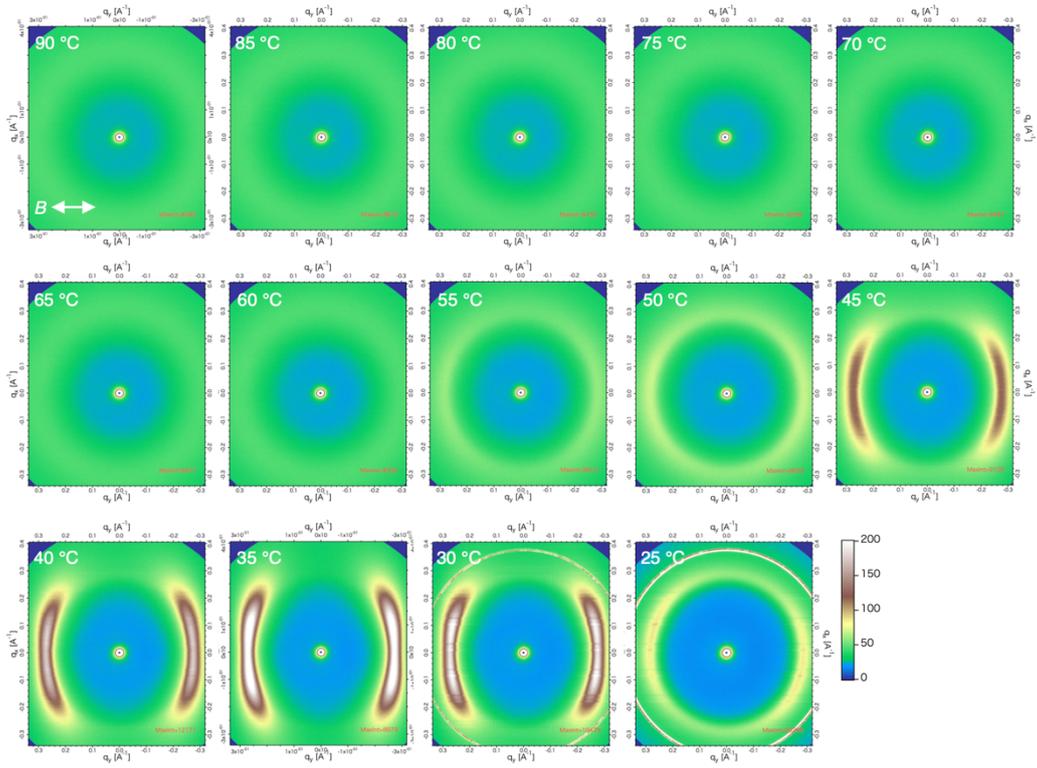

**Supplementary Fig. 20 | 2D XRD profiles of 1/2 mixture with *dr* = 50/50 at various temperatures.** The arrow denotes the direction of the applied magnetic field.

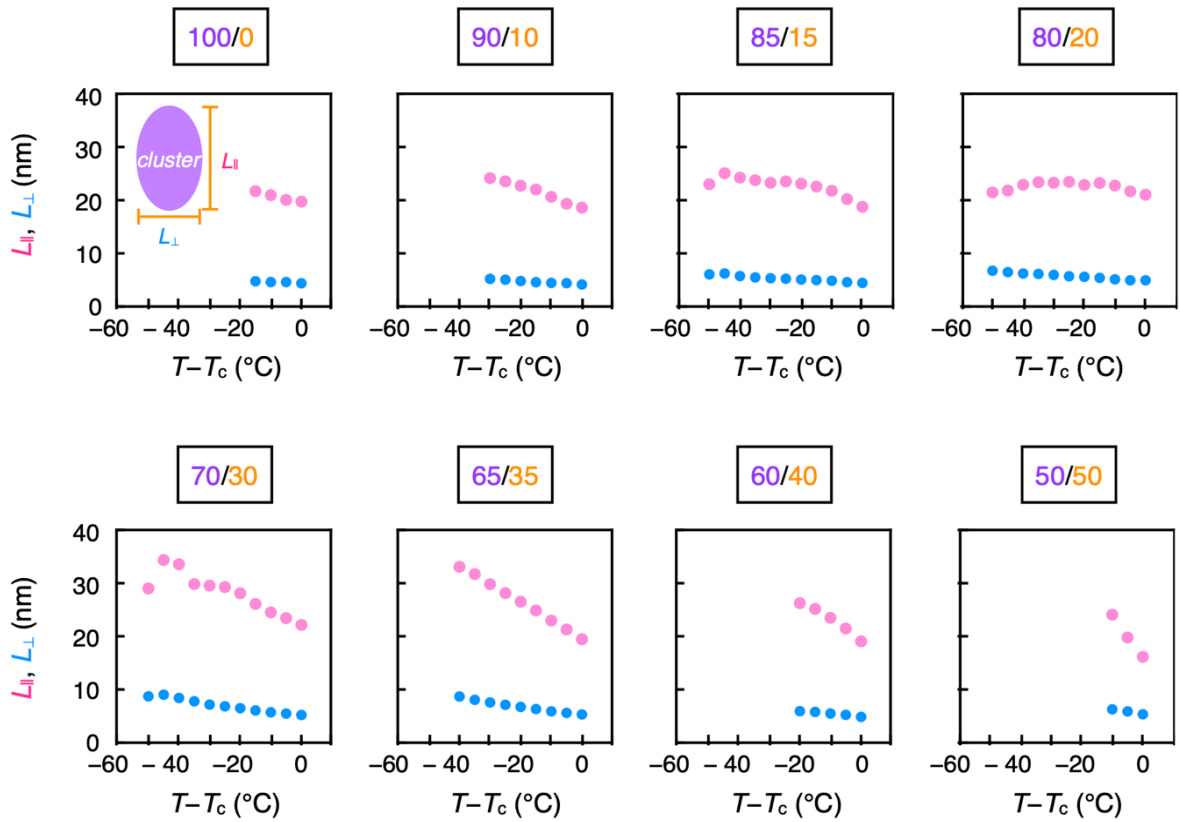

**Supplementary Fig. 21 | The cluster dimension ($L_\parallel$ (Pink) and $L_\perp$ (Blue)) *vs.* temperature for 1/2 mixture with various *dr*.**

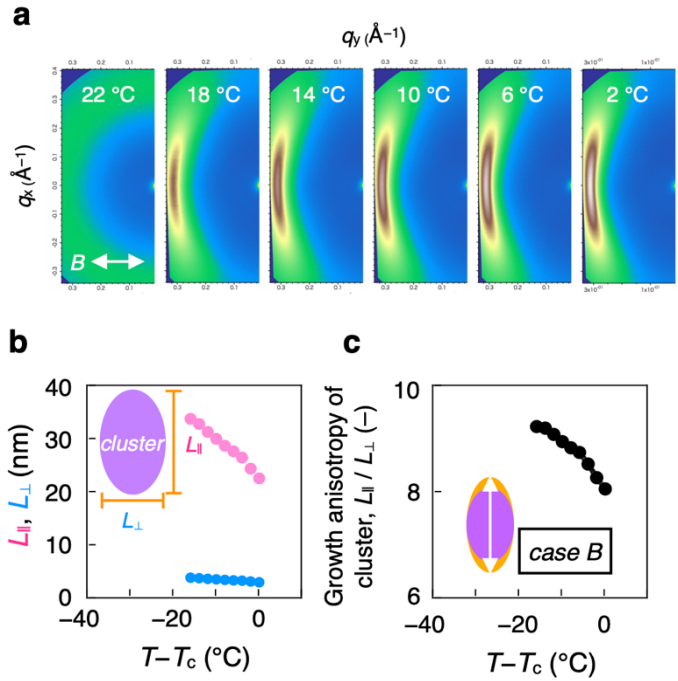

**Supplementary Fig. 22 | Characterization of cybotactic cluster in the $N_F$ phase of UUQU-4-N. a**, Evolution of 2D X-ray diffractogram in various temperature. **b**, The cluster dimension ($L_\parallel$ (Pink) and $L_\perp$ (Blue)) *vs.* temperature. **c**, The relationship between growth anisotropy of cluster and its temperature-dependent. The cluster growth seems to undergo case B (see the panel (**e**) in Fig. 3 in the main text).

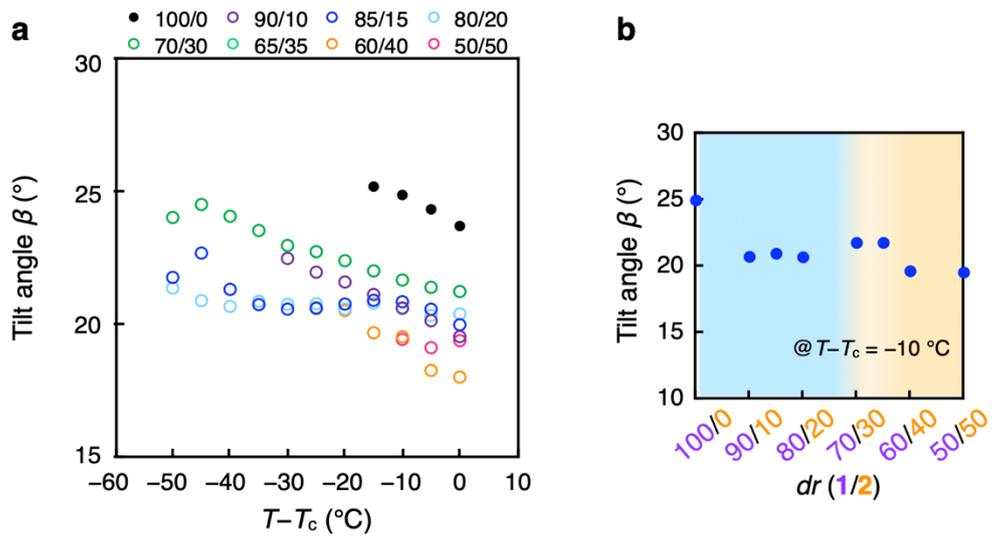

**Supplementary Fig. 23 | Tilt angle *β* *vs.* temperature (a) and *dr* (b) for 1/2 mixture with various *dr*.** See legend for correspondences between samples and colors.

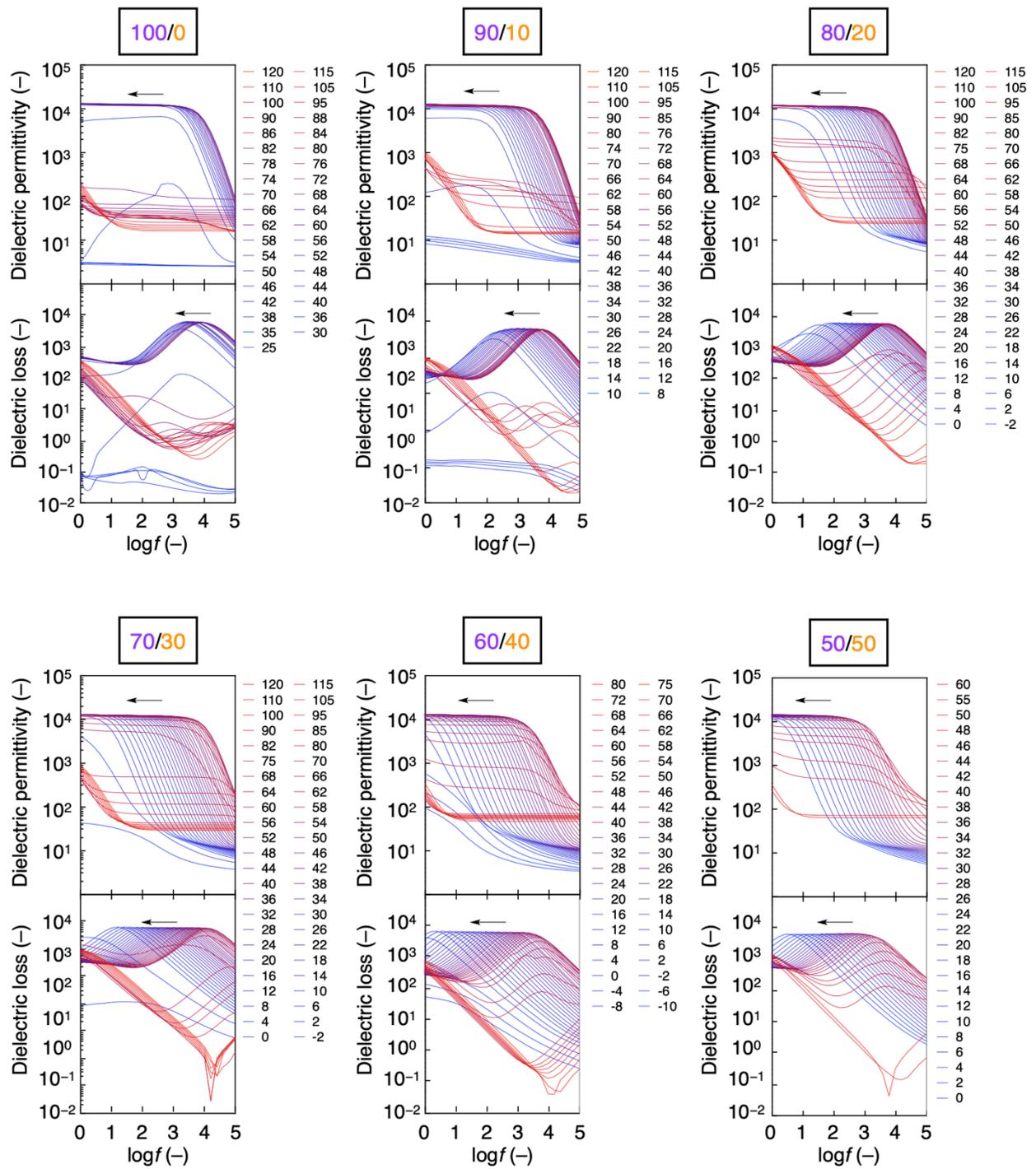

**Supplementary Fig. 24 | Dielectric spectra of 1/2 mixture with various *dr*.** See legend for correspondences between temperatures and colors.

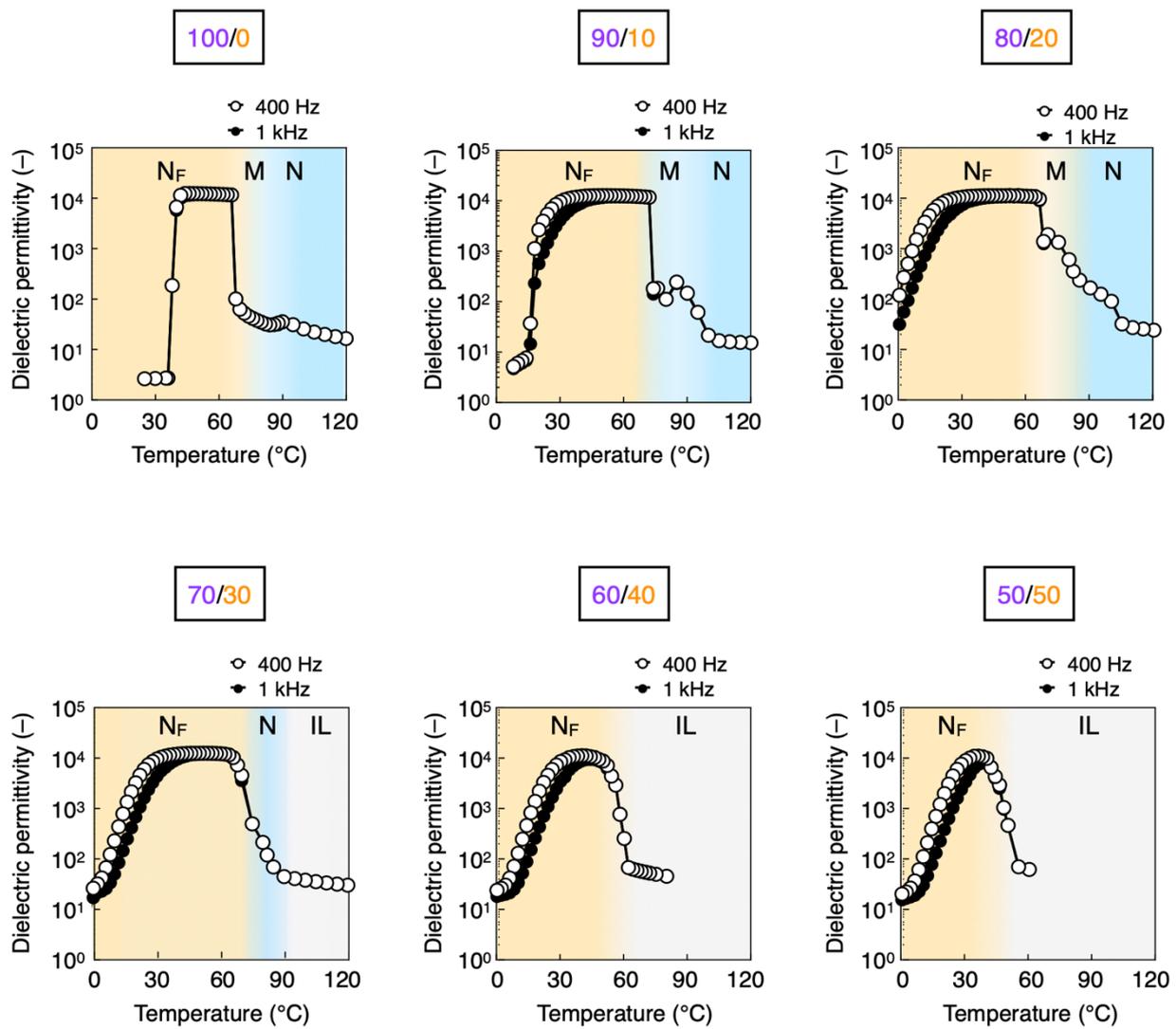

**Supplementary Fig. 25 | Dielectric permittivity *vs.* temperature for 1/2 mixture with various *dr*.** Open and closed circles stand respectively for 400 Hz and 1 kHz.

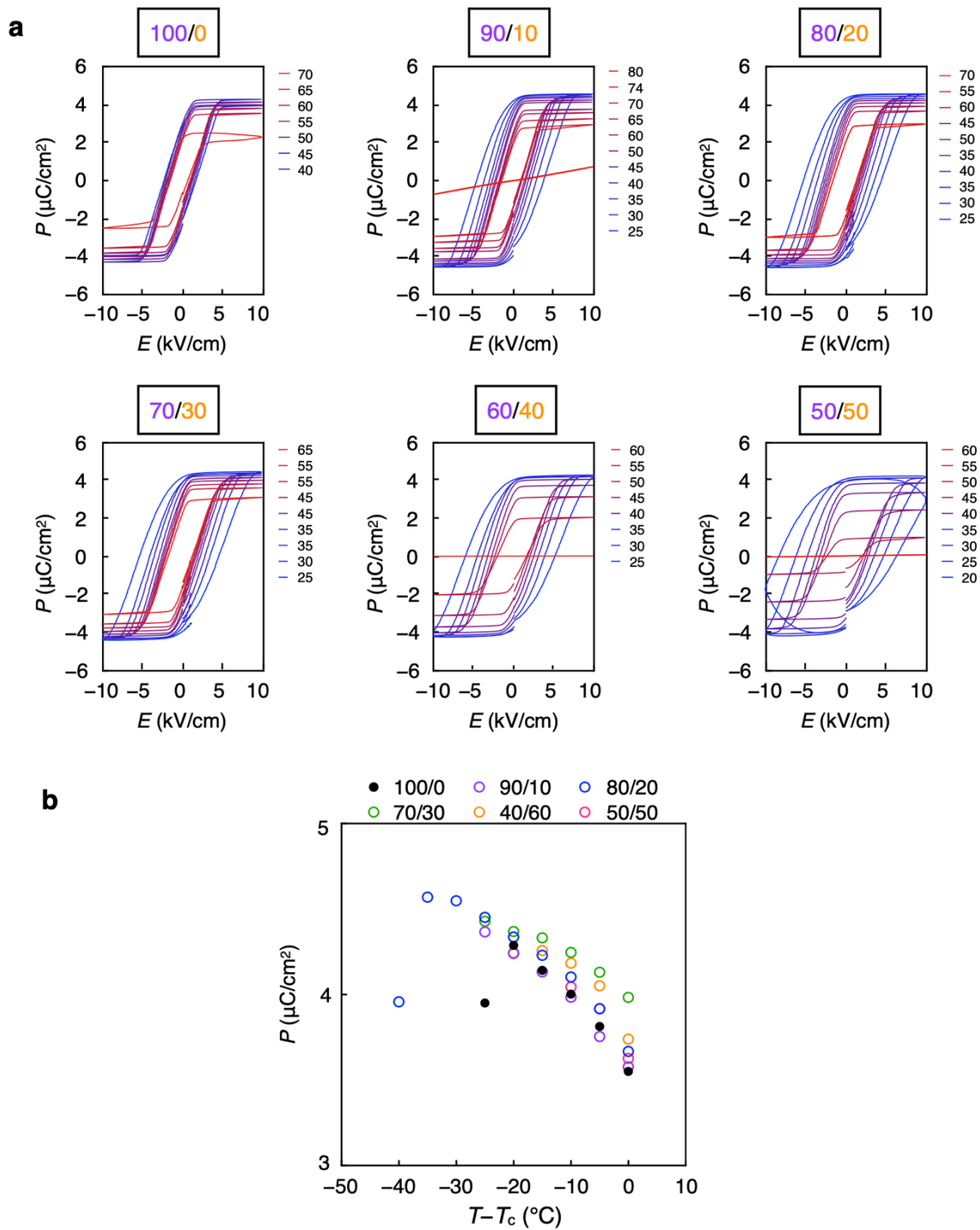

**Supplementary Fig. 26 | *P-E* hysteresis loop for 1/2 mixture with various *dr* (a) and *P vs.* temperature (b).** See legend for correspondences of the colors of the plots.

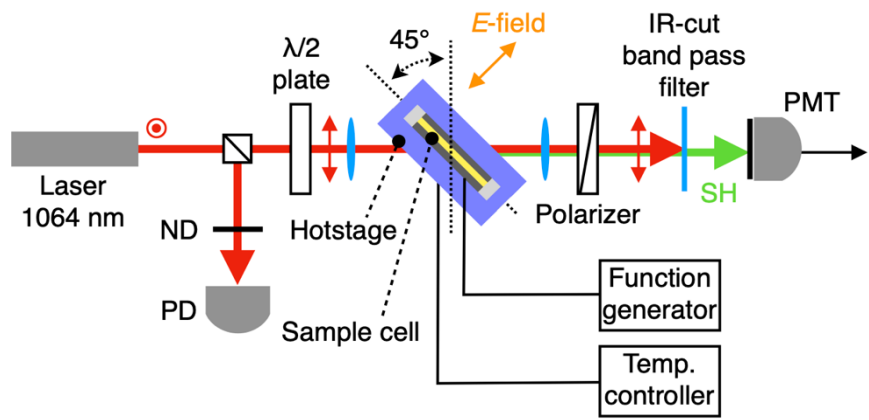

**Supplementary Fig. 27 | The optical setup for SHG studies.** The electric field ($E$-field) was applied normal to the sample cell (thickness: 5 μm), generating the *p*-in/*p*-out polarization combination.

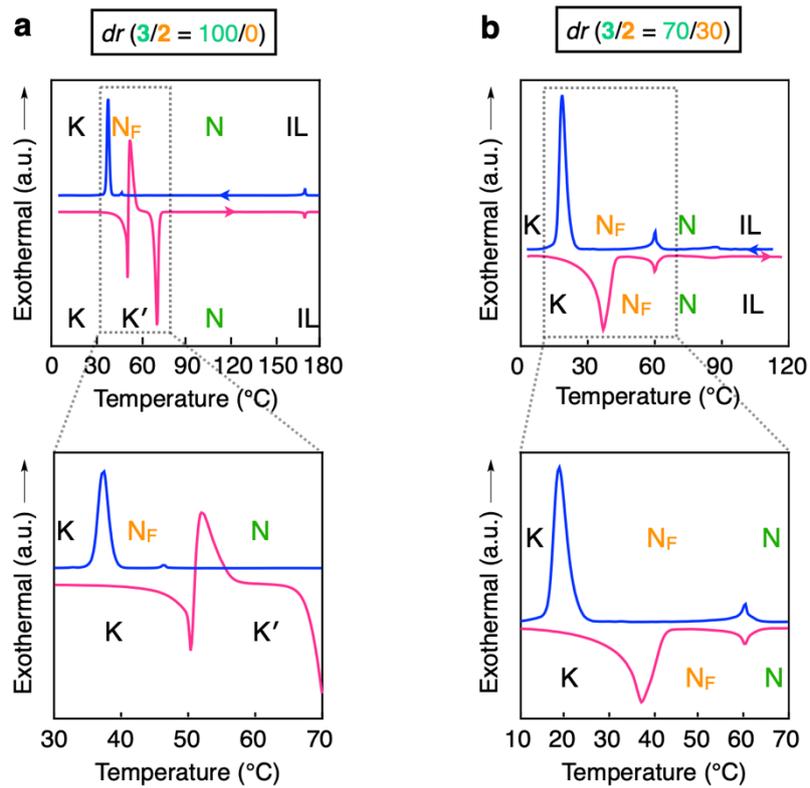

**Supplementary Fig. 28 | DSC curves for 3/2 mixture with *dr* = 100/0 (a) and =70/30 (b).**
Red and blue traces are for heating and cooling, respectively.

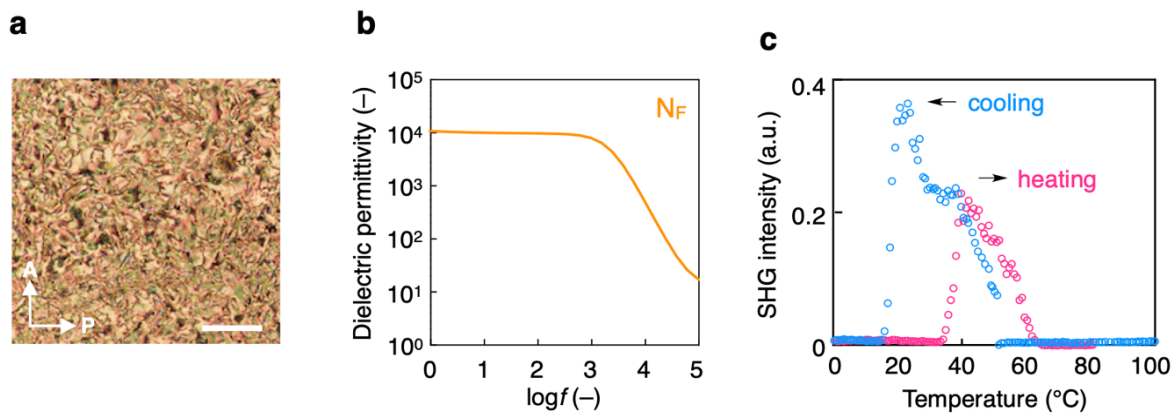

**Supplementary Fig. 29 | Characterization of $N_F$ phase for the 3/2 mixture. a**, POM image at 25 °C taken under crossed polarizers on cooling. Scale bar: 100 μm. **b**, Dielectric permittivity *vs.* frequency at 45 °C. **c**, SHG *vs.* temperature on cooling and heating.

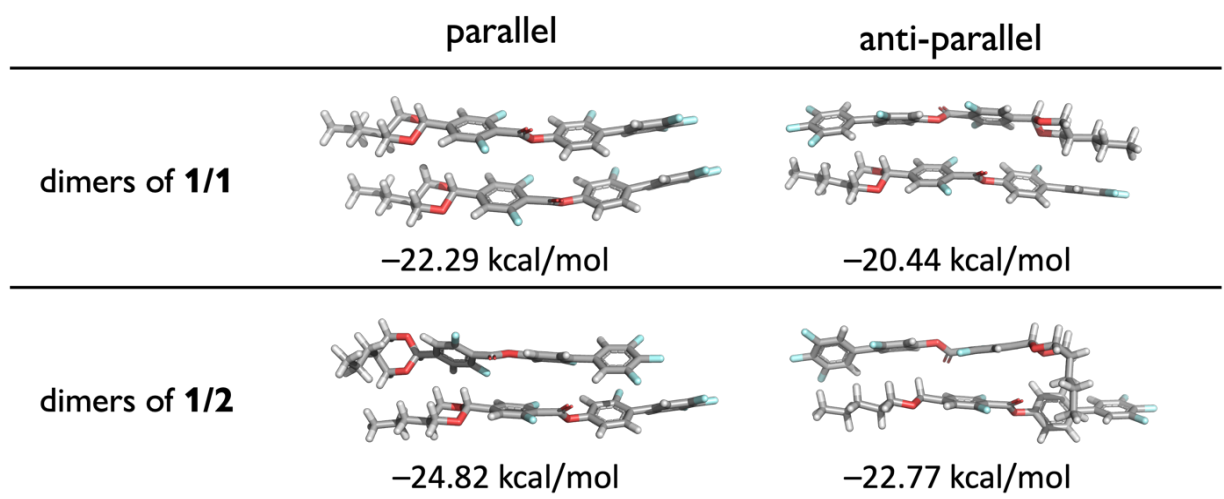

**Supplementary Fig. 30 | Interaction energies of dimers of 1/1 in parallel and anti-parallel and 1/2 in parallel and anti-parallel obtained from DFT calculations.**

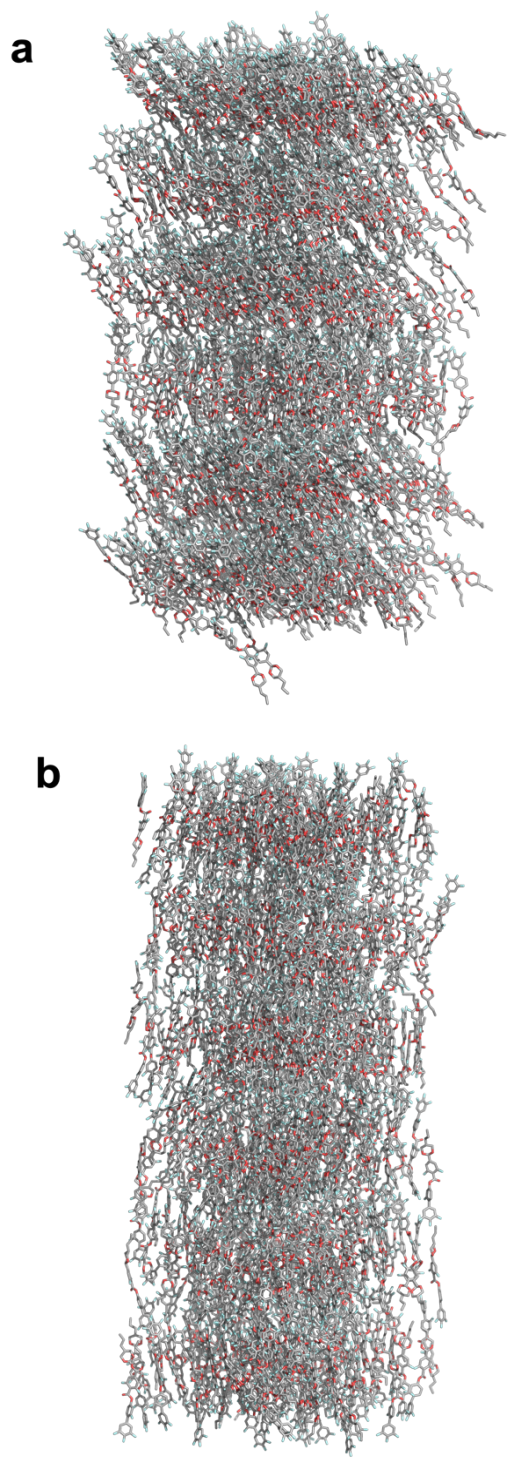

**Supplementary Fig. 31 | MD simulation snapshot of System 1 (a) and System 2 (b) after 300 ns equilibration run.** Hydrogen atoms are not shown for clarity.

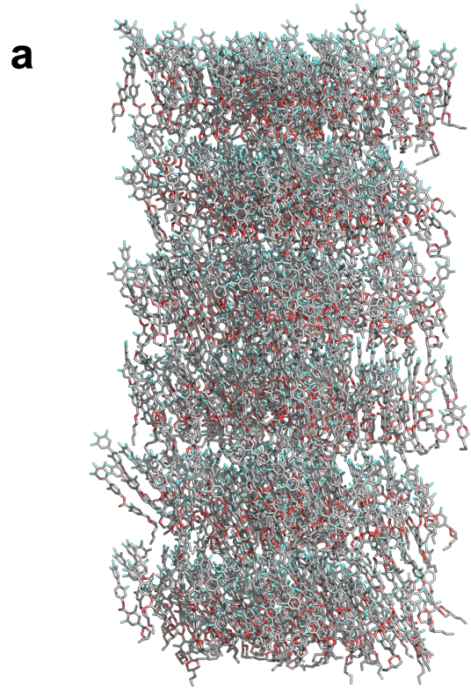

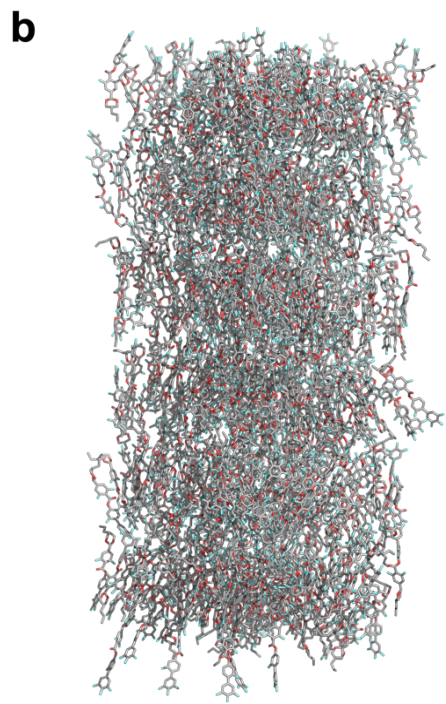

**Supplementary Fig. 32 | MD simulation snapshot of System 3 (a) and System 4 (b) after 300 ns equilibration run.** Hydrogen atoms are not shown for clarity.

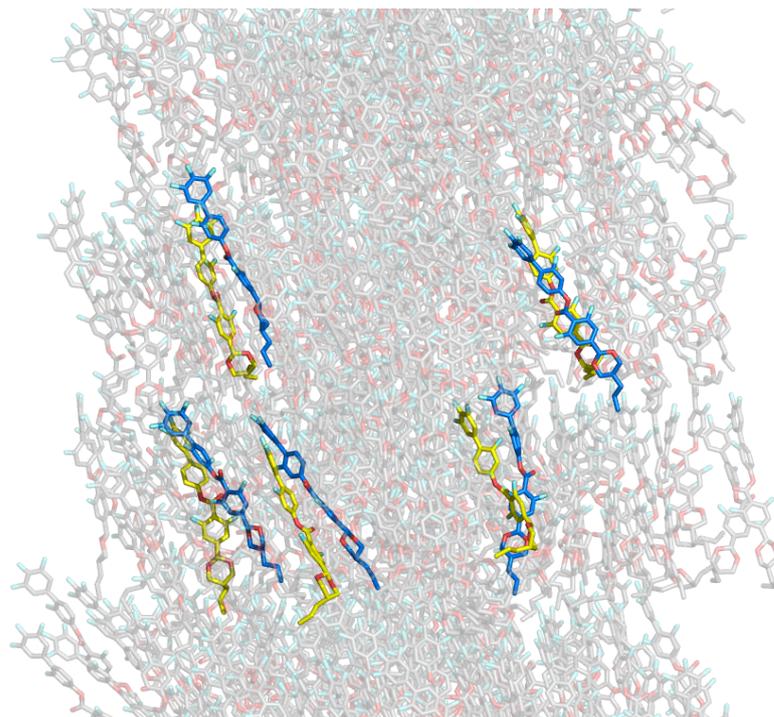

**Supplementary Fig. 33 | Close-up of the MD simulation snapshot of System 3 after 300 ns equilibration run.** Sticks model in blue and yellow represent **1** and **2**, respectively.

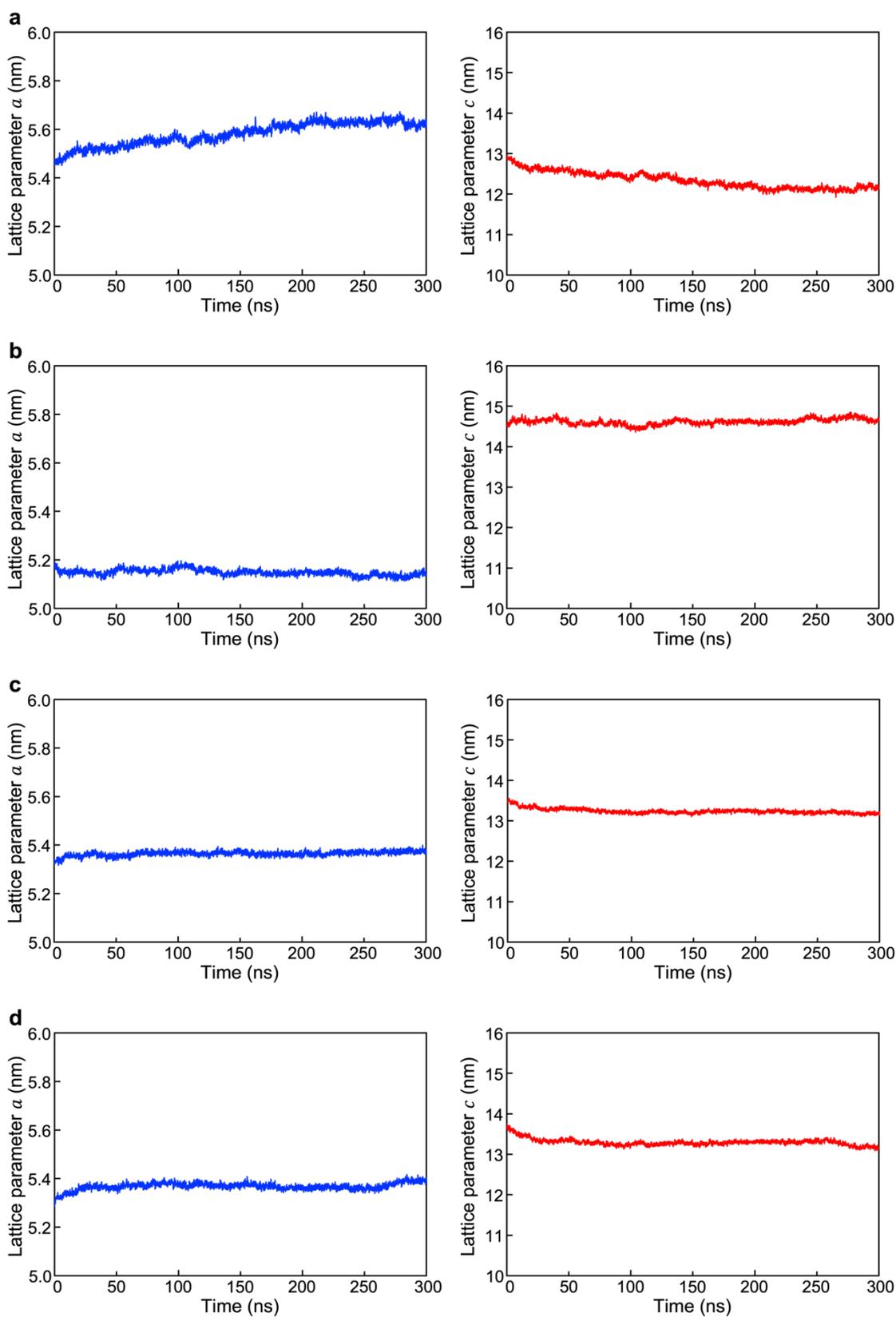

**Supplementary Fig. 34 | Time dependence of lattice parameters of rectangular MD simulation box. a,** the *a* and *c* axes for **System 1**. **b,** the *a* and *c* axes for **System 2**. **c,** the *a* and *c* axes for **System 3**. **d,** the *a* and *c* axes for **System 4**.

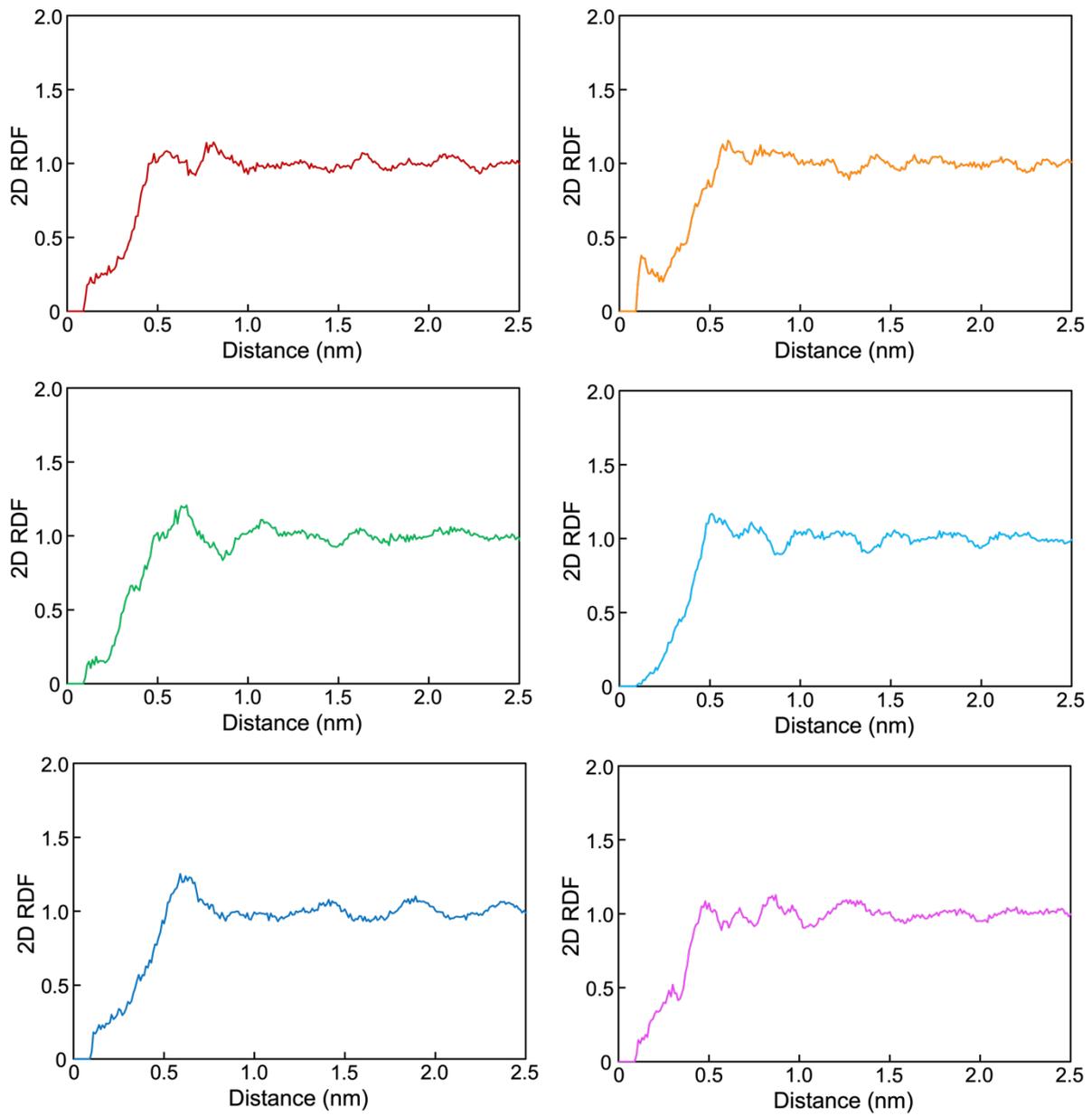

**Supplementary Fig. 35 | Two-dimensional RDF for each layer of System 1.**

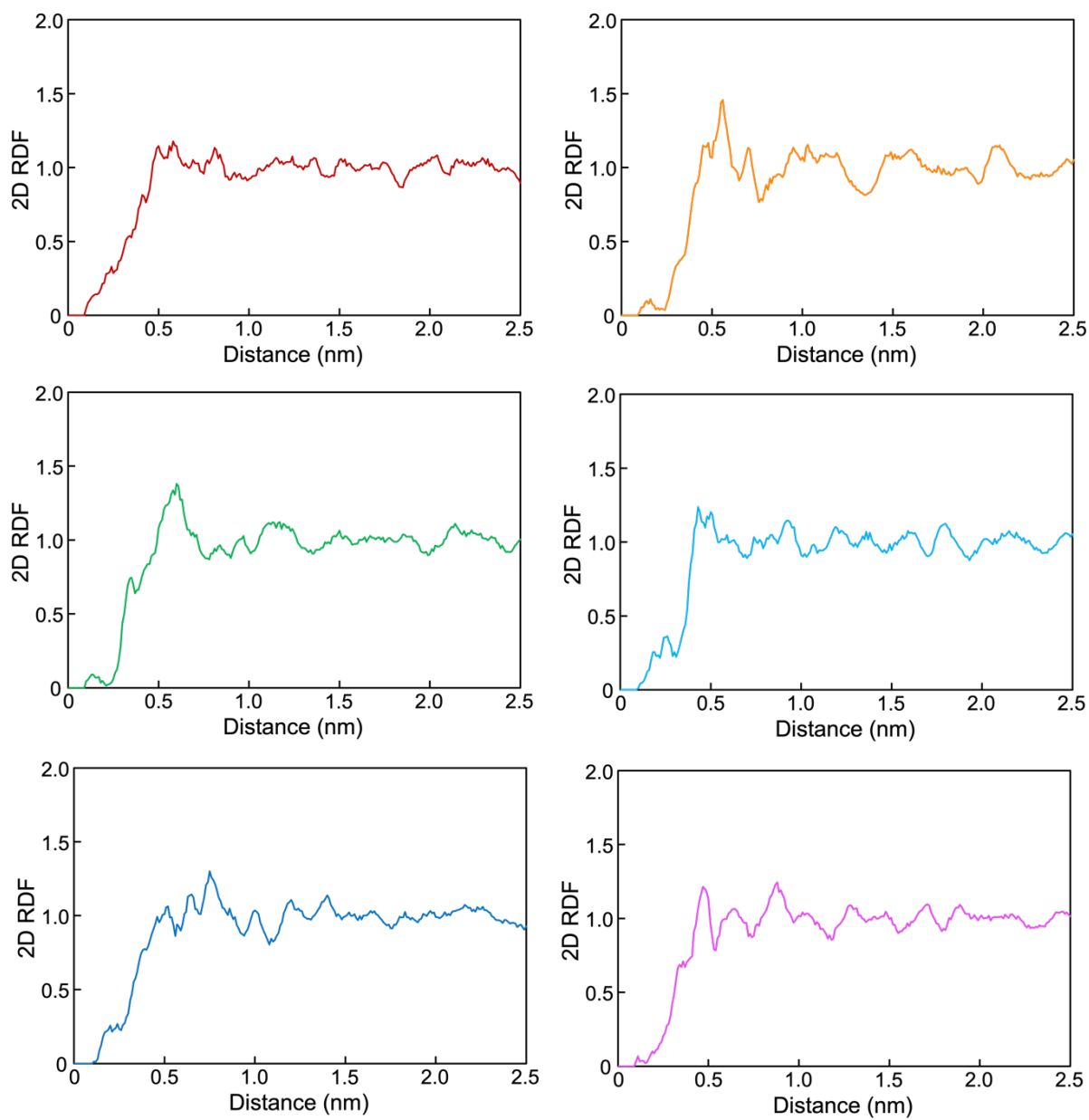

**Supplementary Fig. 36 | Two-dimensional RDF for each layer of System 3.**

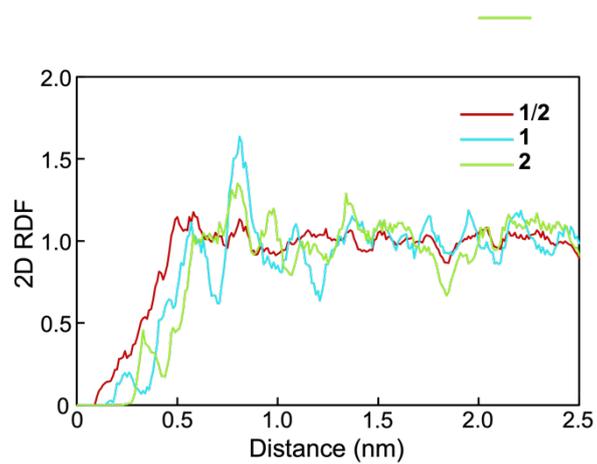

**Supplementary Fig. 37 | Two-dimensional RDF of 1 only, 2 only, and both 1 and 2 in the bottom layer for System 3.**

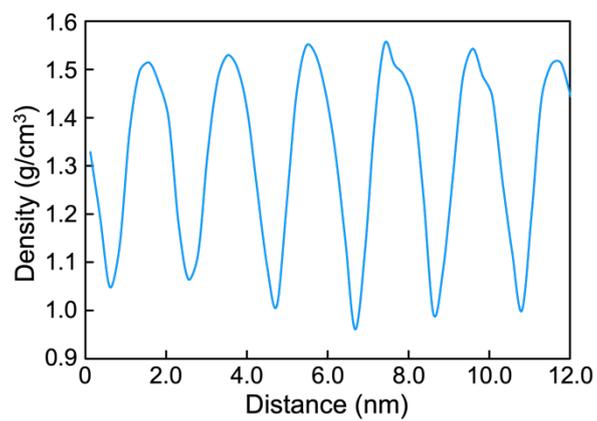

**Supplementary Fig. 38 | Density profile along the layer normal of System 1.**

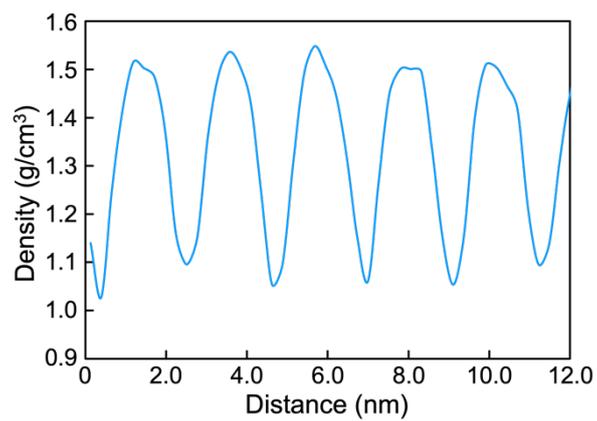

**Supplementary Fig. 39 | Density profile along the layer normal of System 3.**

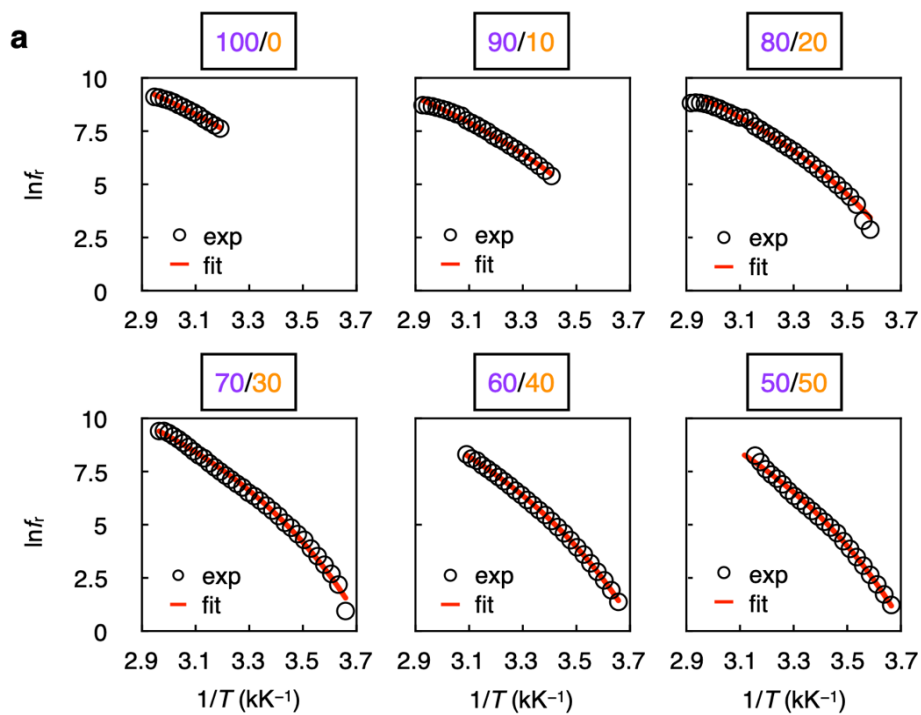
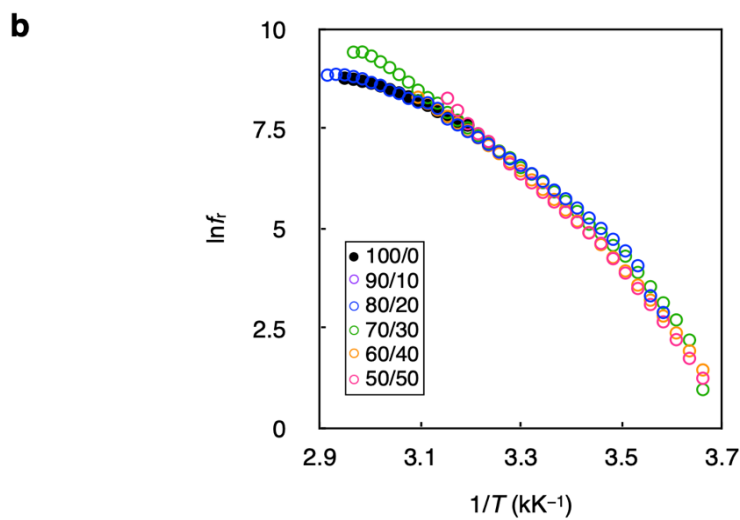

**Supplementary Fig. 40 | Temperature-dependent dielectric studies. a**, Relaxation frequency *vs.* 1/$T$ for **1**/**2** mixture with various *dr*. **b**, Pseudo-master curve construction using data in the panel (**a**).

**Supplementary Table 1 | Thermal properties of the 1/2 mixture with various *dr*.**

| dr | $\Delta H_x$[a] [kJ mol$^{-1}$] | $\Delta H_y$[b] [kJ mol$^{-1}$] | $\Delta H_{tot}$[c] [kJ mol$^{-1}$] |
|---|---|---|---|
| 100/0 | 0.750 | 0.281 | 1.03 |
| 90/10 | 0.480 | 0.526 | 1.01 |
| 85/15 | 0.383 | 0.704 | 1.09 |
| 80/20 | 0.367 | 0.970 | 1.34 |
| 70/30 | 0.219 | 1.34 | 1.56 |
| 65/35 | - | 2.22 | 2.22 |
| 60/40 | - | 2.73 | 2.73 |
| 55/45 | - | 2.95 | 2.95 |
| 50/50 | - | 2.94 | 2.94 |

a) $\Delta H_x$ denotes enthalpy of IL–N phase transition, b) $\Delta H_y$ denotes enthalpy of IL–N$_F$ phase transition, c) $\Delta H_{tot} = \Delta H_x + \Delta H_y$. Note: the enthalpy of N–N$_F$ was omitted because it was negligibly smaller than those of IL–N and IL–N$_F$ phase transition.

**Supplementary Table 2 | Average values of total energy, tilt angle, and layer spacing over the last 20 ns.**

| System | Total energy [kJ mol$^{-1}$] | Tilt angle [deg] | Layer spacing [nm] |
|---|---|---|---|
| 1 | $2.85 \times 10^4$ | 30.8 | 2.03 |
| 2 | $3.05 \times 10^4$ | 1.87 | 2.45 |
| 3 | $-2.88 \times 10^4$ | 18.1 | 2.20 |
| 4 | $-2.64 \times 10^4$ | 5.31 | 2.20 |

**Supplementary References**